\documentclass[aps,prb,twocolumn,longbibliography,floatfix,superscriptaddress]{revtex4-2}
\usepackage{color}
\usepackage{amsmath}
\usepackage{amssymb}
\usepackage{bbold}
\usepackage{graphicx}
\usepackage{natbib}
\usepackage{enumitem}
\usepackage[normalem]{ulem}
\usepackage{subfloat}
\usepackage[caption=false,singlelinecheck=off]{subfig}

\usepackage[colorlinks=true,linkcolor=blue,urlcolor=blue,citecolor=blue]{hyperref}
\hypersetup{colorlinks=true}
\usepackage[all]{hypcap}

\usepackage{upgreek}

\newcommand{\be}{\begin{equation}}
\newcommand{\ee}{\end{equation}}
\newcommand{\bea}{\begin{eqnarray}}
\newcommand{\eea}{\end{eqnarray}}

\begin{document}

\title{\mbox{Contacts, equilibration, and interactions in fractional quantum Hall edge transport}}

\author{C. Sp\r{a}nsl\"{a}tt}
\affiliation{\mbox{Department of Microtechnology and Nanoscience, Chalmers Institute of Technology, S-412 96 Göteborg, Sweden}}
\affiliation{\mbox{Institute for Quantum Materials and Technologies, Karlsruhe Institute of Technology, 76021 Karlsruhe, Germany}}
\affiliation{\mbox{Institut f\"ur Theorie der Kondensierten Materie, Karlsruhe Institute of Technology, 76128 Karlsruhe, Germany}}
\author{Yuval Gefen}
\affiliation{Department of Condensed Matter Physics, Weizmann Institute of Science, Rehovot 76100, Israel}
\affiliation{\mbox{Institute for Quantum Materials and Technologies, Karlsruhe Institute of Technology, 76021 Karlsruhe, Germany}}
\author{I. V. Gornyi}
\affiliation{\mbox{Institute for Quantum Materials and Technologies, Karlsruhe Institute of Technology, 76021 Karlsruhe, Germany}}
\affiliation{\mbox{Institut f\"ur Theorie der Kondensierten Materie, Karlsruhe Institute of Technology, 76128 Karlsruhe, Germany}}
\affiliation{Ioffe Institute, 194021 St.Petersburg, Russia}
\author{D. G. Polyakov}
\affiliation{\mbox{Institute for Quantum Materials and Technologies, Karlsruhe Institute of Technology, 76021 Karlsruhe, Germany}}

\date{\today}

\begin{abstract}
We study electron transport through a multichannel fractional quantum Hall edge in the presence of both interchannel interaction and random tunneling between channels, with emphasis on the role of contacts. The prime example in our discussion is the edge at filling factor 2/3 with two counterpropagating channels. Having established a general framework to describe contacts to a multichannel edge as thermal reservoirs, we particularly focus on the line-junction model for the contacts and investigate incoherent charge transport for an arbitrary strength of interchannel interaction beneath the contacts and, possibly different, outside them. We show that the conductance does not explicitly depend on the interaction strength either in or outside the contact regions (implicitly, it only depends through renormalization of the tunneling rates). Rather, a long line-junction contact is characterized by a single parameter which defines the modes that are at thermal equilibrium with the contact and is determined by the interplay of various types of scattering beneath the contact. This parameter---playing the role of an effective interaction strength within an idealized model of thermal reservoirs---is generically nonzero and affects the conductance. We formulate a framework of fractionalization-renormalized tunneling to describe the effect of disorder on transport in the presence of interchannel interaction. Within this framework, we give a detailed discussion of charge equilibration for arbitrarily strong interaction in the bulk of the edge and arbitrary effective interaction characterizing the line-junction contacts.
\end{abstract}

\maketitle

\section{Introduction}
\label{s1}

The interpretation of electrical conductance measurements in mesoscopic conductors was intensively debated from the very onset of mesoscopic physics up until the late 90s \cite{Beenakker1991,Datta1995}. The discussions mostly revolved around the role of electrical contacts \cite{Landauer1957,*Landauer1989,Fisher1981,Economou1981,Buttiker1986,Glazman1988}, with the focus initially on noninteracting electrons. One conceptually essential point recognized back then concerns the difference between the two- and four-terminal measurements. Specifically, it was understood that, when electrons are not reflected at the contacts (``perfect junctions"), the two-terminal ballistic dc conductance $G$ is measured in units of $e^2/h$ per conducting channel, with the resistance $1/G$ emerging entirely from relaxation processes in the attached contacts (``reservoirs").

As the discussions expanded to cover interacting electrons, the notion of ballistic transport yielded a remarkable result for one-dimensional correlated electrons in a ballistic (conserving both the total electron momentum and the numbers of right- and left-moving electrons) Luttinger liquid (LL) \cite{Giamarchi2003}. Namely, it was established, from various theoretical perspectives \cite{Safi1995,Maslov1995,Ponomarenko1995,Egger1996,*Egger1998,Alekseev1996,*Alekseev1998,Kawabata1996,Oreg1996,Maslov2005}, that when a ballistic LL quantum wire terminates in two Fermi liquid contacts, all signatures of interaction inside the wire vanish from $G$. Under the assumption of  interactions inside the contacts being negligible, $G$ was understood to be universally quantized at $e^2/h$ (per spin), largely in accordance with the experimental observations \cite{Tarucha1995,Yacoby1996}.

Our purpose here is to investigate transport of interacting electrons through a fractional quantum Hall (FQH, fractional QH) edge, with emphasis on the role of contacts and the universality of $G$ for the case when the edge hosts several nonequivalent chiral conducting channels. An FQH edge is a strongly correlated ``chiral LL" \cite{Wen1990b,Wen1992,*Wen1995,Chang2003} that inherits its compositional properties from the topological order of the bulk. For a given bulk filling factor $\nu$, the topological constraint on the edge structure (the number of edge channels and the channel filling factor discontinuities, hence the channel chiralities) allows for multiple specific choices. Which of the choices is realized is determined by the confinement-controlled ``electrostatics" of the edge.

An archetypical example of a multichannel edge, on which we focus in this paper, is the ``hole-conjugate" edge for $\nu=2/3$. The concrete model we discuss corresponds to two counterpropagating channels with filling factor discontinuties 1 and $-$1/3 \cite{MacDonald1990,*Johnson1991}, which is thought to be appropriate for the case of a sufficiently sharp confinement (below, we refer to these channels as channel 1 and channel 1/3, resp.\ modes 1 and 1/3). The key ingredient in our story is interaction between charge densities in these two channels. More complex edge structures with more channels emerge with softening confinement as a result of ``edge reconstruction" \cite{deChamon1994,Meir1994,Chklovskii1995,Wan2002,*Wan2003,Khanna2021}, with the emergence of fractional modes being characteristic of both integer \cite{Chklovskii1995,Khanna2021} and fractional bulk phases, eventually approaching the ``coarse-grained" quasiclassical limit \cite{Chklovskii1992,Geller1994}. Additional channels were argued to play an essential role in certain experiments on the $\nu=2/3$ edge \cite{Wang2013}. The ideas that are central to the description of edge transport within the model we focus on here are equally applicable to these, more involved structures. Experimentally, there has been an immense effort, in the last decade or so, to probe the structure of complex FQH edges, especially at $\nu=2/3$, with evidence pointing towards the existence of counterpropagating edge modes \cite{Bid2009,Bid2010,Deviatov2011,Dolev2011,Gross2012,Venkatachalam2012,Gurman2012,Inoue2014,Grivnin2014,Sabo2017,Banerjee2017,Rosenblatt2017,Banerjee2018,Lafont2019,
Bhattacharyya2019,Cohen2019,Lin2019,Lin2021}.

\begin{figure}[h]
\centering
\includegraphics[width=0.95\columnwidth]{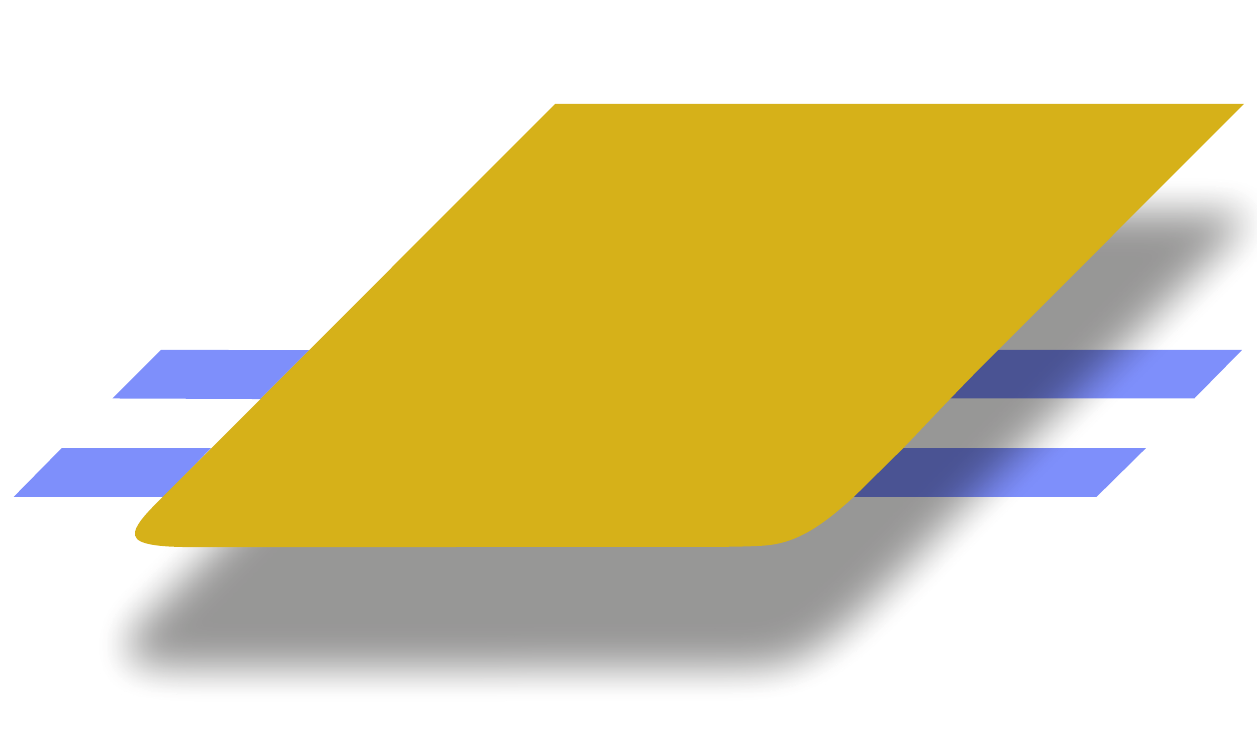}
\caption{Schematic illustration of a contact (orange) side-attached to a chiral (not terminating in the contact) two-channel edge (blue). The contact is connected to an external circuit and coupled to the edge modes through tunneling bridges.}
\label{f1}
\end{figure}

An FQH edge with counterpropagating channels represents an intermediate case between a nonchiral quantum wire (in particular, a conventional LL quantum wire or a symmetric QH line junction \cite{Renn1995,Kane1997a}) and a single-channel Laughlin edge (corresponding to $\nu=1/m$, with $m$ an odd integer), and is different from both in an essential way. Specifically, on the one hand, its chiral nature is manifest in the presence of a ballistic charge mode irrespective of the presence of backscattering disorder inside the edge---in contrast to the LL quantum wire. On the other hand, disorder-induced charge equilibration between the channels generically affects the conductance---in contrast to the single-channel edge. A more subtle difference from the LL quantum wire concerns the nature of contacts. The nonchiral wire can terminate in the contacts, whereas the chiral edge cannot. Therefore, contacts to the QH edge are necessarily ``side attached" (Fig.~\ref{f1}) \cite{Kane1995}.

\subsection{Contacts}
\label{s1a}

For the single-channel case, both the two-terminal conductance $G$ and the Hall conductance $G_H$ show ``fractional quantization," $G=G_H=1/m$ (hereafter, we set $e=\hbar=1$, except for the conductances, which are measured, as in the formula above, in units of $e^2/h$), under the assumption that the side-attached contacts are ``ideal"; otherwise, the quantization of $G$ at $1/m$ is lost \cite{Kane1995,deChamon1997}. An ideal contact is defined as fully absorbing the incident charge current (``black body") and emitting current that is independent of the incident one, with the contact playing the role of a source at equilibrium. For example, a single tunneling link between the edge and the reservoir does not meet these conditions \cite{Kane1995,deChamon1997}, nor does a contact that emits several edge modes that are not at equilibrium with each other \cite{Buttiker1988}. The ideal contact is supposed to be characterized by a certain electrochemical potential $\mu$ and temperature $T$ of electrons. A voltage probe (in its invasive version that ``thermalizes" electrons) is then understood as an ideal contact with no net current flowing through it.

\subsubsection{Nonideal contact}
\label{s1a1}

Calculating the edge conductance with ideal contacts in the clean case (where ``clean" means no charge scattering between channels if the number of channels exceeds one) is thus tantamount to the identification of the edge electron density excitations that are conjugate to $\mu$ \cite{Alekseev1996,*Alekseev1998,Imura2002}. For the single-channel edge, the electron charge density mode is defined uniquely, which yields the abovementioned $G=G_H=1/m$. From this perspective, the quantization of $G=1$ for a clean interacting LL wire \cite{Safi1995,Maslov1995,Ponomarenko1995,Egger1996,*Egger1998,Alekseev1996,*Alekseev1998,Kawabata1996,Oreg1996,Maslov2005} (two chiral channels) stems from $\mu$ being conjugate to the ``bare" (noninteracting) right- or left moving electron modes, and not the chiral, interaction-renormalized electron eigenmodes. That is, the contacts that provide for $G=1$ in interacting LL quantum wires are not ideal: the currents incident on the contacts from the bulk of the wire, which are the eigenmode currents, are not fully absorbed by the contacts, and the currents emitted by the contacts into the bulk are delicate linear combinations of the eigenmodes.

The apparent dichotomy between the requirement of ideal contacts for the quantization of $G$ in the Laughlin edge and the requirement of nonideal contacts for the quantization in nonchiral quantum wires reflects the tacitely assumed absence of interaction between electrons on different segments of the QH edge separated by contacts. Bringing these segments in proximity to each other, in a narrow Hall bar \cite{Wen1991,Oreg1995} or on the sides of a narrow barrier in the QH line-junction setup \cite{Renn1995,Kane1997a}, would also make the ``nonideality" of the contacts---in the same sense as in the LL case---a necessary condition for the quantization of $G$ at $1/m$ for the single-channel edge. Conversely, assuming the current-supplying contacts to be ideal, the interaction between the parts of the Laughlin edge on the opposite sides of the Hall bar yields a nonuniversal value of $G$ \cite{Wen1991}, with $G_H$ still quantized at $1/m$.

Focusing on the multichannel FQH edge in a Hall-bar geometry, we work with the assumption that the interaction between electrons on the edge is sufficiently short-range so as not to act between any points on the edge separated by a contact. However, even with this assumption, the presence of local interchannel interactions both outside the side-attached contact and beneath it---which are not necessarily of the same strength---brings up a question about the influence of these interactions on the conductance, which we will discuss in detail in the main part of the paper.

It was shown in Ref.~\onlinecite{Kane1995b} that generically neither $G$ nor $G_H$ in the clean case is quantized for the $\nu=2/3$ edge with counterpropagating channels, if the contacts are considered as ideal. That is, both $G$ and $G_H$ depend then on the strength of interchannel interaction. This is in similarity to a LL quantum wire with ideal contacts. However, it is a widely held idea that the contacts to a wire that terminates in reservoirs are, as already noted above, generically not ideal---because of the mismatch in the interaction strength inside the wire and in the contacts that are thought to be noninteracting. The mismatch is understood here as a ``sharp" one, with the interaction strength changing nonadibatically fast with respect to the scale of the characteristic wavelength of density excitations. The embodiment of this idea is the model of the LL quantum wire with an inhomogeneous (``vanishing at infinity") strength of interaction \cite{Safi1995,Maslov1995,Ponomarenko1995}.

A direct generalization of this model to the case of a QH edge is the model in which the interacting edges, or the interacting edge channels for that matter, split up from each other and are contacted beyond the region inside which they run parallel (``elongated quantum point contact") \cite{Berg2009,Rosenow2010,Protopopov2017}. Within this construction, the currents in the noninteracting ``spacer" are incident on and emitted by an ideal noninteracting contact. The nonideality of the ``extended contact," which is the contact as such plus the spacer, is then associated with Fresnel-like scattering of density excitations at the boundary between the spacer and the interacting part of the edge. For the model of ideal contacts attached to the noninteracting spacers, the two-terminal conductance of the clean $\nu=2/3$ edge has a universal value independent of interchannel interaction \cite{Protopopov2017}: $G=4/3$, which should be contrasted with the result of Ref.~\onlinecite{Kane1995b}.

This model, however, ignores the possibility of interaction between the edge channels as they run past the contact: indeed, the contact to a QH edge is, as already mentioned above, unavoidably side-attached (Fig.\ref{f1}) \cite{Kane1995}, hence screening of the interchannel interaction by it is not necessarily perfect. This might be particularly clear if one thinks about the contact being attached ``laterally," as opposed to ``on top." If one imagines the limit of no screening, the interaction strength is then homogeneous everywhere along the closed path of the edge. This alone makes a difference between the side-attached contact and the contact terminating a quantum wire, the consequences of which we will explore in the main text. In particular, one of the questions that arise in this connection is whether the two-terminal conductance for the clean $\nu=2/3$ edge depends, instead of being quantized at 4/3, on the strength of interchannel interaction beneath the contacts.

\subsubsection{Line-junction contact}
\label{s1a2}

A very natural model for the side-attached contact that incorporates interchannel interactions is that of a ``line-junction contact" \cite{Kane1995}, which represents a linear sequence of tunnel links connecting the edge and the reservoir. On the phenomenological level, dynamics of electrons in the reservoir is supposed to be fully incoherent and characterized by an infinitely fast equilibration to a given thermal state (``ideal reservoir"). One can supplement the picture by modeling the reservoir as a collection of ``incoherent" chiral noninteracting channels each of which supplies electrons at a given common electrochemical potential \cite{deChamon1997}.

In the limit of an infinite density of infinitesimally weak links with the tunneling rate held fixed, the model is describable by a set of scattering rates between different channels and the reservoir, and between the channels themselves. In this limit, it was argued \cite{Kane1995} that the distinguishing property of a long line-junction contact to the edge with counterpropagating channels---as opposed to that with copropagating channels---is that the current to the reservoir is determined by the relative amplitudes of the nonuniversal partial scattering rates. Moreover, assuming that the interaction strength beneath the contact and outside it is the same, the argument was made \cite{Kane1995} that the conductance should depend only on the scattering rates, but not on, separately, the interaction strength itself, as would be the case for the model with ideal contacts analyzed in Ref.~\onlinecite{Kane1995b}.

By taking this point further, we explicitly calculate $G$ and $G_H$ for the model of line-junction contacts, for arbitrary interaction strengths beneath the contacts and in the rest of the edge. One of the observation we make is that---for arbitrary parameters of the line-junction contact---there exists a set of edge modes that is at equilibrium with a given contact at its end points (the interface between the part of the edge beneath the contact and the ``bulk" of the edge). This means that the contact can be viewed as an ideal one with respect to these modes and characterized by a single parameter describing charge equilibration in the contact region. Remarkably, this parameter plays the same role as the strength of interchannel interaction in the ideal-contact model adopted in Ref.~\onlinecite{Kane1995b}. That is, the line-junction contact is characterized by the strength of an effective interchannel interaction beneath it.

\subsection{Disorder}
\label{s1b}

Apart from categorizing the contacts, another essential ingredient for the framework to describe charge transport through multichannel QH edges is disorder-induced charge scattering between the channels. The role of this scattering is twofold. First, it ``equilibrates" the charge densities in different channels on average. This establishes a ballistic propagation of the total charge density, irrespective of the difference in the values and signs of the channel velocities \cite{Kane1997}. Second, it leads to mesoscopic fluctuations of these densities. For edges with only copropagating channels, disorder-induced charge equilibration plays, in the dc limit, the same role as equilibration by a voltage probe and so does not affect the total edge current, with mesoscopic fluctuations showing up only in the elements of the conductance matrix (in channel indices) \cite{Wen1994}. By contrast, for edges with counterpropagating channels, mesoscopic fluctuations affect the conductance, and so does, generically, charge equilibration, as we outline below.

\subsubsection{Interchannel equilibration}
\label{s1b1}

Charge equilibration
establishes, in the limit of full equilibration, the universal quantization of disorder-averaged $G=\nu$ \cite{Kane1997}, independently of the value of $G$ in the clean limit. The quantization of $G$ at the value of $\nu$, irrespective of the interchannel interaction, in the charge-equilibrated limit was also demonstrated \cite{Kane1995b} (and argued on more phenomenological grounds \cite{Kane1995}) specifically for $\nu=2/3$. Importantly, the charge equilibration-induced quantization of $G$ for $\nu=2/3$ does not rely on the decoupling of the charge and neutral modes at a specific value of the interaction strength, with disorder affecting then only the neutral mode, along the lines of the renormalization-group treatment \cite{Kane1994,Kane1996Book}. The quantization results solely from a combination of the total charge conservation and the local equilibration, in the spirit of hydrodynamics \cite{Kane1997}.

By providing for the conductance quantization at $\nu$, the edge segment consisting of the contact proper and the charge-equilibrated parts of the disordered edge on both sides of it is, as a whole, at equilibrium with the bare (noninteracting) density modes. It was argued \cite{Kane1995} that such a ``compound" contact may be viewed as a realization of B\"uttiker's ideal contact \cite{Buttiker1988} for the edge with counterpropagating channels. It is worth noting, however, that this is only true if there is no interaction between the channels (as was the case in the B\"uttiker construction with copropagating channels). Indeed, as was already mentioned above, the ideal contacts to a clean edge do not yield a quantized conductance for interacting counterpropagating channels \cite{Kane1995b}.

A crossover to the universally quantized $G=\nu$ \cite{Kane1997} as the length of the disordered edge increases was considered for two counterpropagating channels both in the absence \cite{Sen2008,Nosiglia2018} and in the presence \cite{Agarwal2009,Srivastav2020,Protopopov2017} of interchannel interaction. A model of local thermal equilibration \cite{Nosiglia2018}, in which every pair of adjacent tunneling links is separated by voltage and temperature probes in each of the channels, was employed to explore, in the absence of interchannel interaction, also thermal transport and shot noise \cite{Park2019,Spanslatt2019,*Spanslatt2020,Park2020b}. Below, we will study charge equilibration particularly for the line-junction contacts, for arbitrary interactions beneath and outside the contacts.

\subsubsection{Mesoscopic fluctuations}
\label{s1b2}

Turning to mesoscopic fluctuations for the case of counterpropagating channels, it is important to distinguish two essentially different types of the fluctuations. One is not related to the presence or absence of interchannel interaction. Because of the chiral nature of the edge, mesoscopic conductance fluctuations of this type self-average with increasing edge length. For incoherent transport (see Sec.~\ref{s1b3}), they vanish altogether in the Gaussian limit of the Poisson distribution of interchannel tunneling links in space between the contacts. The other relies on spatial inhomogeneity of the interaction strength; in particular, at the contacts. If the interaction strength changes at the contacts and is homogeneous otherwise, then Fresnel-like scattering of the density modes at the contacts creates a Fabry-P\'erot resonator between the contacts, with charge transfer through its facets depending on the particular realization of disorder.

One peculiar situation emerges when the interchannel interaction is strong and the charge and neutral modes decouple \cite{Kane1994,Kane1996Book}, with disorder in interchannel tunneling not affecting the charge mode. For ideal contacts, mesoscopic fluctuations of the two-terminal conductance $G$ are then strictly absent. Otherwise, they are describable in terms of a disordered chiral sine-Gordon ($\chi$sG) model \cite{Naud2001,Rosenow2010,Protopopov2017}. For the $\nu=2/3$ edge, composed of channels 1 and 1/3, with noninteracting spacers at the contacts (see above), fluctuations with varying edge length are strong within this model. Specifically, $G$ fluctuates between 4/3 and 1/3 \cite{Protopopov2017}, or, equivalently, the diagonal element, for channel 1, of the conductance matrix fluctuates between 1 and 1/2 \cite{Rosenow2010}. To the best of our knowledge, this (``coherent") transport regime has not so far been observed experimentally; on the contrary, the dependence of $G$ on the edge length for $\nu=2/3$ was reported to show charge equilibration as the length is increased, with $G$ approaching $2/3$ in an (arguably) smooth manner \cite{Cohen2019}.

\subsubsection{Incoherent transport}
\label{s1b3}

In this paper, we restrict attention to the incoherent model and focus mostly on the case of white-noise (weak and with a vanishing correlation length) disorder in the amplitude of tunneling links. Within the incoherent model with white-noise disorder, $G$ is a smooth function of the edge length with no mesoscopic fluctuations by construction (which seems to be in agreement with the experimental observations \cite{Cohen2019} mentioned above). As part of a brief rationale for the model, let us first comment on the meaning of ``incoherent" in the context of the $\nu=2/3$ edge. A conceptually effective formalism to solve the disordered $\chi$sG model is to map it onto the (pseudo)spin-1/2 dynamics of a chiral fermion in a spatially random Zeeman field \cite{Kane1994}. The coherent $\chi$sG and incoherent models differ then in that the former deals with random rotations of spin over the Bloch sphere, whereas the latter with spin flips between the up and down positions. As such, the incoherent model is fully described by the occupation numbers for spin up and spin down \cite{incoh_model}.

Returning to the original problem, the spin flips correspond to flipping the orientation of a charge dipole between channels 1 and 1/3. As a result, the incoherent model is formalizable in terms of a linear equation of motion for the ``partial" charge densities. This picture is pertinent to the decoupled charge and neutral modes. In the main text, we will discuss the dynamics of charge within the incoherent model, for arbitrary interaction between the channels, in more detail.

From this perspective, a question arises as to the effect of nonzero $T$ on the neutral-mode dynamics within the disordered $\chi$sG model, in particular, whether it may lead to a suppression of the mesoscopic fluctuations. It was argued for the $\nu=2/3$ model \cite{Protopopov2017}---and demonstrated for a related model \cite{Naud2001}, with a proper adjustment for the $\nu=2/3$ case---that the property of transport being coherent is insensitive to temperature as long as the neutral and charge modes are decoupled (otherwise, an incoherent transport regime \cite{incoh_model} emerges if the edge length is sufficiently long). On the other hand, a different scenario was proposed, in which a crossover to the incoherent regime, as $T$ is increased, is governed by the ratio of $T$ and the characteristic energy spacing for the density excitations on the scale of the edge length \cite{Rosenow2010}. Yet another condition for such a crossover was associated with the ratio of $T$ and the energy spacing on the scale of the ``typical distance between scatterers" \cite{Srivastav2020}.

While it is beyond the scope of this paper to go deeper into the story about the coherent regime, let us briefly mention various mechanisms that may indeed suppress mesoscopic fluctuations and justify the incoherent model, irrespective of whether the charge and neutral modes are decoupled or not. In particular, at nonzero $T$, one can think of the mechanisms that are based on weakening the assumptions behind the conventional formulation of the $\chi$sG problem, i.e., introducing perturbations to the coherent model.

For example, this may be nonzero curvature of the density-mode spectrum (related to a finite range of interaction), which leads to a ``self-averaging" of the conductance given by a sum of contributions with different velocities of propagation (to an extent, this is similar to the curvature-induced suppression of interference in a Mach-Zender interferometer \cite{Chalker2007}). Nonzero curvature of the electron spectrum (related to the shape of self-consistent confinement) generically adds decay channels for the edge excitations. Note that the nonlinearity of the spectrum may be enhanced close to the edge reconstruction transition. Or it may be dephasing by the environment. One particular source of it may be temporal fluctuations of the strength of tunneling because of interactions with additional---due to edge reconstruction---channels possibly running close to those taken into account without exchanging charge with them.

Another mechanism of the suppression of mesoscopic flustuations, effective also at $T=0$, may be due to the coherent random interchannel dynamics of charge being extended from the ``bulk" of the edge into the regions beneath the line-junction contacts. Then, fluctuations tend to self-average because the conductance is given by a sum over paths of different length for the density excitations that are emitted and absorbed at different points along the contacts.

Mesoscopic fluctuations may also be suppressed merely because of interchannel interaction being so weak (or so strong, see below) that the edge finds itself, even upon disorder-induced renormalization \cite{Kane1994,Kane1995b}, far away from the point at which the charge and neutral modes decouple. It was argued in Ref.~\onlinecite{Protopopov2017} that mesoscopic fluctuations are suppressed for weak interaction and prominent, for not a too long edge, close to the decoupling point. However, as discussed in Sec.~\ref{s2a}, there is a certain duality between the properties of the edge for weak and strong interchannel interaction. A direct consequence of it is that, for strong interchannel interaction---stronger than required for the decoupling of the charge and neutral modes---disorder should be irrelevant in the renormalization-group sense in the same manner as for weak interaction. Moreover, dephasing of mesoscopic fluctuations at $T\neq 0$ should be characteristic not only of the weak-interaction case, but the opposite case as well.

\subsection{Outline of the results}
\label{s1c}

In Secs.~\ref{s1a} and \ref{s1b}, we discussed the multifaceted issues of contacts and disorder in QH edge transport, by placing them in proper perspective with regard to an FQH edge with counterpropagating channels. With this background in mind, we investigate charge transport through such an edge in the presence of both interchannel interaction and backscattering disorder. We focus on the prime example for these purposes, namely the two-channel edge at $\nu=2/3$ with channels 1 and 1/3. Here, we do so within the incoherent model, defined and rationalized in Sec.~\ref{s1b}.

One of our aims is to gain understanding of how edge transport is affected by interchannel interaction when the interaction strength is different beneath and outside the contacts. The particular model of the contact that we keep in mind in the first place is a line-junction contact. Before proceeding to this model, however, we first set up a general phenomenological framework in which the contact is supposed to be at equilibrium with a particular set of edge modes that are not the eigenmodes outside it. As such, this contact is nonideal---in the precise sense discussed in Sec.~\ref{s1a}.

Our main results, some of which were already mentioned above, can be broadly described as follows.

\begin{itemize}

\item[(1)]{For the contacts represented---first at the model level---as thermal reservoirs for the edge eigenmodes in the contact regions, the conductance of the clean edge does not depend on the strength of interchannel interaction outside these regions. Rather, the conductance depends on the strength of interaction characterizing the modes that are at equilibrium with the contacts. The dependence of transport on the latter is formalized by introducing ``generalized" boundary conditions on the facets of the contacts. The independence of the conductance on interchannel interaction outside the contacts holds irrespective of whether different contacts correspond to the same or different equilibrium modes associated with them. The obtained expressions for the conductance---for various arrangements of the measuring terminals---extend the conventional picture of ideal contacts.}

\item[(2)]{The long line-junction contact is fully characterizable by a single parameter, which brings in the notion of ``universality" in the classification of different contacts attached to the edge. Remarkably, this parameter plays the role of an effective strength of interaction inside the contact within the above model of a thermal reservoir. In the microscopic picture, this effective interaction strength is generically nonzero for the line-junction contact. The parameter ``labelling" a given contact is determined by the interplay of backscattering between the edge channels and scattering between the reservoir and the edge, but not the interaction strength either beneath or outside the contact---apart from the interaction-induced renormalization of the scattering rates.}

\item[(3)]{Disorder-induced tunneling between edge channels is described in terms of charge fractionalization. The framework of fractionalization-renormalized tunneling, which emerges from this approach, yields the dependence of the scattering rates on the interaction strength, formalized in terms of electrostatic screening of charges created by tunneling. This dependence is distinctly different from the conventional renormalization of the tunneling strength by the interaction-induced orthogonality catastrophe. The picture of fractionalization-renormalized tunneling reveals the physics behind the effect of backscattering disorder on the edge eigenmodes, including the charge and neutral modes when these decouple from each other. It also describes the emergence of negative partial scattering rates for sufficiently strong interaction. The thermodynamic constraint on tunneling is framed into the fractionalization picture for an arbitrary strength of interaction.}

\item[(4)]{Charge equilibration and the resulting dependence of the conductance on the edge length are analyzed for arbitrary strength of interchannel interaction both beneath the contacts and outside them. The strength of true interaction is shown to cancel out from the conductance of a disordered edge apart from the renormalization of the scattering rates. Instead, transport is affected by the effective interaction that characterizes equilibration in the long line-junction contacts, making the conductance of a short edge dependent on the strength of this effective interaction. The existence and the properties of the disorder-modified ballistic charge mode, responsible for the universal quantization of the conductance in the limit of a long edge, are discussed for arbitrary strength of interaction. The conventional notions of the contact and bulk contributions to the two-terminal resistance are demonstrated to be inapplicable to the QH edge with counterpropagating channels.}

\end{itemize}

The remainder of the paper is organized in the following way. Section~\ref{s2} is devoted to transport through the clean edge, with emphasis on the formulation of the boundary conditions for the contacts [items (1) and (2) in the above outline of the results]. In Sec.~\ref{s2a}, we specify the model for the $\nu=2/3$ edge. In Sec.~\ref{s2b}, we discuss the generalized boundary condition [item (1)]. In Sec.~\ref{s2c}, we calculate the conductance for this type of the boundary condition with various arrangements of the measuring terminals. In Sec.~\ref{s2d}, we consider the model of the line-junction contact [item (2)]. In Sec.~\ref{s2e}, we map the model of the line-junction contact onto the model of the generalized boundary condition, and obtain the conductance for the line-junction contacts. Section~\ref{s3} deals with fractionalization-renormalized tunneling [item (3)]. In Sec.~\ref{s3a}, we discuss fractionalization upon tunneling into the edge. In Sec.~\ref{s3b}, we turn to fractionalization upon tunneling between edge channels and formulate a general framework to describe fractionalization-renormalized tunneling for the case of a single tunneling link. In Sec.~\ref{s3c}, we analyze the strong-tunneling limit and the thermodynamic constraint on tunneling for an arbitrary strength of interchannel interaction. Section~\ref{s4} covers transport through a disordered edge [item (4)]. In Sec.~\ref{s4a}, we consider the emergence of negative scattering rates for the case of strong interchannel correlations. In Sec.~\ref{s4b}, we discuss the disorder-modified eigenmodes of the edge. In Sec.~\ref{s4c}, we calculate the conductance of the disordered edge for arbitrary parameters of the line-junction contacts and address the question of whether the contact and bulk resistances in the two-terminal setup are generically meaningful notions for a multichannel QH edge. Section \ref{s5} concludes with a succinct summary. In Appendix, we discuss some aspects of the nonchiral LL model from the perspective of the framework formulated for the chiral edge; in particular, the line-junction contact and fractionalization-renormalized tunneling.

\section{Clean edge}
\label{s2}

We begin by considering a clean edge, with no interchannel tunneling, and specifically focus on the $\nu=2/3$ edge within its model discussed at the beginning of Sec.~\ref{s1}, namely the one composed of counterpropagating channels 1 and 1/3.

\subsection{Model for the $\nu=2/3$ edge}
\label{s2a}

The model is defined by the Hamiltonian density
\be
H_0 = \pi\left(v_1n_1^2+3v_2n_2^2+2v_{12}n_1n_2\right)
\label{1}
\ee
and the commutation relation at points $x$ and $x'$ along the edge
\be
\left[\,n_\textrm{i}(x),n_\textrm{j}(x')\,\right]=\frac{i}{2\pi}\delta_\textrm{ij}\,\delta\nu_\textrm{i}\,\partial_x\delta(x-x')
\label{2}
\ee
for the charge densities (in units of $-e$) $n_1$ and $n_2$ in channels 1 and 1/3, respectively. This is equivalent to the Lagrangian formulation in Ref.~\onlinecite{Kane1994}. The constants $v_{1,2}$ are the speeds of propagation of the densities $n_{1,2}$ in the noninteracting limit (here and below, ``noninteracting" means no interchannel interaction, whereas short-range interactions within the channels are incorporated in the velocities $v_{1,2}$). In Eq.~(\ref{1}), the densities $n_1$ and $n_2$ interact with each other via a short-range potential with the zero-momentum Fourier component $2\pi v_{12}$. The filling factor discontinuities $\delta\nu_1=1$ and $\delta\nu_2=-1/3$ in Eq.~(\ref{2}) are of opposite sign, which encodes the property of the modes $n_1$ and $n_2$ propagating in opposite directions for $v_{12}=0$.

In accordance with Eq.~(\ref{1}), we assume everywhere below that the interchannel interaction is point-like. For our purposes in this paper, this is an accurate description of the edge on length scales larger than the screening radius of Coulomb interaction, where screening is provided by a nearby metallic gate. It is worth noting, however, that a nonzero radius of interaction leads to a nonlinear dispersion relation for the density excitations, which may be important on arbitrarily large length scales. For instance, one of the consequences of the nonlinear dispersion, commented upon in Sec.~\ref{s1b3}, is the suppression of mesoscopic conductance fluctuations, which adds to the justification of the incoherent description of the edge dynamics.

A diagonalization of the quadratic form in Eq.~(\ref{1}) represents $H_0$ in terms of the eigenmode (``chiral") charge densities $n_\pm$:
\be
H_0=\frac{\pi v_+}{g_+}n_+^2+\frac{\pi v_-}{g_-}n_-^2~,
\label{3}
\ee
with $n_\pm$ obeying
\be
\left[\,n_\pm(x),n_\pm(x')\,\right]=\pm\frac{i}{2\pi}g_\pm\partial_x\delta(x-x')~.
\label{4}
\ee
Both $n_{1,2}$ and $n_\pm$ are defined as charge densities, so that
\be
n_++n_-=n_1+n_2~.
\label{5}
\ee
The eigenmodes propagate with the velocities $v_\pm$ (both defined positively, as speeds) to the right (+) and to the left ($-$), and are characterized by the dimensionless ``conductances" $g_\pm\geq 0$ \cite{Kane1995b}. Parametrized in terms of the dimensionless interaction strength
\be
\alpha=\frac{2}{\sqrt{3}}\frac{v_{12}}{v_1+v_2}~,
\label{6}
\ee
$v_\pm$ are given by
\be
v_\pm=\frac{1}{2}\left[\,\pm (v_1-v_2)+(v_1+v_2)\sqrt{1-\alpha^2}\,\right]~.
\label{7}
\ee
To represent $g_\pm$ in a compact form, it is convenient to introduce the parameter \cite{Kane1994,Kane1995b}
\be
\Delta=\frac{2-\sqrt{3}\alpha}{\sqrt{1-\alpha^2}}~,
\label{8}
\ee
in terms of which $g_\pm$ are written as
\be
g_\pm =\frac{\Delta\pm 1}{3}~.
\label{9}
\ee

From Eqs.~(\ref{3}) and (\ref{7}), the stability conditions are $\textrm{Im}\,v_\pm=0$, which is $|\alpha|\leq 1$, and $v_\pm\geq 0$, which is \cite{Kane1994}
\be
v_{12}^2\leq 3v_1v_2~.
\label{10}
\ee
The latter is stronger than or identical to the former for arbitrary $v_1$ and $v_2$ (identical for $v_1=v_2$), so that the only condition is Eq.~(\ref{10}). Note that $\Delta\geq 1$ [cf.\ Eq.~(\ref{9})] for all $|\alpha|\leq 1$. The point $\Delta=1$, at which the charge and neutral modes decouple \cite{Kane1994}, corresponds to $\alpha=\sqrt{3}/2$. It is worthwhile to mention that if the velocities $v_{1,2}$ differ strongly enough from each other, namely if $v_1$ is beyond the interval
\be
v_2/3<v_1<3v_2~,
\label{11}
\ee
an instability occurs with increasing $\alpha$ before the point $\Delta=1$ is reached.

It is also worth noticing that the function $\alpha(\Delta)$ [which solves Eq.~(\ref{8})] is double valued for $|\alpha|\leq 1$, with two branches merging at $\Delta=1$. That is, for the case of repulsive interaction ($\alpha>0$), the range of $\Delta$ is not limited to the interval $1\leq\Delta\leq 2$, with the noninteracting point at $\Delta=2$:
\be
0\leq\alpha\leq\sqrt{3}/2\quad\mapsto\quad 1\leq\Delta\leq 2~.
\label{12}
\ee
On the other branch of $\alpha(\Delta)$, with
\begin{align}
\sqrt{3}&/2<\alpha\leq 2\sqrt{v_1v_2}/(v_1+v_2)\nonumber\\
&\mapsto\quad 1<\Delta\leq 2(v_1+v_2-\sqrt{3v_1v_2})/|v_1-v_2|~,
\label{13}
\end{align}
the value of $\Delta$ grows with increasing $\alpha$. This means, in particular, a nonmonotonic dependence of $g_\pm$ on $\alpha$. The largest---within the model (\ref{1})---$\alpha$ in Eq.~(\ref{13}) corresponds to the upper boundary (\ref{10}), at which one of the speeds $v_\pm$ slows down to zero (or both, if $v_1=v_2$).

Note that it is the parameter $\Delta$ that is of prime importance by determining the scaling dimensions of the correlation functions of the model \cite{Kane1994,Kane1995b}. In view of Eqs.~(\ref{12}) and (\ref{13}), the model (\ref{1}) possesses a duality between the cases of small and large $v_{12}$ corresponding to the same value of $\Delta$, with the only difference being the different velocities $v_\pm$ that are determined by $v_{12}$ itself. Perhaps only numerics can say if this captures physics of a more realistic model \cite{remark1}. Having made this cautionary remark, we assume below, for definiteness, that $1\leq\Delta\leq 2$ with $\alpha$ varying within the interval (\ref{12}).

The Heisenberg equations of motion resulting from Eqs.~(\ref{1}) and (\ref{2}), which have the form of continuity equations $\partial_tn_{1,2}+\partial_xj_{1,2}=0$, yield the partial charge currents
\be
j_1=v_1n_1+v_{12}n_2~,\quad j_2 =-(v_2n_2+v_{12}n_1/3)~.
\label{14}
\ee
Similarly, from Eqs.~(\ref{3}) and (\ref{4}), the chiral charge currents $j_\pm$ obey $\partial_tn_\pm+\partial_xj_\pm=0$ with
\be
j_\pm=\pm v_\pm n_\pm~.
\label{15}
\ee
The total edge current $j=j_++j_-=j_1+j_2$.

\subsection{Generalized boundary condition}
\label{s2b}

The equations of motion introduced in Sec.~\ref{s2a} need to be supplied with boundary conditions at the contacts. As outlined in Sec.~\ref{s1c}, we first consider---before turning in Sec.~\ref{s2d} to the line-junction contact---an instructive example in which the contact is at equilibrium with a set of density modes that are, in general, not the eigenmodes outside the contact. Specifically, let the contact be a thermal reservoir for the modes $n_{\pm c}$ [Eqs.~(\ref{3}) and (\ref{4}), with $c$ for ``contact"] corresponding to a given interaction parameter $\Delta_c$ [Eq.~(\ref{8})] not necessarily equal to the parameter $\Delta$ for the edge outside the contact. If $\Delta_c=\Delta$, the contact is, by definition in Sec.~\ref{s1a}, an ideal contact, for which the two-terminal conductance was obtained as \cite{Kane1994,Kane1995,Kane1995b}
\be
G=g_++g_-=\frac{2}{3}\Delta
\label{16}
\ee
[this is a ``cousin'' of the equality $G=K$, where $K$ is the Luttinger constant, for a LL wire with ideal contacts (see Appendix)].
As we will demonstrate in Sec.~\ref{s2e}, this model of a thermal reservoir with a certain $\Delta_c$ is directly related to the line-junction model discussed in Sec.~\ref{s1a2}. Let us, therefore, proceed to the case of $\Delta_c\neq\Delta$.

\begin{figure}[ht]
\centering
\includegraphics[width=0.95\columnwidth]{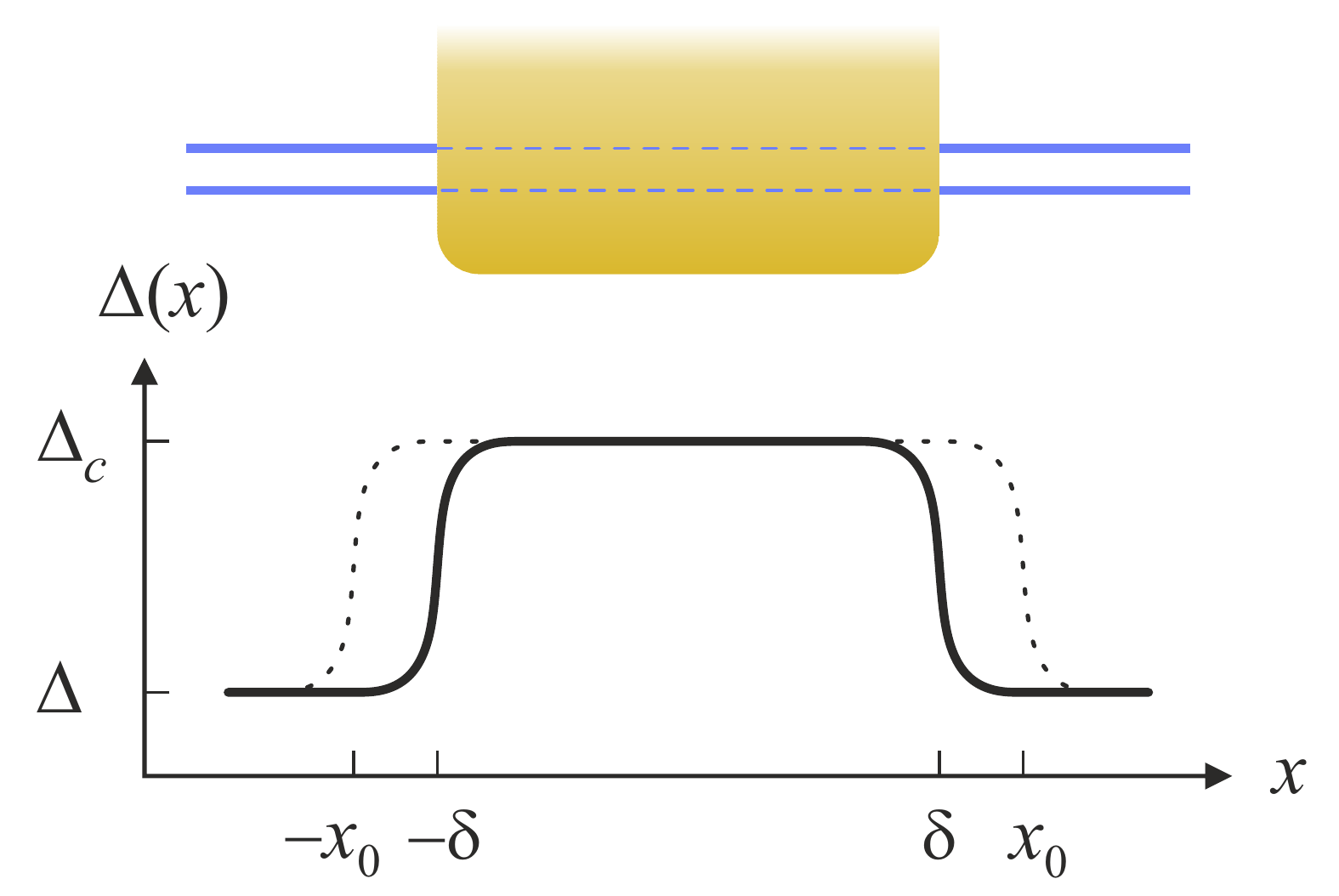}
\caption{Schematic profile (solid line) of the parameter $\Delta(x)$ [Eq.~(\ref{8})] characterizing the strength of interchannel interaction beneath ($\Delta_c$, for $|x|<\delta$) and outside ($\Delta$, for $|x|>\delta$) the side-attached contact. Because of screening of the interaction by the contact, $\Delta_c$ and $\Delta$ are generically not the same. The difference between them depends on the mutual position of the contact and the edge; in particular, the ratio of two characteristic distances: the one between the contact and the edge and the other between the channels. The dashed line for $\Delta(x)$ illustrates the ``point-splitting" procedure discussed below Eqs.~(\ref{22}) and (\ref{23}).}
\label{f2}
\end{figure}

Imagine that the contact ``covers" the part of the edge between $x=\delta$ and $x=-\delta$ (Fig.~\ref{f2}), so that the boundary conditions for the densities outside the contact are set at $|x|=\delta$. The chemical-potential shift $\delta H$ of the Hamiltonian density of the edge, provided by the contact, is
\begin{equation}
\delta H=-\mu n~,
\label{17a}
\end{equation}
where $n$ is the total charge density. According to the above, assume that $\mu$ is conjugate to the densities $n_{\pm c}$. Specifically, writing
\begin{equation}
n=n_{+c}+n_{-c}
\label{17b}
\end{equation}
in $\delta H$ and minimizing $H_0+\delta H$ by varying it with respect to $n_{+ c}$ and $n_{- c}$, the two boundary conditions on the facets of the contact are obtained as
\begin{align}
&\mu=\left(\hat\lambda^T\hat\Pi^{-1}\right)_{++}n_+(\delta)+\left(\hat\lambda^T\hat\Pi^{-1}\right)_{+-}n_-(\delta)~,
\label{17}\\
&\mu=\left(\hat\lambda^T\hat\Pi^{-1}\right)_{-+}n_+(-\delta)+\left(\hat\lambda^T\hat\Pi^{-1}\right)_{--}n_-(-\delta)~.
\label{18}
\end{align}
Here,
\be
\hat\Pi^{-1}=2\pi\left(\begin{array}{cc}
v_+/g_+ & 0 \\
0 & v_-/g_-
\end{array}\right)
\label{19}
\ee
is the inverse thermodynamic compressibility matrix in the bulk of the edge (for $|x|>\delta$) in the $n_\pm$ basis [cf.\ Eq.~(\ref{3})], and the matrix $\hat\lambda$ relates $\textbf{n}_c=(n_{+c},n_{-c})^T$ and $\textbf{n}=(n_+,n_-)^T$:
\be
\textbf{n} =\hat\lambda\,\textbf{n}_c
\label{20}
\ee
($\hat\lambda^T$ denotes transpose of $\hat\lambda$). The matrix $\hat\lambda$ is given by
\begin{align}
\hat\lambda&=\frac{1}{2}\left(\begin{array}{cc}
1+\Delta & 1+\Delta \\
1-\Delta & 1-\Delta
\end{array}\right)\nonumber\\
 &+
\frac{1}{2}\sqrt{\frac{\Delta^2-1}{\Delta_c^2-1}}
\left(\begin{array}{rr}
1-\Delta_c & \,-1-\Delta_c \\
-1+\Delta_c & \,1+\Delta_c
\end{array}\right)~.
\label{21}
\end{align}
By definition, $\hat\lambda^{-1}$ equals $\hat\lambda$ with the exchange $\Delta\leftrightarrow\Delta_c$.

In terms of the currents (\ref{15}), the boundary conditions (\ref{17}) and (\ref{18}) become
\begin{align}
&\frac{\mu}{2\pi}=\frac{j_+(\delta)}{g_+}\,\lambda_{++}-\frac{j_-(\delta)}{g_-}\,\lambda_{-+}~,
\label{22}\\
&\frac{\mu}{2\pi}=\frac{j_+(-\delta)}{g_+}\,\lambda_{+-}-\frac{j_-(-\delta)}{g_-}\,\lambda_{--}~.
\label{23}
\end{align}
Equations (\ref{22}) and (\ref{23}) impose two links on the four currents $j_\pm(\delta)$ and $j_\pm(-\delta)$. Placing boundary conditions of this type on each of the contacts fixes all currents between the contacts on the closed loop along the edge. The boundary conditions (\ref{22}) and (\ref{23}) [or (\ref{17}) and (\ref{18}) for that matter] generalize those for the ideal contact, for which $\hat\lambda=\mathbb{1}$ is the identity matrix.

Although the generalized boundary conditions are significant by themselves, the concern that one might have at this point is over the fact that they are imposed on the facets of the contact where the interaction strength experiences, by construction, a jump, corresponding to the change between $\Delta$ and $\Delta_c$. It is therefore instructive to ``point split" the boundary conditions by placing them at $|x|=x_0+0$, where $x_0>\delta$ (Fig.~\ref{f2}), and assuming that the interaction parameter is given by $\Delta_c$ in the ``spacer" between $|x|=\delta$ and $|x|=x_0$. The purpose is to demonstrate that Eqs.~(\ref{22}) and (\ref{23}) correspond to precisely this procedure with $x_0\to\delta$.

The contact itself (covering, as before the point splitting, $|x|<\delta$) is then ideal, with a simple boundary condition at $|x|=\delta$:
\be
\frac{\mu}{2\pi}=\pm\frac{j_{\pm c}(\pm \delta)}{g_{\pm c}}~,
\label{24}
\ee
where $j_{\pm c}(\pm\delta)$ and $g_{\pm c}$ are the outgoing eigenmode currents (those emitted by the contact) and the eigenmode conductances, respectively, corresponding to the interaction parameter $\Delta_c$. The boundary condition at $|x|=x_0+0$ is thus a combination of the ideal-contact condition (\ref{24}) at $|x|=\delta$ and the matching condition at $|x|=x_0$. The latter relates the densities and currents on the sides of the interface at which the interaction parameter changes from $\Delta_c$ to $\Delta$ with increasing $|x|$.

It is convenient to write the matching condition in the basis that does not change across the interface. Let it be, as a transparent example, the $n_{1,2}$ basis [Eqs.~(\ref{1}) and (\ref{2})]. Around the interface, the continuity equations then read [cf.\ Eq.~(\ref{14})]
\begin{align}
&\partial_tn_1+\frac{1}{2\pi}\,\partial_x\!\left[\left(\hat{\bar\Pi}^{-1}\right)_{11}\!n_1+\left(\hat{\bar\Pi}^{-1}\right)_{12}\!n_2\right]=0~,
\label{25}\\
&\partial_tn_2-\frac{1}{2\pi}\,\frac{1}{3}\,\partial_x\!\left[\left(\hat{\bar\Pi}^{-1}\right)_{21}\!n_1+\left(\hat{\bar\Pi}^{-1}\right)_{22}\!n_2\right]=0~,
\label{26}
\end{align}
where $\hat{\bar\Pi}$ is the $x$ dependent thermodynamic compressibility matrix in the $n_{1,2}$ basis [the bar is put to distinguish $\hat{\bar\Pi}$ from $\hat\Pi$ in Eq.~(\ref{19})], with
\be
\hat{\bar\Pi}^{-1}=2\pi\left(\begin{array}{cc}
v_1 & \,v_{12} \\
v_{12} & \,3v_2
\end{array}\right)~.
\label{27}
\ee
The matching condition is, therefore, the condition of continuity, upon crossing the interface, of the partial ``local chemical potentials"
\begin{align}
&\mu_\textrm{i}=\sum_\textrm{j}\left(\hat{\bar\Pi}^{-1}\right)_\textrm{ij}n_\textrm{j}~:
\label{28}\\
&\mu_\textrm{i}\vert_{|x|=x_0-0}=\mu_\textrm{i}\vert_{|x|=x_0+0}~.
\label{29}
\end{align}
Equivalently, Eq.~(\ref{29}) is the condition of continuity of the currents $j_1=\mu_1$ and $j_2=-\mu_2/3$ that are separately---beyond the total current conservation---continuous across the interface. The continuity of $j_{1,2}$ is achieved by the corresponding jumps in $n_{1,2}$.

By the same token, the continuity of the partial currents is maintained in any basis that is not changed across the interface; in particular, in the basis of $n_{\pm c}$. By representing: (i) $j_{\pm c}(x=\pm x_0\mp 0)$ in terms of $\mu$ from Eq.~(\ref{24}) and (ii) $j_{\pm c}(x=\pm x_0\pm 0)$ in terms of $j_\pm$ at the same $x$ by changing the basis according to Eq.~(\ref{20}), the continuity condition for $j_{\pm c}$ becomes exactly Eqs.~(\ref{22}) and (\ref{23}) after $x_0\to\delta$, which was to be demonstrated.

\subsection{Conductance for the generalized boundary condition}
\label{s2c}

We now turn to the calculation of the two- and four-terminal conductances of a clean $\nu=2/3$ edge for the boundary conditions (\ref{22}) and (\ref{23}), by placing them on each of the contacts, in the dc limit. Nonequilibrium is then created by ``biasing" the current contacts with different chemical potentials (this is a sufficient minimal model to calculate the measured conductance in a QH system with a gapped bulk \cite{Haldane1995,Kane1995}). Assume that all of the contacts are characterized by the same $\Delta_c$ (the case of different contacts will be considered in Sec.~\ref{s4c2}).

\begin{figure}[h]
\centering
\captionsetup[subfigure]{labelfont=bf,labelformat=simple,position=top,justification=raggedright}
\subfloat[]{
\includegraphics[width=0.75\columnwidth]{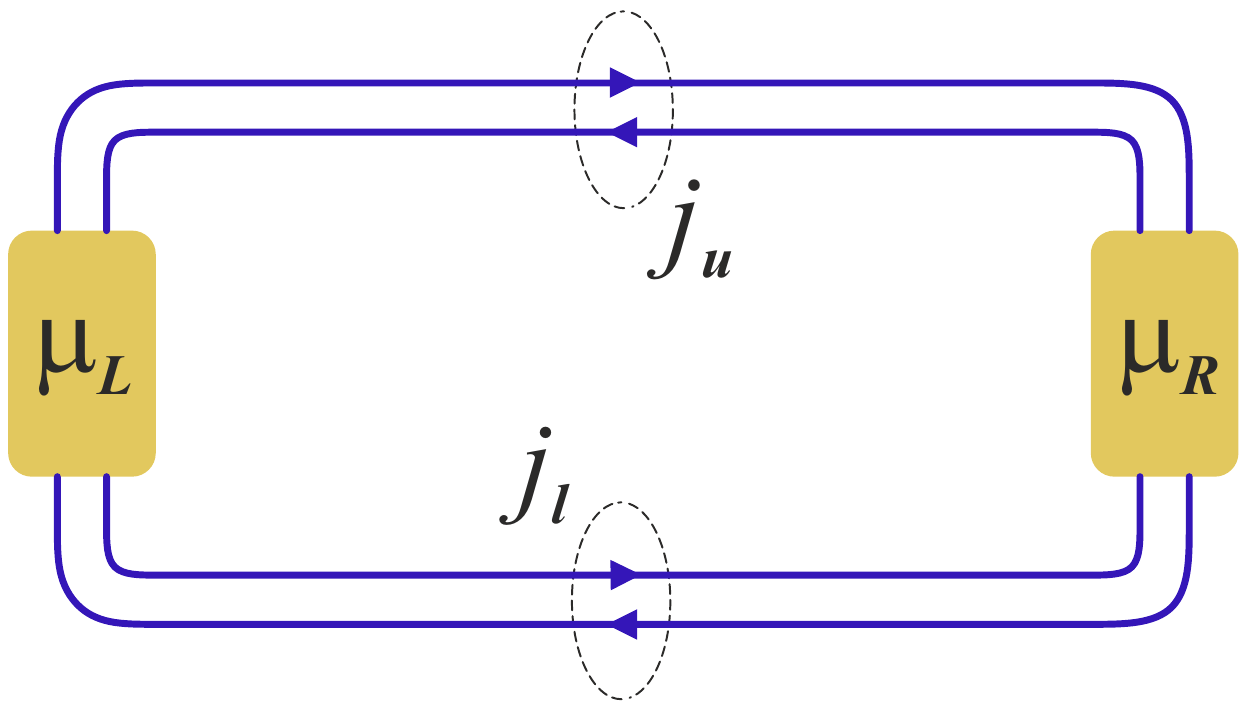}
\label{f3a}}\\
\subfloat[]{
\includegraphics[width=0.75\columnwidth]{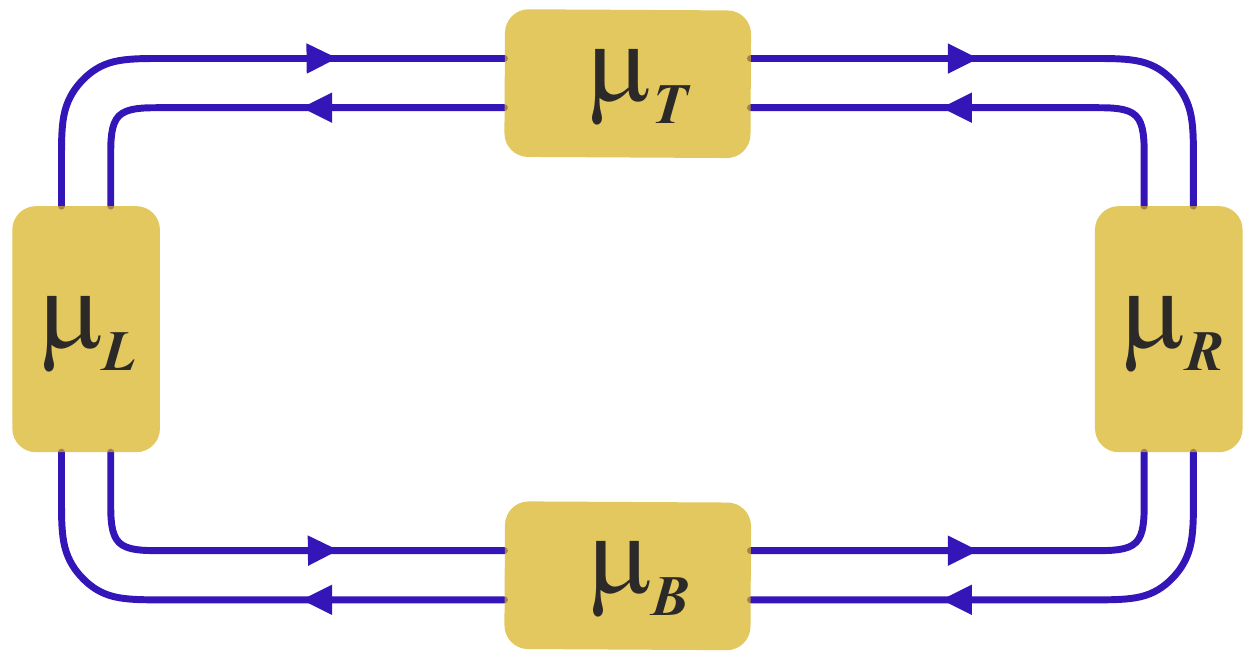}
\label{f3b}}
\caption{Measurement with two (a) and four (b) terminals, characterized by the chemical potentials $\mu_L$ (left), $\mu_R$ (right), $\mu_T$ (top), and $\mu_B$ (bottom), for the edge with two counterpropagating channels. The total edge current in the two-terminal geometry is a sum of the currents in the upper ($j_u$) and lower ($j_l$) parts of the edge. This is so for the four-terminal setup as well, if the top and bottom terminals are the voltage probes.}
\label{f3}
\end{figure}

For the two-terminal setup [Fig.~\ref{f3a}], the currents $j_{u\pm}$ in the upper part of the edge [running to the right (+) and to the left ($-$)] are given, from Eqs.~(\ref{22}) and (\ref{23}), by
\be
\left(\begin{array}{c}
j_{u+}/g_+ \\
-j_{u-}/g_-
\end{array}\right)=\frac{1}{2\pi}\left(\hat\lambda^{-1}\right)^T\left(\begin{array}{c}
\mu_L\\
\mu_R
\end{array}\right)~,
\label{30}
\ee
where $\mu_L$ and $\mu_R$ are the chemical potentials of the left and right contacts, respectively. For the total current at the upper edge $j_u=j_{u+}+j_{u-}$, the combination of $g_\pm$ and $\hat\lambda^{-1}$ reduces to $g_{\pm c}$:
\be
j_u=\frac{1}{2\pi}\left(g_{+c}\mu_L-g_{-c}\mu_R\right)~.
\label{31}
\ee
The total current at the lower edge $j_l$ is obtained from Eq.~(\ref{31}) by exchanging $\mu_L\leftrightarrow\mu_R$, with the total current $j$ between the contacts being $j=j_u-j_l$. The conductance $G=2\pi j/(\mu_L-\mu_R)$ is then found as
\be
G=g_{+c}+g_{-c}=\frac{2}{3}\Delta_c~.
\label{32}
\ee
The strength of interchannel interaction for the edge outside the contacts is thus seen to drop out from the two-terminal conductance (and from each of the currents $j_{u,l}$ separately). That is, as a general rule, the conductance of a clean edge depends only on the characteristics of the modes that are at equilibrium with the contacts [cf.\ Eq.~(\ref{16}) for the ideal contact].

Extending the argument that led to Eq.~(\ref{31}) to compute the four-terminal conductances for the setup in Fig.~\ref{f3b}, the current flowing from the left contact into the edge, $j_L$, is written as
\be
j_L=\frac{1}{2\pi}\left[\,g_{+c}\left(\mu_L-\mu_B\right)-g_{-c}\left(\mu_T-\mu_L\right)\,\right]~,
\label{33}
\ee
where $\mu_L$ is the chemical potential of contact $L$, etc. The currents $j_{T,R,B}$ into contacts $T,R,B$ are obtainable from Eq.~(\ref{33}) by cyclic permutation $L\to T\to R\to B\to L$. For the Hall measurement, the top and bottom contacts are taken as voltage probes with $j_{T,B}=0$, so that the source-drain current $j=j_L=-j_R$. The Hall conductance $G_H$ and the source-drain conductance $G_4$ are then obtained as
\begin{align}
&G_H=\frac{2\pi j}{\mu_T-\mu_B}=\frac{g_{+c}^2+g_{-c}^2}{g_{+c}-g_{-c}}=\frac{1}{3}\left(\Delta_c^2+1\right)~,
\label{34}\\
&G_4=\frac{2\pi j}{\mu_L-\mu_R}=\frac{g_{+c}^2+g_{-c}^2}{g_{+c}+g_{-c}}=\frac{1}{3}\left(\Delta_c+1/\Delta_c\right)~,
\label{35}
\end{align}
with $G_H=\Delta_cG_4$. Similarly to $G$ in Eq.~(\ref{32}), $\Delta$ vanishes from $G_H$ and $G_4$, which depend only on $\Delta_c$.

\subsection{Line-junction contact}
\label{s2d}

\begin{figure}[ht!]
\centering
\includegraphics[width=0.95\columnwidth]{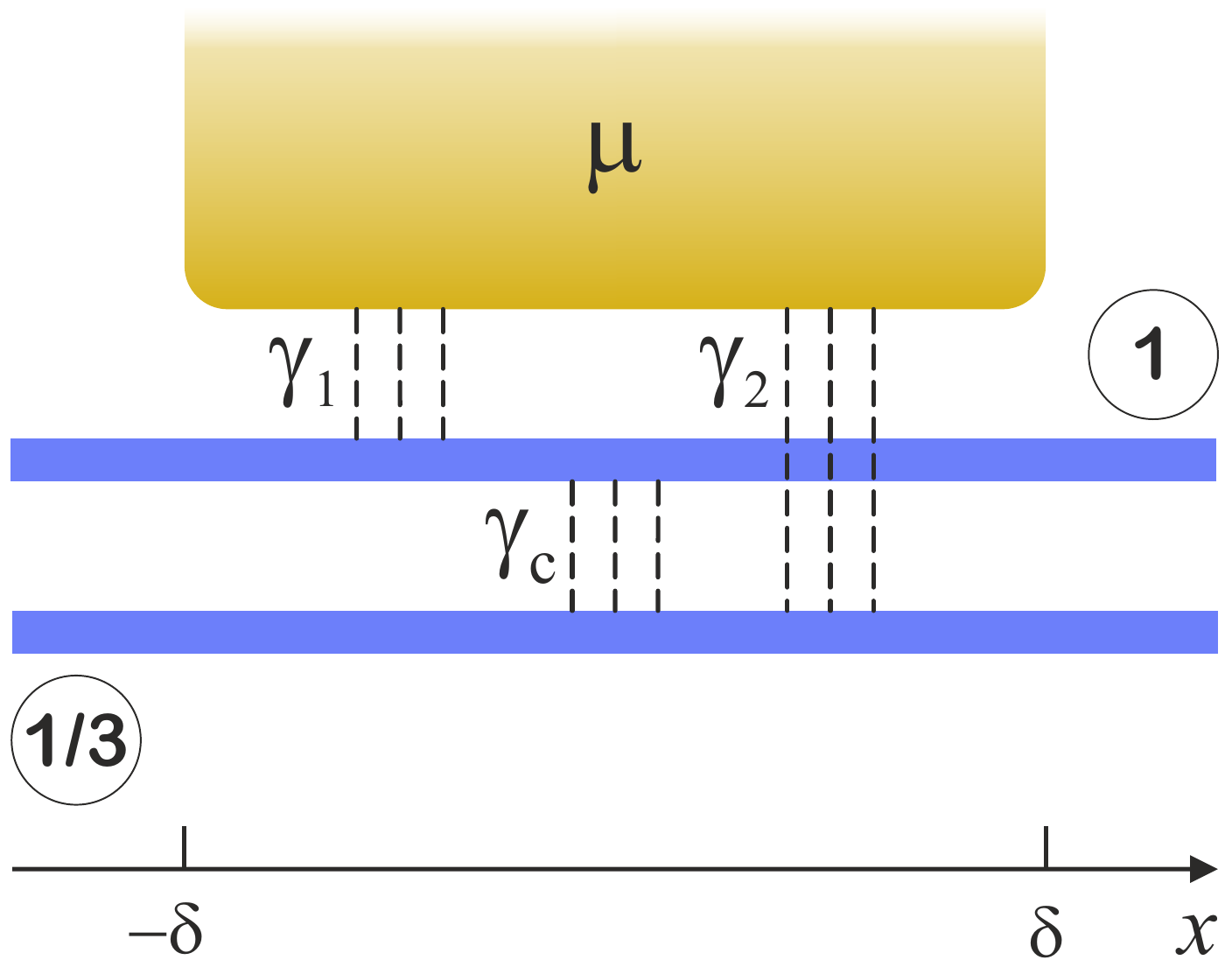}
\caption{Schematic portrait of a line-junction contact. Channels 1 and 1/3 exchange electrons, for $|x|<\delta$, through tunneling links (depicted as dashed lines) with the contact viewed as a thermal reservoir of electrons at the chemical potential $\mu$. The model is characterized by three inverse scattering lengths: $\gamma_1$ and $\gamma_2$ for tunneling between the contact and channels 1 and 1/3, respectively, and $\gamma_c/2$ for tunneling between the channels [Eqs.~(\ref{36}) and (\ref{37})].}
\label{f4}
\end{figure}

Having introduced the phenomenological model for the contacts in Secs.~\ref{s2b} and \ref{s2c}, we now turn to the line-junction contact model \cite{Kane1995}. As will be seen shortly, the two are directly mappable onto each other. As already discussed in Sec.~\ref{s1a2}, the line-junction contact model is formalized in terms of scattering rates at which the edge channels exchange electrons with the reservoir and, in general, also the rates at which the channels exchange electrons among themselves. Specifically, for the $\nu=2/3$ edge, the equations of motion for the currents $j_{1,2}$ beneath the contact (i.e., for $|x|<\delta$) read
\begin{align}
&\partial_xj_1+\frac{1}{2}\gamma_c (j_1+3j_2)-\gamma_1\left(\frac{\mu}{2\pi}-j_1\right)=0~,
\label{36}\\
&\partial_xj_2-\frac{1}{2}\gamma_c (j_1+3j_2)-\gamma_2\left(\frac{\mu}{2\pi}+3j_2\right)=0~,
\label{37}
\end{align}
where $\gamma_1^{-1}$ and $\gamma_2^{-1}$ are the scattering lengths for the exchange of electrons between the reservoir and channels 1 and 1/3, respectively, and $2\gamma_c^{-1}$ is the backscattering length for these channels in the contact region (Fig.~\ref{f4}). The combination $j_1+3j_2=\mu_1-\mu_2$ is given by the difference of the local partial chemical potentials (\ref{28}), a nonzero value of which produces the ``transverse" (between the channels) local current. The combination of $\mu$ and $j_{1,2}$ in the last terms in Eqs.~(\ref{36}) and (\ref{37}) means that the currents $j_1$ and $j_2$ tend to equilibrate, respectively, at $\mu/2\pi$ and $-(1/3)\times\mu/2\pi$, where $\mu$ is the chemical potential of electrons in the piece of metal that constitutes the side-attached contact, viewed as a thermal reservoir for electrons.

The origin of backscattering between the channels beneath the contact is, generically, twofold. Firstly, it may be related to direct disorder-induced tunneling between the channels. Importantly, the disorder strength may be different from that outside the contact region. For example, manufacturing the contact may introduce more disorder to the edge, as a result of which the line-junction contact with disorder-induced $\gamma_c\neq 0$ may be considered as connected to an otherwise clean edge. Secondly, virtual tunneling into and from the contact results in backscattering if the initial channel differs from the final one. However, in the limit of infinitesimally weak individual tunneling links to the contact (Sec.~\ref{s1a2}), these processes do not contribute to $\gamma_c$. Specifically, for the Hamiltonian of a single tunneling link
\be
{\mathcal H}_t^c=t_1\psi_c^\dagger\psi_1+t_2\psi_c^\dagger\psi_2+\textrm{h.c.}~,
\label{38}
\ee
where $\psi_c$ and $\psi_{1,2}$ are the electron operators at the point of tunneling in the contact and in channels 1 and 1/3, respectively, for the tunneling amplitudes $t_{1,2}\to 0$ and the link concentration $n_c\to\infty$ with $n_c|t_{1,2}|^2$ held fixed, the rates of tunneling to and from the contact, related to $\gamma_{1,2}$, are finite, whereas the contribution to $\gamma_c$ vanishes as $n_c|t_1t_2^*|^2\to 0$.

Note that Eq.~(\ref{38}) assumes that the tunneling link connects the reservoir to both channels 1 and 1/3. In a more general modeling of the experimental setup, the links to channels 1 and 1/3 need not be at the same points along the edge and, moreover, their densities may differ. Nevertheless, in the Gaussian limit of infinitesimally weak links, the virtual processes of tunneling into the contact do not contribute to $\gamma_c$ in any case. That said, it may be important to keep in mind that, beyond this limit, $\gamma_c\neq 0$ even if disorder-induced backscattering in the contact region is viewed as negligibly weak.

Equations (\ref{36}) and (\ref{37}) demonstrate a remarkable degree of universality with regard to the strength of interaction. A sufficient, for the purpose of determining the edge conductances, way to think of the source of nonequilibrium is to view it as the difference between the chemical potentials at different contacts represented as thermal reservoirs, as was already mentioned in Sec.~\ref{s2c}. The static effect of possible interaction between electrons inside the contact is then encoded in the difference between the chemical and electrochemical potentials outside the edge and, as such, does not show up explicitly in Eqs.~(\ref{36}) and (\ref{37}). Interaction between the channels beneath the contact manifests itself in the definition of the currents $j_{1,2}$ in Eq.~(\ref{14}), but not explicitly in Eqs.~(\ref{36}) and (\ref{37}). An immediate consequence of this is that, if $\Delta_c=\Delta$, the strength of interchannel interaction---beyond the effects of interaction-induced renormalization of the constants $\gamma_{1,2}$ and $\gamma_c$---drops out from the boundary condition at the contacts and the conductance itself \cite{Kane1995}. Note also that interaction between electrons in the edge and electrons in the contact does not affect the structure of Eqs.~(\ref{36}) and (\ref{37}), including the source terms. Below, we treat $\gamma_{1,2}$ and $\gamma_c$, which are dependent on particular microscopic details of the contact arrangement, as phenomenological parameters of the model \cite{linejunc}.

By introducing $\bar{j}_1=j_1-\mu/2\pi$ and $\bar{j}_2=j_2+(1/3)\times\mu/2\pi$, Eqs.~(\ref{36}) and (\ref{37}) can be rewritten as homogeneous ones:
\be
\left(\partial_x\mathbb{1}+\hat\upgamma_c\right)\left(\begin{array}{c}
\bar{j}_1\\
\bar{j}_2
\end{array}\right)=0~,
\label{39}
\ee
where the matrix of inverse scattering lengths beneath the contact $\hat\upgamma_c$ reads
\be
\hat\upgamma_c=\frac{1}{2}\left[\begin{array}{cc}
\gamma_c+2\gamma_1 & 3\gamma_c \\
-\gamma_c & -3(\gamma_c+2\gamma_2)
\end{array}\right]~.
\label{40}
\ee
The eigenvalues $k_{1,2}$ of $\hat\upgamma_c$ are
\be
k_{1,2}=\frac{\gamma_1-3\gamma_2-\gamma_c}{2}\pm \frac{Q}{2}~,
\label{41}
\ee
where
\begin{align}
Q=(\gamma_1+3\gamma_2-\gamma_c)\left[\,1+\frac{6\gamma_c(\gamma_1+3\gamma_2)}{(\gamma_1+3\gamma_2-\gamma_c)^2}\,\right]^{1/2}~,
\label{42}
\end{align}
with $k_1k_2\leq 0$.

\begin{figure}[t!]
\centering
\includegraphics[width=0.95\columnwidth]{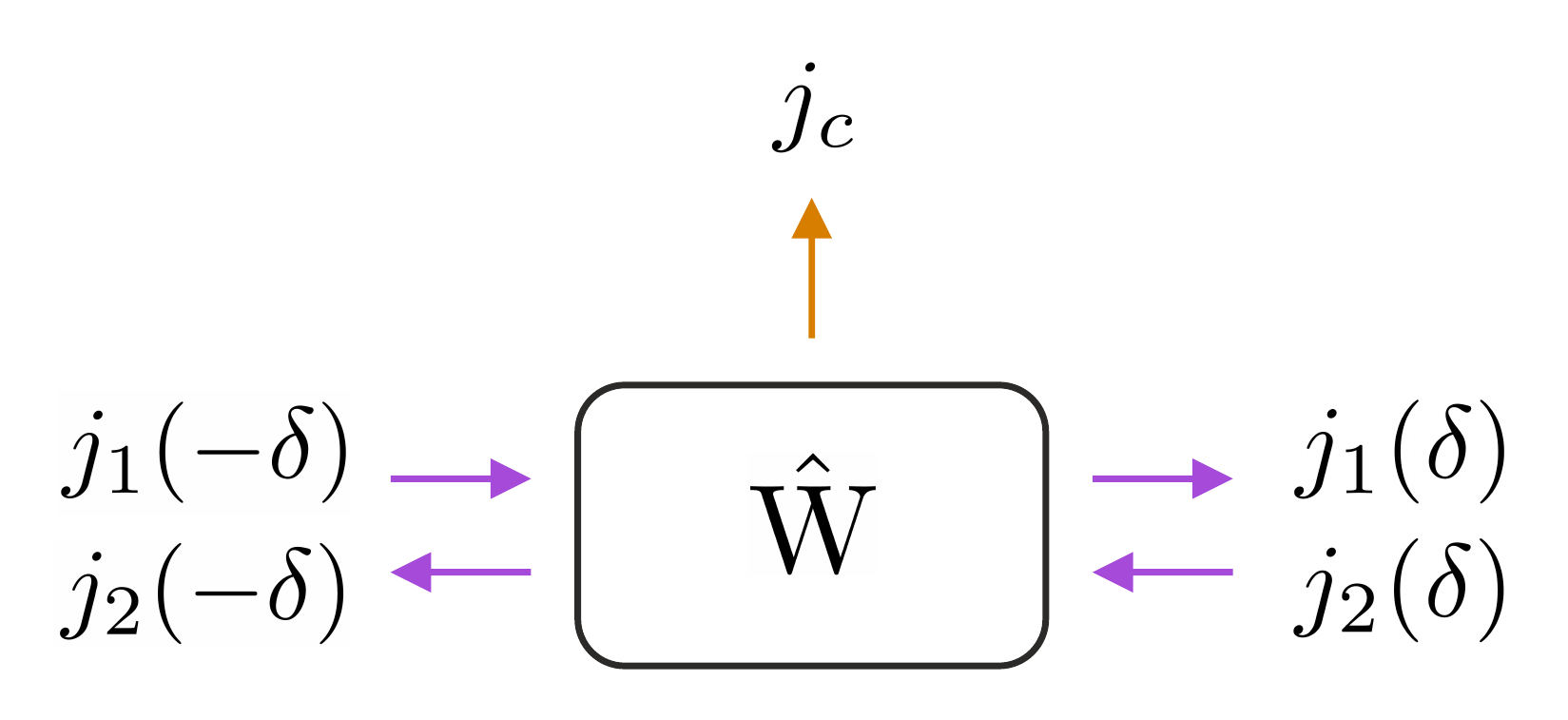}
\caption{Contact as a scatterer. Scattering is parametrized by the leaky scattering matrix $\hat{\textrm W}$. The matrix relates the outgoing [$j_1(\delta)$ and $j_2(-\delta)$] and incoming [$j_1(-\delta)$ and $j_2(\delta)$] currents [Eq.~(\ref{43})], and also expresses the ``leaking" current through the contact $j_c$ in terms of the incoming currents [Eq.~(\ref{45})].}
\label{f5}
\end{figure}

One of the useful ways to represent the solution of Eq.~(\ref{39}) is to view it as a scattering problem (Fig.~\ref{f5}), by expressing the outgoing currents $j_1(\delta)$ and $-j_2(-\delta)$ in terms of the incoming currents $j_1(-\delta)$ and $-j_2(\delta)$:
\be
\left[\begin{array}{c}
\bar{j}_1(\delta) \\
-\bar{j}_2(-\delta)
\end{array}\right]
=\hat{\textrm{W}}
\left[\begin{array}{c}
\bar{j}_1(-\delta) \\
-\bar{j}_2(\delta)
\end{array}\right]~,
\label{43}
\ee
where the leaky current-scattering matrix $\hat{\textrm{W}}$ is given by
\begin{align}
\hat{\textrm{W}}=\,\,\frac{1}{Q \cosh(Q\delta)+\left(\gamma_1+3\gamma_2+2\gamma_c\right)\sinh(Q\delta)}&\nonumber\\
\times\left[\begin{array}{cc}
Qe^{-(\gamma_1-3\gamma_2-\gamma_c)\delta} & 3\gamma_c \sinh(Q\delta)\\
\gamma_c\sinh(Q\delta) & Qe^{(\gamma_1-3\gamma_2-\gamma_c)\delta}
\end{array}\right]&
\label{44}
\end{align}
(note that $Q$ may be of either sign depending on the sign of $\gamma_1+3\gamma_2-\gamma_c$, with $\hat{\textrm{W}}$ being an even function of $Q$). The term ``leaky" in the definition of $\hat{\textrm{W}}$ means that there is a current through the contact $j_c=j_{c1}+j_{c2}$ (Fig.~\ref{f5}), with
\be
\left(\begin{array}{c}
j_{c1} \\
j_{c2}
\end{array}\right)
=\left(\mathbb{1}-\hat{\textrm{W}}\right)
\left[\begin{array}{c}
\bar{j}_1(-\delta) \\
-\bar{j}_2(\delta)
\end{array}\right]~.
\label{45}
\ee
In the limit of $\hat{\textrm{W}}\to 0$,
\be
j_c=\bar{j}_1(-\delta)-\bar{j}_2(\delta)=j_1(-\delta)-j_2(\delta)-\frac{4}{3}\frac{\mu}{2\pi}~,
\label{46}
\ee
which means complete absorption of the currents $j_1(-\delta)$ and $-j_2(\delta)$ that enter the contact.

\subsection{Conductance for the line-junction contacts}
\label{s2e}

In the limit of $\hat{\textrm{W}}\to 0$ [Eq.~(\ref{46})], combining two contacts at the chemical potentials $\mu_L$ and $\mu_R$, respectively, as shown in Fig.~\ref{f3a}, the two-terminal conductance of a clean edge is obtained as
\be
G=4/3~,
\label{47}
\ee
independently of the interchannel interaction strength beneath the line-junction contacts. Equation (\ref{47}) relies on the currents $j_{1,2}$ being continuous at the interface between the parts of the edge beneath the contacts, where the currents are described by Eqs.~(\ref{36}) and (\ref{37}), and the bulk of the edge. Importantly, this condition is fulfilled irrespective of the relation between $\Delta_c$ and $\Delta$, as demonstrated at the end of Sec.~\ref{s2b}. That is, Eq.~(\ref{47}) is valid for the clean edge with the line-junction contacts in the limit of $\hat{\textrm{W}}\to 0$ for arbitrary $\Delta_c$ and $\Delta$, being independent of both of them, also for $\Delta_c\neq\Delta$. This is in contrast to $G$ in Eq.~(\ref{32}) for the generalized boundary condition, where $G$ is independent of $\Delta$, but depends on $\Delta_c$. The independence of $G$ in Eq.~(\ref{47}) on $\Delta$ is thus in line with Eq.~(\ref{32}), but the line-junction contact with $\hat{\textrm{W}}\to 0$ corresponds to the contact described in Sec.~\ref{s2b} if one puts an effective parameter
\be
\Delta_c^\textrm{eff}=2~,
\label{48}
\ee
regardless of the actual one, in the generalized boundary conditions (\ref{17}) and (\ref{18}), or (\ref{22}) and (\ref{23}). Recall that, as was already mentioned in Sec.~\ref{s1a1}, Eq.~(\ref{47}) holds also for the model of noninteracting reservoirs [Eq.~(\ref{32}) with $\Delta_c=2$] attached to noninteracting spacers \cite{Protopopov2017}.

The independence of $G$ for the line-junction contacts on the interaction strength beneath them is a direct consequence of the fact that electron tunneling between the contact and the edge in Eqs.~(\ref{36}) and (\ref{37}) occurs between the contact and channels 1 and 1/3 separately \cite{chem-electrochem}. If tunneling occurred between the contact and channels $\pm$, i.e., the charges that constitute the eigenmode were tunneling to the contact from modes 1 and 1/3 simultaneously (and similarly for tunneling from the contact), the boundary conditions on the contact facets in the limit $\hat{\textrm{W}}\to 0$ would be those from Sec.~\ref{s2b}. The conductance $G$ would then be given by Eq.~(\ref{32}), instead of Eq.~(\ref{47}). By contrast, local tunneling from either mode 1 or 1/3 to the contact creates two density pulses running along the edge beneath the contact in opposite directions, i.e., excites both eigenmodes (the fractionalization dynamics will be considered in more detail in Sec.~\ref{s3}).

The mixing of modes $\pm$ by tunneling between modes 1 or 1/3 and the contact is the reason for the nonideality, as is manifest from Eq.~(\ref{47}), of the line-junction contact. Indeed, the condition $\hat{\textrm{W}}=0$ means that the contact is at equilibrium with modes 1 and 1/3, with $\bar{j}_1(\delta)=\bar{j}_2(-\delta)=0$. The outgoing eigenmode currents $j_+(\delta)$ and $j_-(-\delta)$ obey, then, Eqs.~(\ref{22}) and (\ref{23}) with the matrix $\hat\lambda$ corresponding to $\Delta_c=2$ [Eq.~(\ref{48})], i.e., with $\hat\lambda\neq\mathbb{1}$ unless $\Delta=2$ as well. Nonzero off-diagonal elements $\lambda_{+-,-+}\neq 0$ produce partial reflection of the eigenmodes incident on the contact by admixing them to the outgoing eigenmodes, while the difference of the diagonal elements from unity, $\lambda_{++,--}\neq 1$, means that the emitted eigenmodes are not at equilibrium with the contact. Both the former and the latter signify nonideality of the contact.

The limit of $\hat{\textrm{W}}\to 0$ can be achieved by first putting $\gamma_c\to 0$, which gives a diagonal matrix $\hat{\textrm{W}}$:
\be
\hat{\textrm{W}}\rightarrow\left(\begin{array}{cc}
e^{-2\gamma_1 \delta} & 0 \\
0 & e^{-6\gamma_2 \delta}
\end{array}\right)~,
\label{49}
\ee
and then taking the limit of $\delta\to\infty$. Equation (\ref{49}) simply describes an edge not equilibrated with the contact because of the finite size of the latter. More interesting---and essential to our discussion---is the limit of a long contact, $\delta\to\infty$, taken without making any assumption about $\gamma_c$ or $\gamma_{1,2}$. In this limit, $\hat{\textrm{W}}$ is purely nondiagonal:
\be
\hat{\textrm{W}}\to c\left(\begin{array}{cc}
0 & 3 \\
1 & 0
\end{array}\right)~,
\label{50}
\ee
characterized by a single parameter \cite{asymptW}
\be
c=\frac{1}{2+\zeta+\sqrt{(2+\zeta)^2-3}}~,
\label{51}
\ee
where
\be
\zeta=\frac{\gamma_1+3\gamma_2}{\gamma_c}~.
\label{52}
\ee
The constant $c$ is constrained by $0\leq c\leq 1/3$, with the limiting values of $c=0$ and $c=1/3$ obtained for $\zeta\to\infty$ and $\zeta\to 0$, respectively.

With $\hat{\textrm{W}}$ from Eq.~(\ref{50}), the currents $j_{1,2}(\pm\delta)$ on the facets of the left [Fig.~\ref{f3a}] contact are obtained, for the clean edge outside the contact regions, as
\begin{align}
j_1(\delta)&=\frac{1}{2\pi}\,\frac{(1-c)\mu_L+c(1-3c)\mu_R}{1-3c^2}~,
\label{53}\\
j_2(\delta)&=-\frac{1}{2\pi}\,\frac{c(1-c)\mu_L+\frac{1}{3}(1-3c)\mu_R}{1-3c^2}~,
\label{54}
\end{align}
and
\be
j_{1,2}(-\delta)=j_{1,2}(\delta)\vert_{\mu_L\leftrightarrow\mu_R}~.
\label{55}
\ee
The resulting conductance $G=2\pi j_c/(\mu_L-\mu_R)$, with the current through the contact $j_c=j_1(\delta)+j_2(\delta)-(\delta\to -\delta)$ [or, equivalently, Eq.~(\ref{45})], is then given by
\be
G=\frac{2}{3}\,\Delta_c^\textrm{eff}~,
\label{56}
\ee
where
\be
\Delta_c^\textrm{eff}=1+\frac{(1-3c)^2}{1-3c^2}~.
\label{57}
\ee
For $c=0$, Eq.~(\ref{56}) reduces to Eq.~(\ref{47}). As $\zeta$ [Eq.~(\ref{52})] decreases with increasing strength of backscattering beneath the contact, $G$ changes from 4/3 for $c=0$ to 2/3 for $c=1/3$.

As follows from Eqs.~(\ref{56}) and (\ref{57}), there exists a surface in space of $\gamma_1$, $\gamma_2$, and $\gamma$ on which
\be
\Delta_c^\textrm{eff}=\Delta~.
\label{58}
\ee
On this surface, the contact is ideal, with the emitted currents being at equilibrium with the contact, namely $j_\pm(\pm\delta)=\pm g_\pm\mu/2\pi$ [cf.\ Eq.~(\ref{24})]. At the same time, the incident eigenmode currents $j_\pm(\mp\delta)$ are completely absorbed. This construction is thus the embodiment of the notion of an ideal contact---for an arbitrary strength of interchannel interaction both beneath the contact and outside it.

Away from the surface on which Eq.~(\ref{58}) holds, the contact is at equilibrium with the eigenmodes corresponding to the effective interaction parameter $\Delta_c^\textrm{eff}$ from Eq.~(\ref{57}). As a consequence, the calculation in Secs.~\ref{s2b} and \ref{s2c} for the generalized boundary condition applies directly to the line-junction model, with $\textbf{n}_c$ understood as the eigenmodes for $\Delta_c$ substituted by $\Delta_c^\textrm{eff}$. The purpose of making a ``digression" to discuss the generalized boundary condition in Secs.~\ref{s2b} and \ref{s2c} has now become clear. As a starting point, it demonstrated in a concise and precise manner that the strength of interaction outside the contacts drops out from the conductance of a clean edge [Eqs.~(\ref{32}), (\ref{34}), and (\ref{35})]. Perhaps more importantly, when combined with the calculation for the line-junction, it serves as a basis for the physical picture in which the conductance does not depend on either $\Delta$ or $\Delta_c$, but is determined by the effective strength of interaction beneath the side-attached contact, parametrized by $\Delta_c^\textrm{eff}$ [Eq.~(\ref{56})].

From this perspective, the conductance $G$ varies between 2/3 and 4/3---with the value of $\Delta_c^\textrm{eff}$ varying between 1 and 2---being controlled by the single parameter (\ref{51}) that reflects the interplay of tunneling between the edge channels beneath the contact and tunneling between the contact and the edge. This establishes a significant degree of universality of the model of long line-junction contacts. Instead of contrasting, as in Ref.~\onlinecite{Kane1995}, the model of ideal contacts on the one hand and the model of line-junction contacts on the other, this viewpoint emphasizes that the two models are not at all different from each other in terms of universality. If a contact of either type is viewed as a ``black box" (with arbitrary microscopic ``content"), there is one single parameter that characterizes the contact as far as the current flowing through it is concerned.

\section{Fractionalization-renormalized tunneling}
\label{s3}

As an interlude between the discussions of the clean (Sec.~\ref{s2}) and disordered (Sec.~\ref{s4}) edges, it is instructive to formulate the ``electrostatic" approach to disorder-induced tunneling for the $\nu=2/3$ edge. In essential terms, we mean electrostatics of tunneling that combines two effects: (i) the creation of screening charges in channels 1 and 1/3 by the tunneling electron and the hole that it leaves behind and (ii) the ensuing fractionalization of the created charges, resolved with respect to both the edge channels and chirality. This is quite apart from the renormalization of the tunneling strength that is due to the interaction-induced orthogonality catastrophe upon tunneling (``zero-bias anomaly" in the tunneling density of states in the presence of a scatterer) \cite{Kane1995b,Moore1998}. The latter is formalizable in terms of the infrared-singular correlator of quantum fluctuations of the electron density around the tunneling link. By contrast, the fractionalization-induced renormalization of the tunneling strength describes ultraviolet---short-range in real space---correlations between the charges created by the tunneling event. These correlations are crucial for determining the $\Delta$ dependent relation between the tunneling rates for different channels. In particular, they are behind the emergence of the neutral mode at $\Delta=1$ \cite{Kane1994}, irrespective of the strength of disorder, from the point of view of electrostatics, as will be seen below.

\subsection{Fractionalization upon tunneling into the edge}
\label{s3a}

Let us first consider fractionalization in the $\nu=2/3$ edge upon addition, locally, of a unit charge to channel 1 or 1/3, without tunneling between them. The matrix $\hat\Lambda$ that expresses the eigenmodes $\textbf{n}=(n_+,n_-)^T$ in terms of $\bar{\textbf n}=(n_1,n_2)^T$ for the interaction parameter $\Delta$,
\be
\textbf{n}=\hat\Lambda\,\bar{\textbf n}~,
\label{59}
\ee
is given by $\hat\Lambda=\hat\lambda\vert_{\Delta_c=2}$, with $\hat\lambda$ from Eq.~(\ref{21}). Note that $\hat\Lambda$ preserves the total charge [Eq.~(\ref{5})]. The ratio $\eta_+$ of $n_2$ and $n_1$ in mode + and the ratio $\eta_-$ of $n_1$ and $n_2$ in mode $-$~ are, then, written as
\begin{align}
\eta_+&=\left(n_2/n_1\right)_+=\left.\left(\hat\Lambda^{-1}\right)_{21}\middle/\left(\hat\Lambda^{-1}\right)_{11}~,\right.
\label{60}\\
\eta_-&=\left(n_1/n_2\right)_-=\left.\left(\hat\Lambda^{-1}\right)_{12}\middle/\left(\hat\Lambda^{-1}\right)_{22}~,\right.
\label{61}
\end{align}
where
\begin{align}
\hat\Lambda^{-1}=&\,\,\frac{1}{2}\left(\begin{array}{rr}
3 & 3 \\
-1 & -1
\end{array}\right)\nonumber\\
+&\,\,\frac{1}{2}\sqrt{\frac{3}{\Delta^2-1}}
\left(\begin{array}{rr}
1-\Delta & \,-1-\Delta\\
-1+\Delta & \,1+\Delta
\end{array}\right)~.
\label{62}
\end{align}

The constants $\eta_\pm$ have the meaning of screening charges; specifically, $\eta_+$ is the charge in channel 1/3 induced by a unit charge in channel 1, when both run to the right (mode +), and $\eta_-$ is the charge in channel 1 induced by a unit charge in channel 1/3, when both run to the left (mode $-$). They are related to each other as
\be
\frac{\eta_+}{\eta_-}=\frac{1}{3}~,
\label{63}
\ee
with
\be
\eta_+=-\frac{\sqrt{(\Delta+1)/3}-\sqrt{\Delta-1}}{\sqrt{3(\Delta+1)}-\sqrt{\Delta-1}}~.
\label{64}
\ee
While keeping in mind the relation (\ref{63}), the fractionalization picture is most clearly formulated in terms of both $\eta_+$ and $\eta_-$.

For $\Delta=2$, the screening charges vanish: $\eta_\pm=0$. For $\Delta=1$, they are given by
\be
\eta_-=3\eta_+=-1~.
\label{65}
\ee
Note that screening is perfect, with the screening charge $\eta_-=-1$ being exactly the mirror charge, for $\Delta=1$ in mode $-$, which becomes then the ``neutral mode" identified in Ref.~\onlinecite{Kane1994}. The charge carried by mode $-$ vanishes at $\Delta=1$ as
\be
n_-\to\sqrt{\frac{\Delta-1}{6}}\,(n_1+3n_2)~,
\label{66}
\ee
where $n_1+3n_2\to 2n_2$ according to Eq.~(\ref{65}). In mode +, which becomes for $\Delta=1$ the ``charge mode" with $n_+=n_1+n_2$, the ``screening coefficient" $\eta_+=-1/3$, so that the unit charge propagating to the right at $\Delta=1$ is split into the charge 3/2 in mode 1 and the charge $-1/2$ in mode 1/3 \cite{rmrk2}.

\begin{figure*}[ht!]
\centering
\includegraphics[width=0.95\columnwidth]{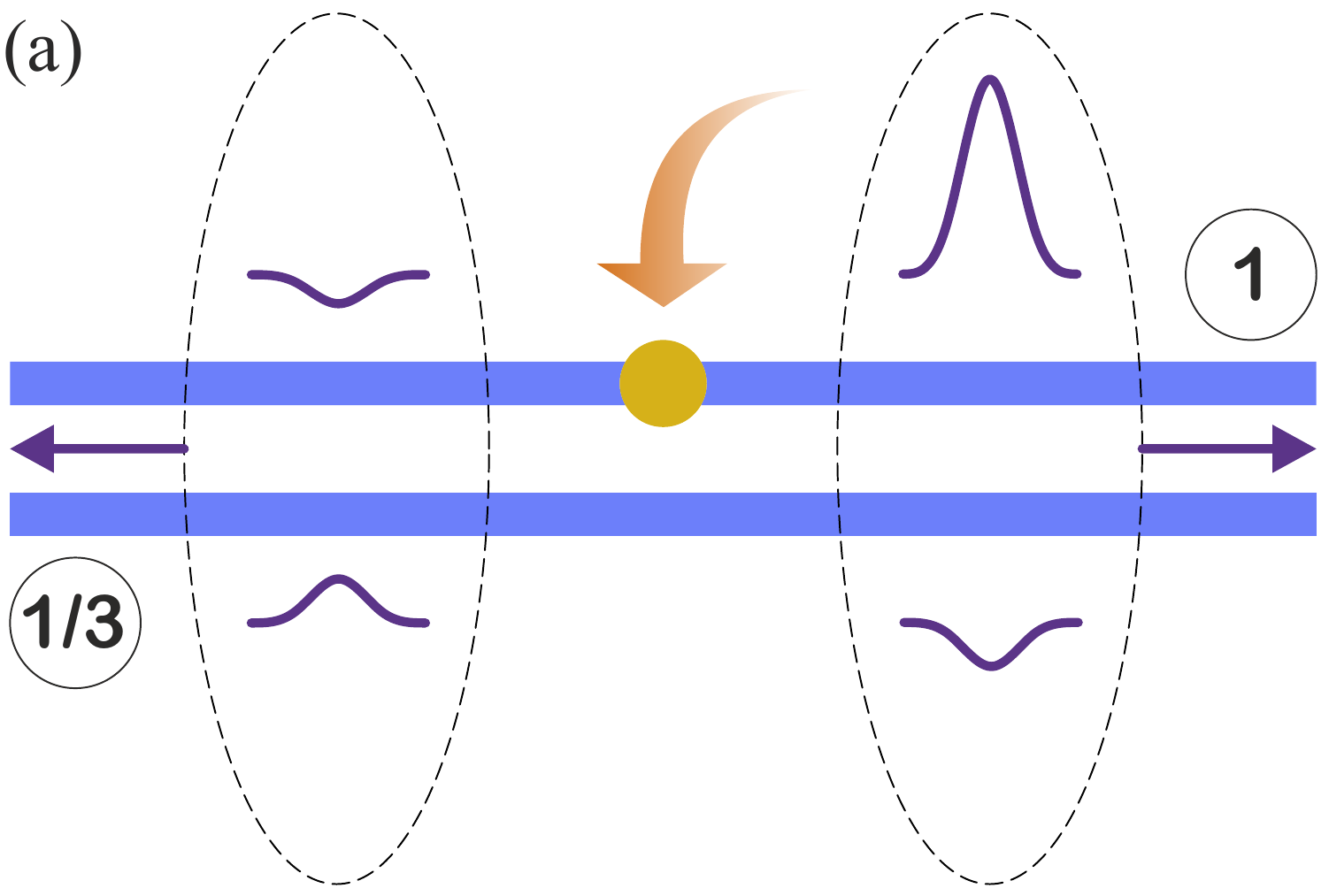}\qquad\quad
\includegraphics[width=0.95\columnwidth]{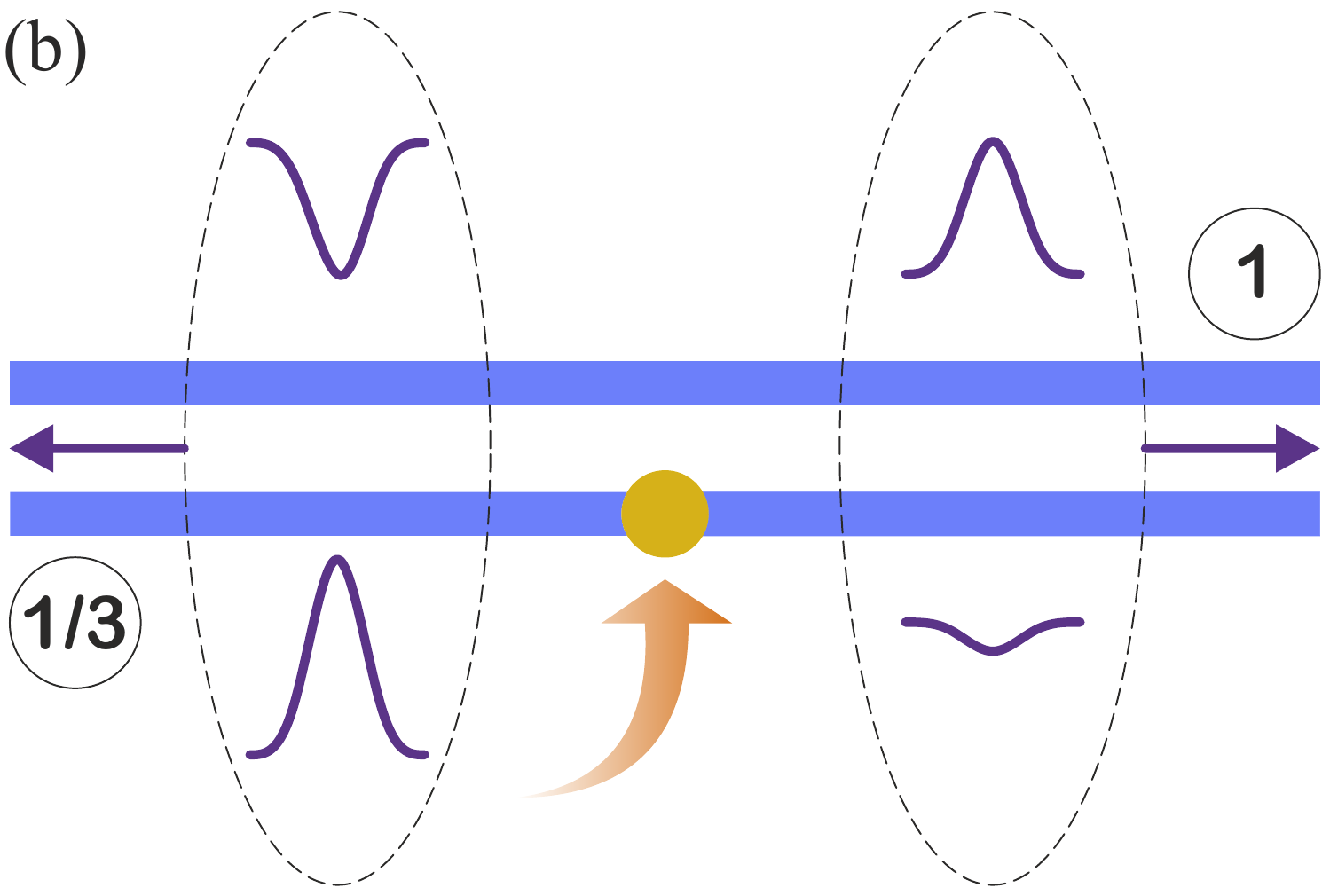}
\caption{Charge fractionalization upon addition of a compact charge to channel 1 (a) or channel 1/3 (b). The relative amplitudes of the fractionalized charge pulses are shown for $\Delta=26/23\simeq 1.13$ (which corresponds to the screening coefficients $\eta_+=-2/9$ and $\eta_-=-2/3$). The charge distribution is described by the fractionalization matrix given by Eqs.~(\ref{67}) and (\ref{68}) for cases (a) and (b), respectively.}
\label{f6}
\end{figure*}

The fractionalization process in a clean edge is illustrated in Fig.~\ref{f6}. It is convenient to introduce the ``fractionalization matrix" $\hat q$, whose elements are $q_{\pm,\textrm{i}}$ with $\textrm{i}=1,2$, where $q_{+,1}$ is the charge running in channel 1 in mode +, etc., upon insertion of a unit charge in channel 1. From Eqs.~(\ref{60})-(\ref{62}), $\hat q$ is obtained as
\begin{widetext}
\bea
\hat{q}&=&\frac{1}{1-\eta_+\eta_-}\left(\begin{array}{cc}
1 & \eta_+ \\
-\eta_+\eta_- & -\eta_+
\end{array}\right)\nonumber\\ \nonumber\\
&=&\frac{1}{4}\left(\begin{array}{rr}
\left[\sqrt{3(\Delta+1)}-\sqrt{\Delta-1}\,\right]^2 & \,\,\,-\left[\sqrt{3(\Delta+1)}-\sqrt{\Delta-1}\,\right]\left[\sqrt{(\Delta+1)/3}-\sqrt{\Delta-1}\,\right] \\
-\left[\sqrt{\Delta+1}-\sqrt{3(\Delta-1)}\,\right]^2 & \,\,\,\left[\sqrt{3(\Delta+1)}-\sqrt{\Delta-1}\,\right]\left[\sqrt{(\Delta+1)/3}-\sqrt{\Delta-1}\,\right]
\end{array}\,\right)~.
\label{67}
\eea
The unit local charge added to channel 1 thus splits, for $\Delta\neq 2$, into two parts running in opposite directions and creates a growing dipole in channel 1/3. The fractional charges obey charge conservation with $q_{+,1}+q_{-,1}=1$ and $q_{+,2}+q_{-,2}=0$. Similarly, when a unit local charge is inserted into channel 1/3, the fractionalization matrix $\hat{\bar q}$ (where the bar is put to distinguish it from the fractionalization matrix for insertion into channel 1) reads
\bea
\hat{\bar q}&=&\frac{1}{1-\eta_+\eta_-}\left(\begin{array}{cc}
-\eta_- & -\eta_+\eta_- \\
\eta_- & 1
\end{array}\right)\nonumber\\ \nonumber\\
&=&\frac{3}{4}\left(\begin{array}{rr}
\left[\sqrt{\Delta+1}-\sqrt{3(\Delta-1)}\,\right]\left[\sqrt{\Delta+1}-\sqrt{(\Delta-1)/3}\,\right] & \,\,\,-\left[\sqrt{(\Delta+1)/3}-\sqrt{\Delta-1}\,\right]^2 \\
-\left[\sqrt{\Delta+1}-\sqrt{3(\Delta-1)}\,\right]\left[\sqrt{\Delta+1}-\sqrt{(\Delta-1)/3}\,\right] & \,\,\,\left[\sqrt{\Delta+1}-\sqrt{(\Delta-1)/3}\,\right]^2
\end{array}\,\right)~,
\label{68}
\eea
\end{widetext}
with $\bar{q}_{+,1}+\bar{q}_{-,1}=0$ and $\bar{q}_{+,2}+\bar{q}_{-,2}=1$. Note also the relation $q_{\pm,1}=\bar{q}_{\mp,2}$.

For $\Delta=1$, the matrix $\hat q$ takes a simple form, as discussed around Eqs.~(\ref{65}) and (\ref{66}):
\be
\hat{q}=\frac{1}{2}\left(\begin{array}{rr}
3 & -1 \\
-1 & 1
\end{array}\right)~,
\label{69}
\ee
whereas $\hat{\bar q}$ for $\Delta=1$ is given by
\be
\hat{\bar q}=\frac{1}{2}\left(\begin{array}{rr}
3 & -1 \\
-3 & 3
\end{array}\right)~.
\label{70}
\ee
In Eqs.~(\ref{69}) and (\ref{70}), perfect screening in the neutral mode is apparent in that the sum of the matrix elements in the lower row is zero in both cases. Note that the amplitude of the charge mode excited by adding a unit charge at $\Delta=1$ to channel 1 is the same as by adding it to channel 1/3. By contrast, the amplitude of the neutral mode is a factor of 3 larger in the latter case.

The phenomenon of charge fractionalization has two facets, inherently related to each other. Specifically, at the single-particle level, it refers to the factorization of a single-particle propagator in the space-time representation into a product of parts moving with different velocities (with proper care taken in dealing with the ultraviolet cutoff \cite{Solyom1979,*Voit1995}), as is the case, e.g., for electrons in a LL. At the level of two-particle correlations, this results in splitting of a compact density pulse into parts characterized by different velocities of propagation. The above picture of fractionalization in the $\nu=2/3$ edge is a generalization---formalized in purely electrostatic terms---of the fractionalization picture in a LL \cite{Pham2000,Steinberg2008,LeHur2008,Leinaas2009} to the case of two nonequivalent counterpropagating channels. Here, ``nonequivalent counterpropagating" signifies that two channels together constitute a chiral system---the one characterized by $g_+\neq g_-$.

It is worth mentioning that, in the original sense, the notion of charge fractionalization in a two-channel (spinless) LL refers to splitting of the density pulse into two chiral parts \cite{Pham2000,Steinberg2008,LeHur2008} (in a spinful LL, the ``chiral" separation is complemented with spin-charge separation). In Eqs.~(\ref{67})-(\ref{70}), the charge splits into four parts (spatially separated both in the longitudinal and transverse directions), with fractionalization ``resolved" with respect to not only chirality but also channels. This is similar to the charge-fractionalization picture in the single-channel edge for $\nu=1$, when counterpropagating parts of the edge are brought in proximity to each other and form together a nonchiral system akin to the LL \cite{Berg2009,Leinaas2009,*Horsdal2011,Kamata2014,Brasseur2017}, or in two copropagating channels in the edge for $\nu=2$ \cite{Berg2009,Neder2012,Bocquillon2013,Milletari2013,Inoue2014a,Hashisaka2017,Lin2021}. In the latter case, the fractionalized parts of the density pulse run in the same direction, with the decoupling of the charge and neutral modes being equivalent to spin-charge separation. For $\nu=2/3$, charge fractionalization was probed experimentally \cite{Lin2021} through time-resolved scattering of current pulses off the interface between regions with zero and nonzero strength of interchannel interaction (cf.\ the time-resolved measurements of this type of scattering in an artificial LL \cite{Kamata2014,Brasseur2017} and in the two-channel edge for $\nu=2$ \cite{Lin2021}).

\subsection{Fractionalization upon intermode tunneling}
\label{s3b}

\begin{figure*}[ht!]
\centering
\includegraphics[width=1.5\columnwidth]{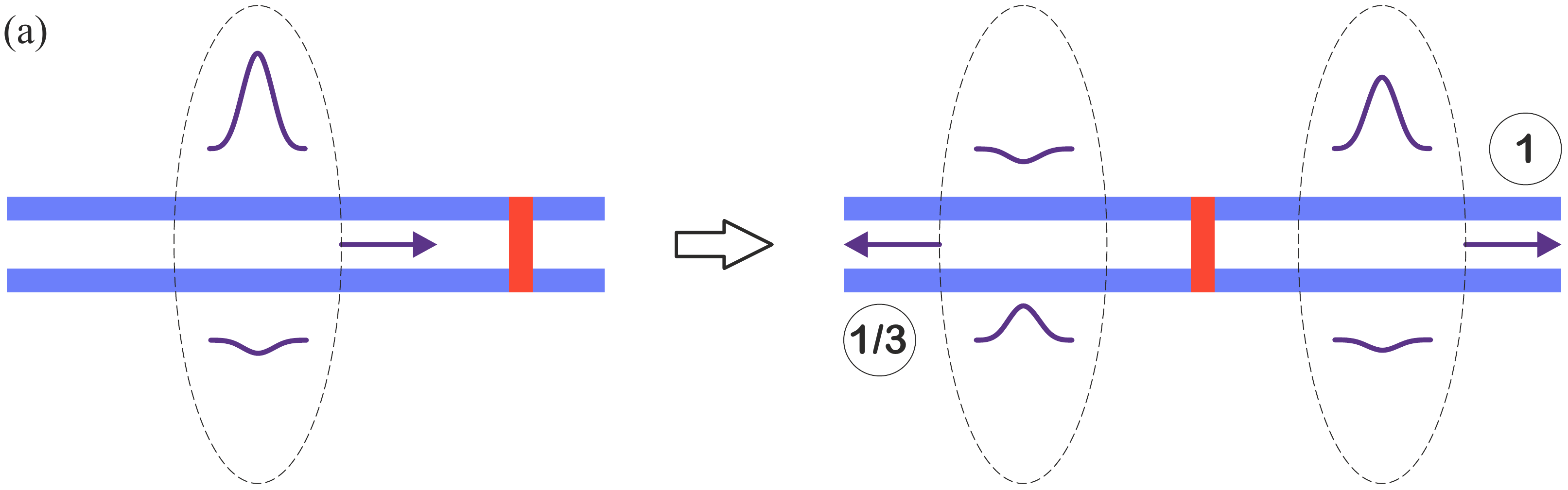}\vspace{5mm}
\includegraphics[width=1.5\columnwidth]{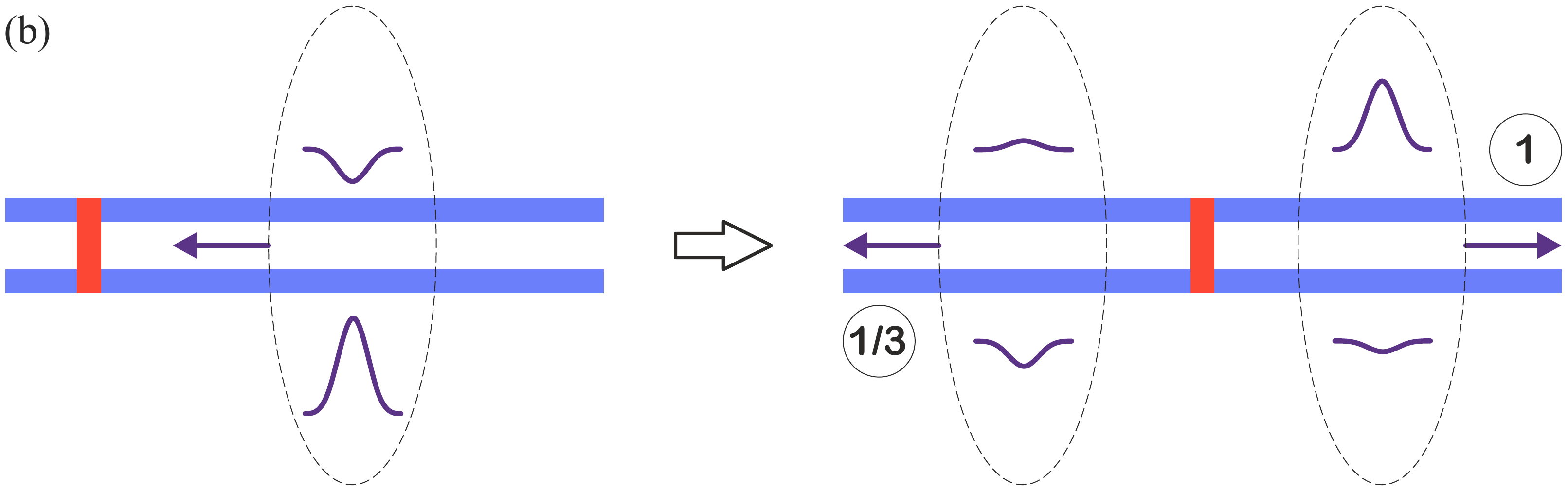}
\caption{Fractionalization-renormalized tunneling. Scattering of a charge pulse incident on a tunneling link (marked in red) from the left (a) and from the right (b). Electron tunneling occurs between the right-moving channel 1 and left-moving channel 1/3. The relative amplitudes of the components of the incoming and outgoing pulses in channels 1 and 1/3 are shown for $\Delta=3/2$ (which corresponds to the screening coefficients $\eta_+\simeq -0.1$ and $\eta_-\simeq -0.3$). The outgoing pulses are shown for $R_0=3/2$ [Eq.~(\ref{89})] (which corresponds for $\Delta=3/2$ to the reflection coefficients $R_+=1/4$ and $R_-=5/4$), with their scattering amplitudes given by Eqs.~(\ref{87}) and (\ref{88}).}
\label{f7}
\end{figure*}

Imagine now that a compact unit charge in channel 1 is incident on a tunneling link in mode + (Fig.~\ref{f7}). As follows from Sec.~\ref{s3a}, this charge is accompanied by the screening charge $\eta_+$ in channel 1/3. Either of the two can tunnel upon hitting the tunneling link. Scattering of the right-moving composite object consisting of the charges 1 and $\eta_+$ is thus a combination of four processes: (i) tunneling of the charge 1 to channel 1/3 and its ensuing fractionalization are accompanied by (ii) fractionalization of the charge $-1$ left in channel 1 and, similarly, (iii) tunneling of the charge $\eta_+$ to channel 1 and its fractionalization are accompanied by (iv) fractionalization of the charge $-\eta_+$ left in channel 1/3.

Specifically, to describe fractionalization in the process of tunneling, it is instructive to first consider the matrix $\hat Q$ with elements $Q_{\pm,\textrm{i}}$, where $Q_{+,1}$ is the charge running in channel 1 in mode +, etc.\ [cf.\ Eqs.~(\ref{67}) and (\ref{68})]. Scattering of the charge pulse in mode + is then formalized by
\be
\hat{Q}_\textrm{out}-\hat{Q}_\textrm{in}=\left(W_1^+-\eta_+W_2^+\right)\left(\hat{\bar q}-\hat{q}\right)~,
\label{71}
\ee
where
\be
\hat{Q}_\textrm{in}=\left(\begin{array}{cc}
1 & \eta_+ \\
0 & 0
\end{array}\right)
\label{72}
\ee
is the matrix of incoming right-moving charges, $\hat{Q}_\textrm{out}$ is the matrix of outgoing charges, and $W_1^+$ and $W_2^+$ are the scattering coefficients in mode + for tunneling from channels 1 and 1/3, respectively. The matrix of scattered charges, proportional to the combination $\hat{\bar q}-\hat{q}$, contains four terms, each of which describes one of the four processes mentioned above. Namely, the part proportional to $\hat{\bar q}$ stands for creation (the term with $W_1^+$) of the charge 1 and annihilation ($W_2^+$) of the charge $\eta_+$ in channel 1/3, while the part proportional to $\hat q$ denotes annihilation ($W_1^+$) of the charge 1 and creation ($W_2^+$) of the charge $\eta_+$ in channel 1. For scattering of a charge pulse in mode $-$, with scattering coefficients $W^-_1$ and $W^-_2$ for tunneling from channels 1 and 1/3 in mode $-$, respectively, the analog of Eq.~(\ref{71}) is given by
\be
\hat{Q}_\textrm{out}-\hat{Q}_\textrm{in}=\left(W_2^--\eta_-W_1^-\right)\left(\hat{q}-\hat{\bar q}\right)~,
\label{73}
\ee
where $\hat{Q}_\textrm{in}$ is chosen as the matrix of incoming left-moving charges:
\be
\hat{Q}_\textrm{in}=\left(\begin{array}{lc}
0 & \!0 \\
\eta_- & \!1
\end{array}\right)~.
\label{74}
\ee
The explicit form of the scattering coefficients $W^\pm_{1,2}$ will be discussed below [Eqs.~(\ref{82}) and (\ref{83})].

Having described the mechanism of fractionalization in the process of tunneling in a detailed and pictorial way by means of Eqs.~(\ref{71}) and (\ref{73}), we now combine them in the form of scattering theory for an arbitrary two-vector of the eigenmode charges $\textbf{Q}=(Q_+,Q_-)^T$, where $Q_+=Q_{+,1}+Q_{+,2}=Q_{+,1}(1+\eta_+)$ and $Q_-=Q_{-,1}+Q_{-,2}=Q_{-,2}(1+\eta_-)$. Specifically, the relation between the vectors of incoming ($\textbf{Q}_\textrm{in}$) and outgoing ($\textbf{Q}_\textrm{out}$) charges is written as
\be
\textbf{Q}_\textrm{out}=\left(\mathbb{1}+\hat{\mathcal T}\right)\textbf{Q}_\textrm{in}~,
\label{75}
\ee
where the $T$ matrix for scattering at a single tunneling link reads
\be
\hat{\mathcal T}=\left(\begin{array}{rr}
-R_+ & R_- \\
R_+ & -R_-
\end{array}\right)~,
\label{76}
\ee
with $R_\pm$ being the reflection coefficients for the charge pulses in modes $\pm$. The matrix structure of Eq.~(\ref{76}) signifies charge conservation in the tunneling process for arbitrary screening charges. From Eqs.~(\ref{71}) and (\ref{73}),
\begin{align}
&R_+=(q_{+,1}-\bar{q}_{+,1})\left(W_1^+-\eta_+W_2^+\right)~,
\label{77}\\
&R_-=(\bar{q}_{-,2}-q_{-,2})\left(W_2^--\eta_-W_1^-\right)~,
\label{78}
\end{align}
where [Eqs.~(\ref{67}) and (\ref{68})]
\begin{align}
&q_{+,1}-\bar{q}_{+,1}=\frac{\sqrt{\Delta-1}}{2}\left[\sqrt{3(\Delta+1)}-\sqrt{\Delta-1}\,\right]~,
\label{79}\\
&\bar{q}_{-,2}-q_{-,2}=\frac{\sqrt{\Delta+1}}{2\sqrt{3}}\left[\sqrt{3(\Delta+1)}-\sqrt{\Delta-1}\,\right]~.
\label{80}
\end{align}

Note that the matrix (\ref{76}), which relates the compact charges, i.e., the time integrals of the current pulses, is also understood, by extension, as the current-splitting matrix that relates the incoming and outgoing eigenmode currents $j_\pm$ on the sides of the tunneling link in a stationary state by
\be
\left(\begin{array}{r}
j_+\\
-j_-
\end{array}\right)_\textrm{out}=\left(\mathbb{1}+\hat{\mathcal T}\right)\left(\begin{array}{r}
j_+\\
-j_-
\end{array}\right)_\textrm{in}~.
\label{81}
\ee
In the stationary case, the currents $j_\pm$ are spatially homogeneous on the sides of the link and experience a jump at it, as given by Eq.~(\ref{81}), with the sum $j_++j_-$ being continuous across the link. This is how the fractionalization physics, encoded in the interaction-renormalized matrix $\hat{\mathcal T}$ through Eqs.~(\ref{77}) and (\ref{78}), affects the time-averaged current.

Now turn to the scattering coefficients $W^\pm_{1,2}$. Let, for concreteness, the Hamiltonian for a single tunneling link be of the form ${\mathcal H}_t=t_0\psi_1^\dagger\psi_2+\textrm{h.c.}$ In the Born approximation with respect to tunneling, $W^\pm_{1,2}$ read
\begin{align}
&W^\pm_1=2\pi|\tilde{t}_0|^2\frac{1}{v_\pm}\!\left(\bar{\Pi}_{22}+\bar{\Pi}_{21}\right)~,
\label{82}\\
&W^\pm_2=2\pi|\tilde{t}_0|^2\frac{1}{v_\pm}\!\left(\bar{\Pi}_{11}+\bar{\Pi}_{12}\right)~,
\label{83}
\end{align}
where $\hat{\bar\Pi}$ is the compressibility matrix already introduced by defining the local chemical potentials in Eqs.~(\ref{27}) and (\ref{28}), and the tilde in $\tilde{t}_0$ denotes the renormalization of $t_0$ by the interaction-mediated orthogonality catastrophe \cite{Kane1995b,Moore1998}. In the limit of no interchannel interaction, $W^+_1=W^-_2/3=|\tilde{t}_0|^2/3v_1v_2$ [these two coefficients are then the only ones that enter Eqs.~(\ref{77}) and (\ref{78}), with $R_+=W^+_1$ and $R_-=W^-_2$].

The factor $1/v_\pm$ in Eqs.~(\ref{82}) and (\ref{83}) comes from the amplitude of the incident current, while the factors in the brackets are the total thermodynamic compressibilities for placing a charge in channel 1/3 ($W^\pm_1$) and 1 ($W^\pm_2$). Substituting Eq.~(\ref{27}) in Eqs.~(\ref{82}) and (\ref{83}), and the resulting expressions for $W^\pm_{1,2}$ in Eqs.~(\ref{77}) and (\ref{78}), taking account of Eqs.~(\ref{79}) and (\ref{80}), the reflection coefficients $R_\pm$ reduce to
\be
R_\pm=|\tilde{t}_0|^2\frac{1}{3v_+v_-}(\Delta\mp 1)~.
\label{84}
\ee
Equation (\ref{84}) is also representable in the form
\be
R_\pm=|\tilde{t}_0|^2\frac{g_\mp}{v_+v_-}=2\pi|\tilde{t}_0|^2\frac{1}{v_\pm}\Pi_{\mp\mp}~,
\label{85}
\ee
where $\Pi_{++}=g_+/2\pi v_+$ and $\Pi_{--}=g_-/2\pi v_-$ are the elements of the eigenmode compressibility matrix (\ref{19}). Note that $R_+$ and $R_-$ are universally related to each other: $R_+/R_-=(\Delta-1)/(\Delta+1)$, with this universality being entirely due to fractionalization upon tunneling.

It is worth emphasizing that two types of interaction-induced renormalization of $R_\pm$, one by fractionalization and the other by the orthogonality catastrophe, are sharply distinguished in Eq.~(\ref{85}). Specifically, the former is inherently linked to the behavior of the thermodynamic compressibility $\Pi_{\mp\mp}$ with varying $\Delta$, whereas the latter is encoded in the modification of $|\tilde{t}_0|^2$ by a factor that represents the difference between the thermodynamic compressibility and the single-particle tunneling density of states. While the simple form of Eq.~(\ref{85}) is suggestive, it is the ``unfolding" of Eq.~(\ref{85}) in terms of two factors in Eqs.~(\ref{77}) and (\ref{78}) that reveals the mechanism of fractionalization-renormalized tunneling that is behind the emergence of the factor $\Delta\mp 1$ in Eq.~(\ref{84}).

The tunneling conductance for a weak tunneling link connecting channels 1 and 1/3 for arbitrary $\Delta$, which conforms with Eq.~(\ref{84}), was also obtained by a direct perturbation theory for the tunneling current in Ref.~\onlinecite{Srivastav2020}. It is worth noting that, for weak tunneling, the factor $|\tilde{t}_0|^2/3$ in Eq.~(\ref{84}) reduces exactly to $|t_0|^2$ renormalized by the orthogonality catastrophe upon tunneling in a nonchiral LL (assuming the same ultraviolet cutoff) with a substitution of $\Delta$ for the Luttinger constant (supplemented with a straightforward substitution of $v_+$ and $v_-$ for the plasmon velocity in the right- and left-moving parts of the tunneling operator, respectively). In particular, for a given infrared cutoff of the renormalization, say, the temperature $T$, the factor $|\tilde{t}_0|^2$ for a weak tunneling link scales with $T$ as $T^{2(\Delta-1)}$. What is, however, especially revealing in the structuring of $R_\pm$ into the product of $|\tilde{t}_0|^2$ and the fractionalization-induced factor is that the former does not show any peculiar behavior near $\Delta=1$ (apart from the vanishing of the scaling exponent in the power-law renormalization, in exact correspondence with the noninteracting case in a LL), in contrast to the latter, as we discuss next.

One significant feature of the dependence of $R_\pm$ on $\Delta$ is the vanishing of $R_+$ at $\Delta=1$. The insensitivity of the + mode to disorder for $\Delta=1$ is a manifestation of the charge and neutral mode decoupling \cite{Kane1994}. Equations~(\ref{77}), (\ref{79}), and (\ref{82}) show precisely how the decoupling occurs from the point of view of the underlying physics. Specifically, the factor $\Delta-1$ in Eq.~(\ref{84}) is a product of two factors $\sqrt{\Delta-1}$ of distinctly different origin. The factor $\sqrt{\Delta-1}$ in Eq.~(\ref{79}) reflects partial cancellation, to the right of the tunneling link, of the screening charge ($\bar{q}_{+,1}$) for a fractionalized electron that has tunneled to channel 1/3 and the charge of the fractionalized hole in channel 1 ($-q_{+,1}$) that the electron left behind. At $\Delta=1$, the cancellation is exact and, as a result, no mode $-$ is created upon mode + hitting the tunneling link, so that mode + passes through the link without distortion. The other factor $\sqrt{\Delta-1}$ stems from partial cancellation of the tunneling current of the charge transferred from channel 1 to channel 1/3 ($W^+_1$) by the backflow tunneling current of the screening charge from channel 1/3 to channel 1 ($-\eta_+W^+_2$). Again, at $\Delta=1$, the cancellation is exact.

From Eqs.~(\ref{73}), (\ref{78}), and (\ref{80}), the $-$ mode for $\Delta=1$ is scattered off a tunneling link according to
\be
\hat{Q}_\textrm{out}=(1-R_-)\left(\begin{array}{rr}
0 & 0 \\
-1 & 1
\end{array}\right)~,
\label{86}
\ee
where the matrix on the right hand side stands for the incident $-$ mode [Eq.~(\ref{74})] and the term proportional to $R_-$ corresponds to flipping the neutral-mode dipole, as was already discussed in Sec.~\ref{s1b3}. Scattering in Eq.~(\ref{86}) is purely in the forward direction. For $\Delta\neq 1$, the forward scattering of the pulse in mode $-$, which is no longer neutral, is accompanied by ``firing" off a charge pulse in the opposite direction, the amplitude of which scales for $\Delta\to 1$ as $1+\eta_-\propto\sqrt{\Delta-1}$:
\be
\hspace{-1mm}\hat{Q}_\textrm{out}=(1-R_-)\left(\begin{array}{cc}
0 & 0 \\
\eta_- & 1
\end{array}\right)
+R_-\frac{1+\eta_-}{1+\eta_+}\left(\begin{array}{cc}
1 & \eta_+ \\
0 & 0
\end{array}\right)~.
\label{87}
\ee
Conversely, for $\Delta\neq 1$, the charge pulse in mode + emits, upon hitting the tunneling link, a charge pulse in the opposite direction:
\be
\hat{Q}_\textrm{out}=(1-R_+)\left(\begin{array}{cc}
1 & \eta_+ \\
0 & 0
\end{array}\right)
+R_+\frac{1+\eta_+}{1+\eta_-}\left(\begin{array}{cc}
0 & 0 \\
\eta_- & 1
\end{array}\right)~.
\label{88}
\ee
The amplitude of the backscattered charge pulse in Eq.~(\ref{88}) scales for $\Delta\to 1$ as $R_+\propto\Delta-1$.

A rather nontrivial point about tunneling between interacting channels is that the scattering coefficients $W^\pm_{1,2}$ in Eqs.~(\ref{82}) and (\ref{83}) are not necessarily positive. In the absence of interchannel interaction, they are positively defined. However, as an indication of the highly collective nature of tunneling in the presence of interaction, one of them---but not both---may become negative if interaction is strong enough, reflecting the emergence of a negative value for one of the partial thermodynamic compressibilities [the factors in the brackets in Eqs.~(\ref{82}) and (\ref{83})]. As we will see shortly, in Sec.~\ref{s4a}, this has ramifications for the scattering rates (per unit time) which describe tunneling from channel 1 to channel 1/3 and vice versa, with one of them becoming negative for sufficiently large $v_{12}$.

\subsection{Strong-tunneling limit}
\label{s3c}

The difference of the reflection coefficients for the charge incident from the left and from the right ($R_+\neq R_-$) is a direct consequence of the inequality $g_+\neq g_-$ [Eq.~(\ref{85})], which precisely encodes the property of the two-channel edge being a chiral system. The difference shows up also in the maximum value the reflection coefficient can reach with increasing strength of tunneling. Going beyond the Born approximation for tunneling [Eqs.~(\ref{82}) and (\ref{83})] results in a substitution of $|\tilde{t}_0|^2/v_+v_-$ in Eqs.~(\ref{84}) and (\ref{85}) by a certain number $R_0$,
\be
R_\pm=R_0\,\frac{\Delta\mp 1}{3}~,
\label{89}
\ee
where $R_0$ is bounded from above by conservation of energy upon tunneling.

In terms of the matrix (\ref{76}), the thermodynamic constraint on the current scattering matrix with no interchannel interaction in the incoming and outgoing currents ($\Delta=2$ outside the scattering region) \cite{Wen1994,Chklovskii1998,Sen2008,Protopopov2017} corresponds to
\be
R_+\leq 1/2~,\quad R_-\leq 3/2~,
\label{90}
\ee
with $R_+/R_-=1/3$. The condition (\ref{90}), when applied to a tunneling link between noninteracting channels 1 and 1/3 ($R_0\leq 3/2$), says that mode 1/3 can be fully reflected ($R_-=1$ is allowed), whereas mode 1 cannot ($R_+\leq 1/2$). The explicit expression for the tunneling conductance $G_t$ between channels 1 and 1/3 for $\Delta=2$ from Ref.~\onlinecite{deChamon1997}, when taken in the strong-tunneling limit, gives $G_t=1/2$, which complies with the maximum value of $R_0$ equal to $3/2$ for $\Delta=2$. This upper bound on $R_0$ follows directly from the duality between weak and strong tunneling \cite{deChamon1997,Sandler1998}.

The latter inequality in Eq.~(\ref{90}) allows for $R_-$ to be larger than 1. In the interval $1<R_-\leq 3/2$ (or, equivalently, when represented in terms of the conductance matrix \cite{Protopopov2017}, for the negative value of the diagonal, for channel 1/3, element of it), the system behaves in a nontrivial way. Namely, in the two-terminal setting, it can perform as a step-up transformer \cite{Wen1994,Chklovskii1998} (in particular, in the form of an ``adiabatic junction"  \cite{Chklovskii1998}) if the source for channel 1 has a higher electrochemical potential. Or, if the source for channel 1/3 is biased in this way, the charge current in channel 1/3 can be ``sucked in" from---not absorbed by---the nominally drain reservoir for channel 1/3 (``nominally" means that the drain reservoir is at a lower electrochemical potential than the source reservoir), so that both reservoirs simultaneously supply current to channel 1/3. The suck-in effect in tunneling ($R_->1$) is phenomenologically similar to Andreev reflection \cite{Sandler1998}.

The negative values of $W^\pm_{1,2}$, mentioned at the end of Sec.~\ref{s3b}, and the values of $R_-$ in Eq.~(\ref{90}) larger than 1 are two startling manifestations of the strongly correlated nature of tunneling, both of which carry nontrivial connotations from the point of view of thermodynamics. It is worth emphasizing, however, that they are distinctly different in nature and, as such, not directly related to each other. The former occurs if interchannel interaction is strong enough, irrespective of the strength of tunneling. By contrast, the latter occurs if tunneling is strong enough, also (as is the case discussed above) in the limit of no interchannel interaction.

In the scattering problem described by Eq.~(\ref{75}), in contrast to the assumption made in Refs.~\onlinecite{Wen1994,deChamon1997,Chklovskii1998,Sandler1998,Sen2008,Protopopov2017}, the incoming and outgoing currents in modes 1 and 1/3 are generically interacting with each other. The interchannel interaction strength is supposed to be homogeneous in the vicinity of the tunneling link. To generalize the thermodynamic constraint of the type (\ref{90}) to $\Delta\neq 2$, assume that the incoming currents are at thermal equilibrium with $T=0$. Denote the incoming and outgoing energy currents associated with charges in modes $\pm$ by $j^\epsilon_{\pm,\textrm{in}}$ and $j^\epsilon_{\pm,\textrm{out}}$, respectively. These obey
\be
j^\epsilon_{+,\textrm{in}}+j^\epsilon_{-,\textrm{in}}\geq j^\epsilon_{+,\textrm{out}}+j^\epsilon_{-,\textrm{out}}~,
\label{91}
\ee
where the inequality accounts for the possibility of tunneling-induced inelastic scattering which produces chargeless excitations inside channel + or $-$, which carry energy away from the tunneling link. The charge-related energy currents $j^\epsilon_\pm$ and the charge currents $j_\pm$, both incoming and outgoing, are related to each other by $j^\epsilon_{\pm}=j_\pm\mu_\pm/2$, where $\mu_\pm=\pm 2\pi j_\pm/g_\pm$ are the local chemical potentials of modes $\pm$ [cf.\ Eq.~(\ref{28}) for the local chemical potentials in channels 1 and 1/3], i.e.,
\be
j^\epsilon_\pm=\pm\frac{\pi}{g_\pm}j_\pm^2~.
\label{92}
\ee

The incoming and outgoing energy currents $j^\epsilon_{\pm,\textrm{in}}$ and $j^\epsilon_{\pm,\textrm{out}}$ are related to each other by the same current-scattering matrix as the charge currents in Eq.~(\ref{81}):
\be
\left(\begin{array}{r}
j^\epsilon_+\\
-j^\epsilon_-
\end{array}\right)_\textrm{out}=\left(\mathbb{1}+\hat{\mathcal T}\right)\left(\begin{array}{r}
j^\epsilon_+\\
-j^\epsilon_-
\end{array}\right)_\textrm{in}~.
\label{93}
\ee
Substituting Eqs.~(\ref{92}) and (\ref{93}) in Eq.~(\ref{91}), together with using Eqs.~(\ref{81}) and (\ref{89}), produces the constraint on $R_\pm$ for arbitrary $\Delta$:
\be
0\leq R_\pm\leq\frac{\Delta\mp 1}{\Delta}~,
\label{94}
\ee
with $R_0\geq 0$ generically attaining any value up to $3/\Delta$:
\be
0\leq R_0\leq 3/\Delta~.
\label{95}
\ee
For $R_0=3/\Delta$, at which $R_+$ and $R_-$ simultaneously reach their maximum allowed values, tunneling is ``dissipationless" in the sense that the sign between the two parts of Eq.~(\ref{91}) becomes ``equals," i.e., no chargeless excitations are created upon tunneling. Note that the maximum values of $R_+$ and $R_-$ behave differently as $\Delta$ varies from $\Delta=2$ to $\Delta=1$: the former decreases (down to zero), while the latter grows. Note also that $R_+$ is always smaller than 1, in contrast to $R_-$.

For $\Delta=1$, the maximum value for $R_-$, according to Eq.~(\ref{94}), is 2. From Eq.~(\ref{86}), it follows that the neutral mode incident on the tunneling link with $R_-=2$ deterministically changes sign upon passing through it:
\be
\hat{Q}_\textrm{out}=\left(\begin{array}{rr}
0 & 0 \\
1 & -1
\end{array}\right)=-\hat{Q}_\textrm{in}~.
\label{96}
\ee
The strong-tunneling limit for $\Delta=1$ thus corresponds to flipping the neutral-mode dipole with probability 1. The conductance $G$ in this limit only depends on $N_u$ and $N_l$, where $N_{u,l}$ is the number of the tunneling links in the upper ($u$) and lower ($l$) parts of the edge [Fig.~\ref{f3a}]. Specifically, $G$ can only take one of three values. If both $N_u$ and $N_l$ are even, tunneling does not affect either the charge mode or the neutral mode, so that $G$ is the same as for a clean edge [Eq.~(\ref{56})]. If $N_u$ is odd, the current $j_u$ in the upper part of the edge is obtained from Eqs.~(\ref{22}) and (\ref{23}) by imposing the condition that the ratio of $j_-$ at the right terminal and $j_-$ at the left terminal is $-1$ (odd number of the dipole flips), which yields \cite{sing}
\be
j_u=\frac{1}{2\pi\Delta_c^\textrm{eff}}(g_{+c}\mu_L+g_{-c}\mu_R)~,
\label{97}
\ee
instead of Eq.~(\ref{31}). If $N_l$ is also odd, the current $j_l$ in the lower part of the edge follows from Eq.~(\ref{97}) by exchanging $\mu_L\leftrightarrow\mu_R$; otherwise, this exchange should be performed on Eq.~(\ref{31}). The three values of the conductance, $G_{e-e}$, $G_{o-o}$, and $G_{e-o}$, corresponding to the even-even, odd-odd, and even-odd configurations, respectively, are then given by
\begin{align}
&G_{e-e}=\frac{2\Delta_c^\textrm{eff}}{3}~,\quad G_{o-o}=\frac{2}{3\Delta_c^\textrm{eff}}~,\nonumber\\
&G_{e-o}=\frac{1}{3}\left(\Delta_c^\textrm{eff}+\frac{1}{\Delta_c^\textrm{eff}}\right)~.
\label{98}
\end{align}

Note that, for $\Delta_c^\textrm{eff}=2$, the maximum and minimum values of $G$ from Eq.~(\ref{98}) are 4/3 and 1/3---the same as mentioned above in connection with the $\chi$sG model with weak disorder \cite{Naud2001,Rosenow2010,Protopopov2017} (Secs.~\ref{s1b2} and \ref{s1b3}). It is also worth mentioning that scattering for $R_-=1$ corresponds to flipping half of the dipole current, which, on average, annihilates the neutral mode.

\section{Disordered edge}
\label{s4}

Having considered the contacts and the clean edge in Sec.~\ref{s2}, and scattering at a single tunneling link in Sec.~\ref{s3}, we are now prepared to discuss transport through a disordered edge. As was already mentioned in the introductory overview, we restrict our attention here to incoherent transport (Sec.~\ref{s1b3}). Between the contacts, the equation of motion for the disorder-averaged densities $n_\pm$ in the presence of random tunneling between channels 1 and 1/3 then reads
\be
(\partial_t\pm v_\pm\partial_x)n_\pm\pm I_t=0~,
\label{99}
\ee
with the ``collision" term
\be
I_t=\Gamma_+n_+-\Gamma_-n_-
\label{100}
\ee
and the scattering rates
\be
\Gamma_\pm=n_tv_\pm R_\pm~,
\label{101}
\ee
where $n_t$ is the concentration of the tunneling links. The reflection coefficients $R_\pm$ are given by Eq.~(\ref{84}) for weak tunneling through a single tunneling link, or by Eq.~(\ref{89}) otherwise. Below, we focus on the Gaussian limit of $n_t\to\infty$ and $t_0\to 0$ with $n_t|t_0|^2$ held constant, where $n_\pm$ in Eq.~(\ref{99}) are the exact density distributions with vanishing mesoscopic fluctuations.

\subsection{Negative scattering rates}
\label{s4a}

Before proceeding to the solution of Eq.~(\ref{99}), let us return to the conceptually significant point, mentioned at the very end of Sec.~\ref{s3b}, about the signs of the scattering rates for tunneling between channels 1 and 1/3. Recall also that this peculiarity is substantially different from another nontrivial, from the point of view of thermodynamics, point, namely the one about $R_->1$, as commented in Sec.~\ref{s3c}. The equation of motion for the densities $n_{1,2}$ is obtained from Eq.~(\ref{99}) by using the transformation (\ref{59}):
\begin{align}
&\partial_tn_1+\partial_x(v_1n_1+v_{12}n_2)+\bar{I}_t=0~,
\label{102}\\
&\partial_tn_2-\partial_x\left(v_2n_2+\frac{1}{3}v_{12}n_1\right)-\bar{I}_t=0~,
\label{103}
\end{align}
where the collision term
\be
\bar{I}_t=\Gamma_1n_1-\Gamma_2n_2
\label{104}
\ee
is proportional to that in Eq.~(\ref{99}). Specifically,
\be
\bar{I}_t=\sqrt{\frac{3}{\Delta^2-1}}\,I_t~,
\label{105}
\ee
and $\Gamma_{1,2}$ are written as
\be
\Gamma_1=\frac{1}{3}\gamma_0(v_1-v_{12})~,\quad\!\!\Gamma_2=\frac{1}{3}\gamma_0(3v_2-v_{12})~,
\label{106}
\ee
where
\be
\gamma_0=\frac{n_t|\tilde{t}_0|^2}{v_+v_-}
\label{107}
\ee
(with $v_+v_-=v_1v_2-v_{12}^2/3$). Note that $\bar{I}_t$ does not describe the total-current relaxation (which distinguishes the chiral edge from a LL):
\be
\bar{I}_t=\frac{1}{3}\gamma_0(j_1+3j_2)
\label{108}
\ee
[cf.\ Eqs.~(\ref{36}) and (\ref{37})].

Remarkably, the scattering rates $\Gamma_{1,2}$ need not be both positive within the stability range of $v_{12}$ from Eq.~(\ref{10}), in contrast to $\Gamma_\pm$. In particular, for $\Delta=1$, i.e., for
\be
v_{12}=\frac{3}{4}(v_1+v_2)~,
\label{109}
\ee
the ratio $\Gamma_1/\Gamma_2$ is universal (independent of $v_{1,2}$) and given by
\be
\frac{\Gamma_1}{\Gamma_2}=-\frac{1}{3}~.
\label{110}
\ee
Since $\Gamma_1/\Gamma_2$ is positive at $\Delta=2$ and negative at $\Delta=1$, it changes sign in between, with, as follows from Eq.~(\ref{106}), either $\Gamma_1=0$ (for $v_1<3v_2$) or $\Gamma_2=0$ (for $v_1>3v_2$) at that point. If the point $\Delta=1$ is reachable with increasing $v_{12}$ [the condition for this to happen is Eq.~(\ref{11})], it is $\Gamma_1$ that changes sign at $v_{12}=v_1$. At the stability threshold [Eq.~(\ref{10})], $\Gamma_1/\Gamma_2=-\sqrt{v_1/3v_2}<0$.

The negative value of one of the scattering rates $\Gamma_{1,2}$ does not spell instability, or contradict causality for that matter, because causality only requires that
\be
\Gamma=\Gamma_1+\Gamma_2=\Gamma_++\Gamma_-\geq 0~.
\label{111}
\ee
This follows from the characteristic equation for the dynamics of $n_{1,2}$, or $n_\pm$,
\be
(\omega-v_+q)(\omega+v_-q)+i\omega\Gamma-iq(\Gamma_-v_+-\Gamma_+v_-)=0~,
\label{112}
\ee
where the linear-in-$\omega$ term contains $\Gamma_{1,2}$ in the combination $\Gamma=\Gamma_1+\Gamma_2>0$ [the stability condition (\ref{10}) is stricter than $v_{12}<(v_1+3v_2)/2$, which guarantees the positive sign of $\Gamma$ within the range of $v_{12}$ from Eq.~(\ref{10})].

It is worth noting that the ratio of $\Gamma_1/\Gamma_2$ (or $\Gamma_+/\Gamma_-$ for that matter) is obtainable on rather general grounds. In Sec.~\ref{s3}, we presented a microscopic picture of fractionalization-renormalized tunneling that produced both the explicit dependence of $R_+/R_-$ on $\Delta$ [Eq.~(\ref{84})] and the expression of $R_\pm$ in terms of the thermodynamic compressibility matrix [Eq.~(\ref{85})]. Alternatively, one might have relied on gauge invariance in ``constructing" the collision terms in phenomenological equations of motion for the densities \cite{Kane1997}. Specifically, the collision terms should remain intact upon shifting the densities
\be
\bar{\textbf n}=\left(\begin{array}{l}
n_1\\
n_2
\end{array}\right)\to\bar{\textbf n}+\hat{\bar\Pi}\left(\begin{array}{l}
1\\
1
\end{array}\right)\delta\mu
\label{113}
\ee
for arbitrary $\delta\mu$ \cite{Kane1997}, or similarly for the densities in any other basis. Stating that $\hat{\bar\Pi}(1,1)^T$ is the zero mode of the collision operator (whose structure in terms of $\Gamma_{1,2}$ follows from conservation of the total charge),
\be
\left(\begin{array}{rr}
\Gamma_1 & -\Gamma_2\\
-\Gamma_1 & \Gamma_2
\end{array}\right)
\hat{\bar\Pi}
\left(\begin{array}{l}
1\\
1
\end{array}\right)=0~,
\label{114}
\ee
Eq. (\ref{114}) fixes the ratio $\Gamma_1/\Gamma_2$ in terms of the elements of the compressibility matrix [cf.\ Eqs.~(\ref{27}) and (\ref{106})].

Importantly, $\hat{\bar\Pi}$ in Eq.~(\ref{114}) is the thermodynamic compressibility, not the single-particle tunneling density of states affected by the orthogonality catastrophe. It is this distinction, as was already mentioned in Sec.~\ref{s3b}, that makes the concept of fractionalization-renormalized tunneling precise, by cleanly separating it from that of the renormalization of tunneling by the orthogonality catastrophe.

\subsection{Tunneling-modified eigenmodes}
\label{s4b}

As an essential part of the physical picture, let us discuss how tunneling affects dynamics of the edge eigenmodes. It is the last (linear in $q$) term in Eq.~(\ref{112}) that encodes the property of the edge being chiral. Specifically, from Eq.~(\ref{101}), the combination of $\Gamma_+$ and $\Gamma_-$ within the brackets can be represented as
\be
\Gamma_-v_+-\Gamma_+v_-=n_tv_+v_-(R_--R_+)~,
\label{115}
\ee
which is nonzero only if the reflection coefficients for the currents incident on the tunneling link from the left and from the right are not the same [for the $\nu=2/3$ edge, it is also representable as $(2/3)n_t|\tilde{t}_0|^2$].

In the limit of small $q$ (specifically, for $|q|\ll\Gamma/\max\{v_1,v_2\}$ for any $\Delta$ between $\Delta=2$ and $\Delta=1$), the solution of Eq.~(\ref{112}) gives two low-energy eigenmodes of the edge in the presence of random tunneling. One is a weakly damped (if $\Delta\neq 1$) ballistic mode with
\be
\omega=\omega_\rho\to v_\rho q-iD_bq^2~,
\label{116}
\ee
where the velocity of charge propagation (hence the subscript $\rho$) $v_\rho$ is given by
\be
v_\rho=\frac{\Gamma_-v_+-\Gamma_+v_-}{\Gamma}=\frac{2}{3}\,\frac{3v_1v_2-v_{12}^2}{v_1+3v_2-2v_{12}}~.
\label{117}
\ee
This mode is propagating ballistically only because of $R_+\neq R_-$ [Eq.~(\ref{115})]. The direction of propagation coincides with that for the + mode. Note that, unless $\Delta=1$, the velocity of the right-moving eigenmode in the small-$q$ limit is modified (ballistic propagation is slowed down) by tunneling:
\be
v_\rho\leq v_+~.
\label{118}
\ee

The designation ``weakly damped" above Eq.~(\ref{116}) means a slow diffusive broadening in real space on top of the ballistic propagation, with the diffusion coefficient
\be
D_b=\frac{1}{\Gamma}(v_+-v_\rho)(v_-+v_\rho)~.
\label{119}
\ee
Again, even when one of the scattering rates $\Gamma_1$ or $\Gamma_2$ becomes negative with increasing $v_{12}$, they enter $D_b\geq 0$ in the combination $\Gamma=\Gamma_1+\Gamma_2\geq 0$. The other eigenmode in the small-$q$ limit is gapped, with the gap determined, similarly to $D_b$, by the total scattering rate:
\be
\omega=\omega_n\to -i\Gamma-v_nq~,
\label{120}
\ee
where
$v_n=(\Gamma_-v_--\Gamma_+v_+)/\Gamma=v_\rho-v_1+v_2$ and the subscript $n$ stands for ``neutral" (the meaning of which will become clear in the next paragraph). At the stability threshold [Eq.~(\ref{10})], both $v_\rho$ and $D_b$ vanish to zero, with $\Gamma\to\infty$ [the numerator of $D_b$ in Eq.~(\ref{119}) also vanishes].

The property of the tunneling-modified eigenmodes $\omega_\rho$ and $\omega_n$ being propagating and localized, respectively, shows up in splitting of a compact charge pulse into two parts: mobile (charged) and immobile (neutral). Consider the evolution in time of $\bar{\textbf n}=(n_1,n_2)^T$ with the initial condition at $t=0$ fixing a certain density profile $\bar{\textbf n}_q(t=0)$ in $q$ space. Assume that compact charges $Q_{1,2}$ are created at the point $x=0$ at $t=0$, i.e., $\bar{\textbf n}_q(t=0)=(Q_1,Q_2)^T$. Solution to the equation of motion in the limit of small $|q|\ll\Gamma/\max\{v_1,v_2\}$ gives, for $\bar{\textbf n}$ in $(\omega,q)$ space:
\begin{align}
\bar{\textbf n}_{\omega q}\to &\,\,\frac{i}{\omega-v_\rho q+iD_bq^2}\,\frac{Q_1+Q_2}{\Gamma}\left(\begin{array}{l}
\Gamma_2\\
\Gamma_1
\end{array}\right)\nonumber\\
+&\,\,\frac{i}{\omega+i\Gamma}\,\frac{\Gamma_1Q_1-\Gamma_2Q_2}{\Gamma}
\left(\begin{array}{r}
1\\
-1
\end{array}\right)~.
\label{121}
\end{align}
The physical picture that emerges from Eq.~(\ref{121}) is that the total charge $Q_1+Q_2$ is all taken away by the ballistic mode. What is left behind is a localized dipole (with opposite charges in channels 1 and 1/3) that is stuck to the starting point $x=0$ and decays ``on the spot." More precisely, the extinction of the dipole on distances exceeding the backscattering length from the starting point is accompanied by a ballistic run at speed $v_n$ [Eq.~(\ref{120})]. Importantly, this ``charge-neutral" separation occurs for arbitrary $\Delta$. The existence of the nondecaying ballistic---in the limit of small $q$---charge mode (\ref{116}) is fundamental to the conductance quantization in the limit of charge equilibration (Sec.~\ref{s1b1}), a crossover to which from ballistic transport is discussed in detail below, in Sec.~\ref{s4c}.

Tunneling modifies the screening properties of the ballistic mode: $n_1/n_2$ at the same spatial point in the running away diffusion-broadened charge pulse in Eq.~(\ref{121}) is given by
\be
n_1/n_2=\Gamma_2/\Gamma_1~,
\label{122}
\ee
which nullifies $\bar{I}_t$, instead of Eq.~(\ref{60}) for the clean edge. This is because the tunneling-renormalized ballistic mode is a combination of $n_+$ and $n_-$. Note that $n_1/n_2$ in Eq.~(\ref{122}) can be of either sign (Sec.~\ref{s4a}).

\subsection{Conductance}
\label{s4c}

Now let us turn to the conductance of a disordered edge. In the dc limit, the nonzero root $\partial_x\to\gamma\neq 0$ of the characteristic equation to Eq.~(\ref{99}) (the other is trivially zero, because of conservation of $j_++j_-$) is
\be
\gamma=\frac{\Gamma_-}{v_-}-\frac{\Gamma_+}{v_+}=n_t(R_--R_+)~,
\label{123}
\ee
or, in terms of $\gamma_0$:
\be
\gamma=\frac{2}{3}\gamma_0~.
\label{124}
\ee
The nonvanishing of $\gamma$ is only due to chirality of the edge ($R_+\neq R_-$). Remarkably, the fractionalization-induced $\Delta$-dependent factors in $R_\pm$ (\ref{84}) cancel out in the combination $R_--R_+$ in Eq.~(\ref{123}), so that the scattering length $2\gamma^{-1}$ depends on $\Delta$ only through the renormalization of $t_0$ (which, in particular, does not show any singular behavior for $\Delta\to 1$, as was already discussed in Sec.~\ref{s3b}). This can also be clearly seen from Eqs.~(\ref{102}) and (\ref{103}), which---in view of Eq.~(\ref{108})---can be rewritten as [cf.\ Eq.~(\ref{39})]
\be
\left(\partial_x\mathbb{1}+\hat\upgamma\right)\left(\begin{array}{c}
j_1\\
j_2
\end{array}\right)=0~,
\label{125}
\ee
where the matrix $\hat\upgamma$ does not explicitly depend on $\Delta$:
\be
\hat\upgamma=\frac{1}{2}\,\gamma\left(\begin{array}{rr}
1 & 3\\
-1 & -3
\end{array}\right)~.
\label{126}
\ee

\subsubsection{Identical contacts}
\label{s4c1}

For the line-junction contacts (Secs.~\ref{s2d} and \ref{s2e}), solution of Eq.~(\ref{99}) in the dc limit, with the boundary conditions (\ref{43}) and the current-scattering matrix (\ref{50}), gives for the currents $j_{u\pm}$ in the upper part of the edge in Fig.~\ref{f3a}
\begin{align}
\left(\begin{array}{r}
j_{u+} \\
-j_{u-}
\end{array}\right)&=\frac{1}{6\pi}\,\dfrac{1}{e^{\gamma L}-\chi}\left[\,\left(\mu_Le^{\gamma L}-\mu_R\,\chi\right)\!\left(\begin{array}{r}
\Delta+1\\
\Delta-1
\end{array}\right)\right.\nonumber\\
&\left.-\,(\mu_L-\mu_R)e^{\gamma x}\sqrt{(\Delta^2-1)\,\chi}\left(\begin{array}{r}
1 \\
1
\end{array}\right)\,\right]~,
\label{127}
\end{align}
where
\be
\chi=\dfrac{\Delta^\textrm{eff}_c-1}{\Delta^\textrm{eff}_c+1}
\label{128}
\ee
and $L$ is the distance between the left and right contacts, which are, respectively, at the chemical potentials $\mu_L$ and $\mu_R$. In Eq.~(\ref{127}), both contacts are assumed to be identical and characterized by the same effective-interaction parameter $\Delta_c^\textrm{eff}$ (\ref{57}). For $\Delta=2$ (noninteracting edge) and $\Delta^\textrm{eff}_c=2$ (ideal contact for $\Delta=2$), Eq.~(\ref{127}) agrees with the expressions for the currents from Refs.~\onlinecite{Sen2008,Nosiglia2018,Spanslatt2019,*Spanslatt2020}.

The first term in the square brackets in Eq.~(\ref{127}) does not depend on $x$, whereas the second one is proportional to $e^{\gamma x}$. Note that the currents in Eq.~(\ref{127}) are $x$ independent not only if $\Delta=1$, but also if $\Delta_c^\textrm{eff}=1$. For $\gamma L\gg 1$, the first term is dominant for almost the whole length of the edge for arbitrary $\Delta$ and $\Delta_c^\textrm{eff}$, and describes the counterpropagating charge currents equilibrated with each other at the chemical potential $\mu_L$. Equilibration, described by the second term, occurs then in a narrow region within the distance of the order of $\gamma^{-1}$ near the right contact (emitting mode 1/3 at $\Delta_c^\textrm{eff}=2$). Specifically, for $\gamma L\gg 1$,
\begin{align}
&\left(\begin{array}{r}
j_{u+} \\
-j_{u-}
\end{array}\right)\simeq\frac{1}{2\pi}\left[\,\mu_L\!\left(\begin{array}{r}
g_+ \\
g_-
\end{array}\right)\right.\nonumber\\
&\quad\,\,\,\left.-\,e^{-\gamma (L-x)}\sqrt{g_+g_-\chi}\,(\mu_L-\mu_R)\!\left(\begin{array}{r}
1 \\
1
\end{array}\right)\right]~.
\label{129}
\end{align}

\begin{figure}[ht]
\centering
\includegraphics[width=0.95\columnwidth]{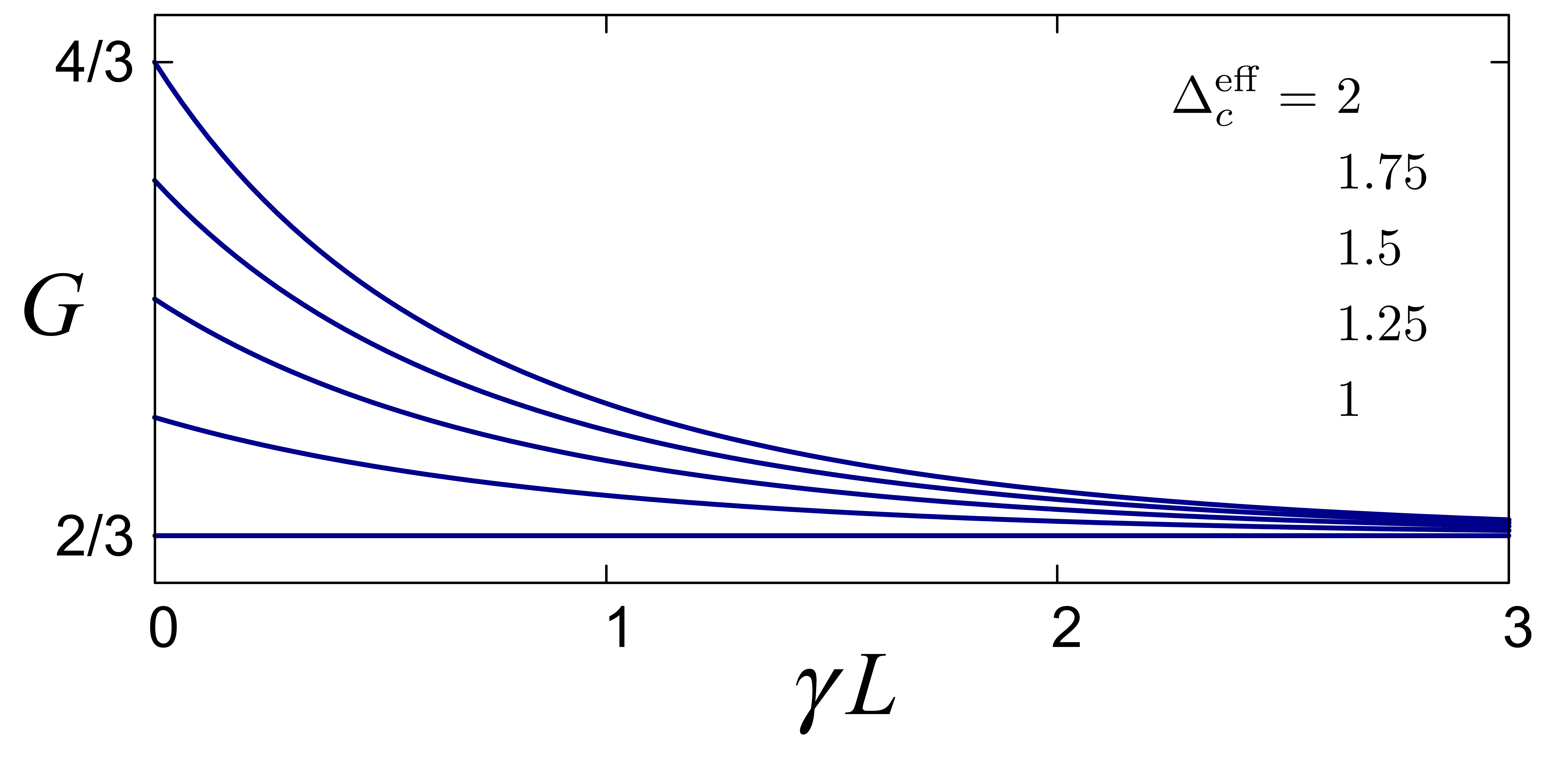}
\caption{Two-terminal conductance $G$ of the $\nu=2/3$ edge as a function of $\gamma L$, where $L$ is the edge length and $2/\gamma$ is the backscattering length, for different parameters $\Delta_c^\textrm{eff}$ [Eq.~(\ref{57})] characterizing the effective interaction strength beneath the contacts. Top to bottom: $\Delta_c^\textrm{eff}=2$, 1.75, 1.5, 1.25, 1. The conductance does not depend on either the interaction strength outside the contacts ($\Delta$) or the true interaction strength beneath the contacts ($\Delta_c$), apart from the interaction-induced renormalization of $\gamma$ and the scattering lengths defining $\Delta_c^\textrm{eff}$. In the charge-equilibrated limit of $\gamma L\to\infty$, the conductance $G\to 2/3$ irrespective of the value of $\Delta_c^\textrm{eff}$.}
\label{f8}
\end{figure}

Either of the currents $j_{u+}$ and $j_{u-}$, and either of two terms in Eq.~(\ref{127}), depends on both $\Delta$ and $\Delta_c^\textrm{eff}$. Remarkably, the total current $j_u=j_{u+}+j_{u-}$ depends only on $\Delta_c^\textrm{eff}$ (apart from the dependence of $\gamma$ on $\Delta$):
\be
j_u=\frac{1}{2\pi}\,\frac{2}{3}\,\dfrac{\,\mu_L-e^{-\gamma L}\,\chi\,\mu_R\,}{1-e^{-\gamma L}\,\chi}~.
\label{130}
\ee
The total current $j_l$ in the lower part of the edge follows from Eq.~(\ref{130}) by exchanging $\mu_L\leftrightarrow\mu_R$ (assuming that the upper and lower parts of the edge have the same length $L$ and are characterized by the same strength of disorder). The two-terminal conductance of the edge $G=2\pi j/(\mu_L-\mu_R)$, with $j=j_u-j_l$, is then given by
\be
G=\frac{2}{3}\,\frac{1+e^{-\gamma L}\,\chi}{1-e^{-\gamma L}\,\chi}~,
\label{131}
\ee
as illustrated in Fig.~\ref{f8}. The dependence on $\Delta$ (besides that which is encoded in $\gamma$) cancels out from $G$---as it does separately for $j_u$ and $j_l$. Recall that also the dependence on $\Delta_c$ cancels out for the line-junction model of the contacts, assumed in Eq.~(\ref{131}), with $\Delta_c^\textrm{eff}$ in Eq.~(\ref{128}) being the effective interaction strength beneath the contacts (Secs.~\ref{s2d} and \ref{s2e}). For $\Delta_c^\textrm{eff}=2$, Eq.~(\ref{131}) agrees with that from Ref.~\onlinecite{Srivastav2020}. In the clean limit, Eq.~(\ref{131}) reduces to $G=2\Delta_c^\textrm{eff}/3$ [Eq.~(\ref{56})]. In the limit of large $\gamma L$, the two-terminal conductance is quantized at $G=2/3$, in accordance with the universal result $G=\nu$ for the limit of full charge equilibration \cite{Kane1997}, as was already discussed in Sec.~\ref{s1b1}. In particular, no decoupling between the charge and neutral modes is needed for the quantization. Equation (\ref{131}) shows that the quantization holds irrespective of not only the value of $\Delta$, but also the value of $\Delta_c^\textrm{eff}$. The quantization is inherently related to the existence of the ballistic mode (\ref{116}), which carries over charge, as discussed below Eq.~(\ref{121}).

It may be worth mentioning---particularly with regard to the experimental results \cite{Cohen2019}, which have demonstrated the quantization of $G\simeq 2/3$ for large $L$ and a crossover to larger values of $G$ for smaller $L$---that varying the control parameter of the experiment for given $L$ (primarily the magnetic field in Ref.~\onlinecite{Cohen2019}) may change not only $\gamma$ and $\Delta$, but also $\Delta_c^\textrm{eff}$. Within the picture described by Eq.~(\ref{131}), the variation of $\Delta$ may only show up indirectly, through the dependence of $\gamma$ on $\Delta$, whereas the variation of $\Delta_c^\textrm{eff}$ manifests itself in the limiting value of $G$ for small $L$. Recall also that the value of $\Delta_c^\textrm{eff}$ is determined by the interplay of tunneling processes beneath the contact and may possibly be controlled in a similar way as the tunneling length $1/\gamma$ outside the contact. In Ref.~\onlinecite{Cohen2019}, $G\simeq 4/3$ was observed for small $L$, which corresponds to $\Delta_c^\textrm{eff}\simeq 2$, for relatively large equilibration lengths (stronger magnetic fields), while the data for smaller equilibration lengths (weaker magnetic fields) may be compatible with the limiting value of $G=2\Delta_c^\textrm{eff}/3$, as $L$ is decreased, substantially smaller than 4/3.

Following the same route of cyclic permutation as for the derivation of Eqs.~(\ref{34}) and (\ref{35}), the four-terminal conductances $G_H$ (Hall) and $G_4$ (source-drain) [Fig.~\ref{f3b}] for the line-junction contacts (all four assumed to be identical) in the presence of disorder are obtained as
\begin{align}
&G_H=\frac{2}{3}\,\frac{1+e^{-\gamma L}\chi^2}{\left(1-e^{-\gamma L/2}\chi\right)^2}~,
\label{132}\\
&G_4=\frac{2}{3}\,\frac{1+e^{-\gamma L}\chi^2}{1-e^{-\gamma L}\chi^2}~.
\label{133}
\end{align}
In Eqs.~(\ref{132}) and (\ref{133}), for ease of presentation, each pair of adjacent contacts is assumed to be separated by the same distance $L/2$. Similarly to $G$ in Eq.~(\ref{131}), both $G_H$ and $G_4$ are equal to 2/3 in the limit of large $\gamma L$, for arbitrary $\Delta$ and $\Delta_c^\textrm{eff}$, which generalizes the result $G=\nu$ \cite{Kane1997} to the case of four-terminal measurements \cite{fourterm}.

\begin{figure}[ht]
\centering
\includegraphics[width=0.95\columnwidth]{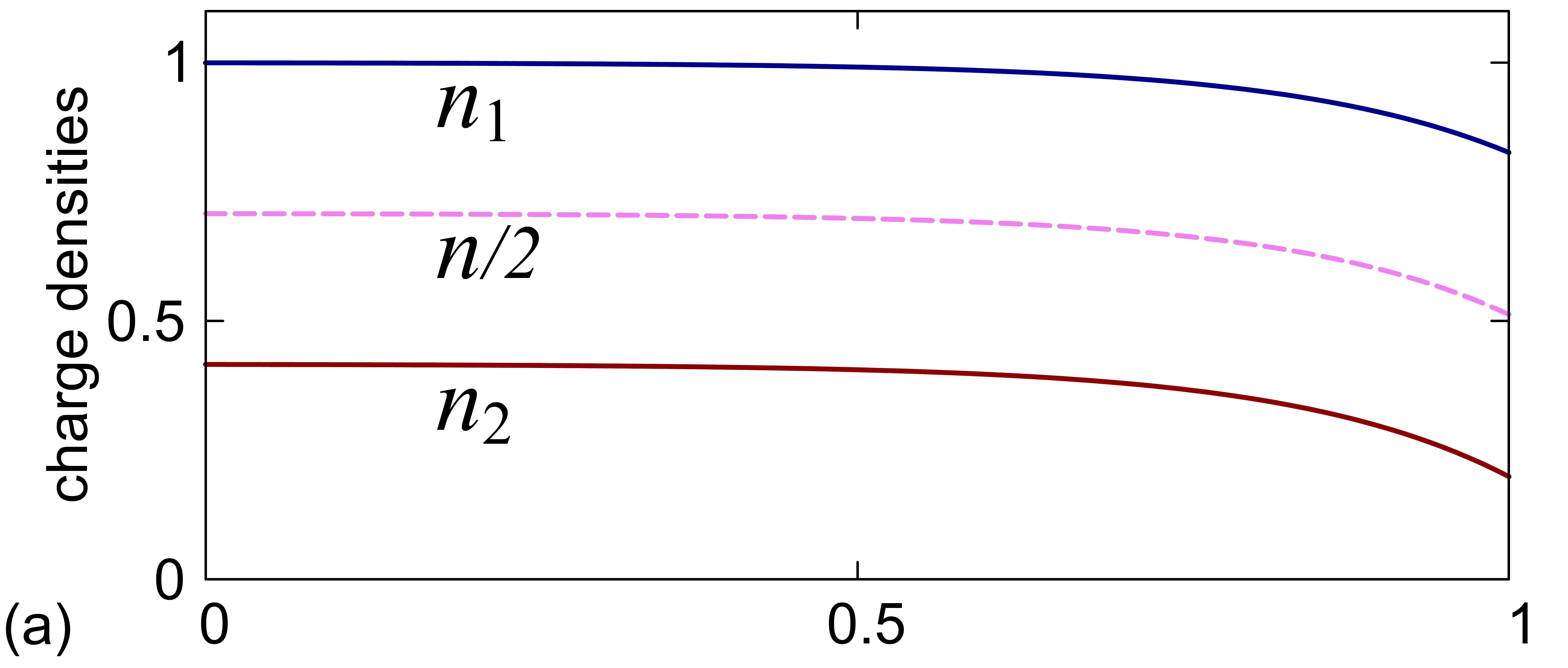}
\includegraphics[width=0.95\columnwidth]{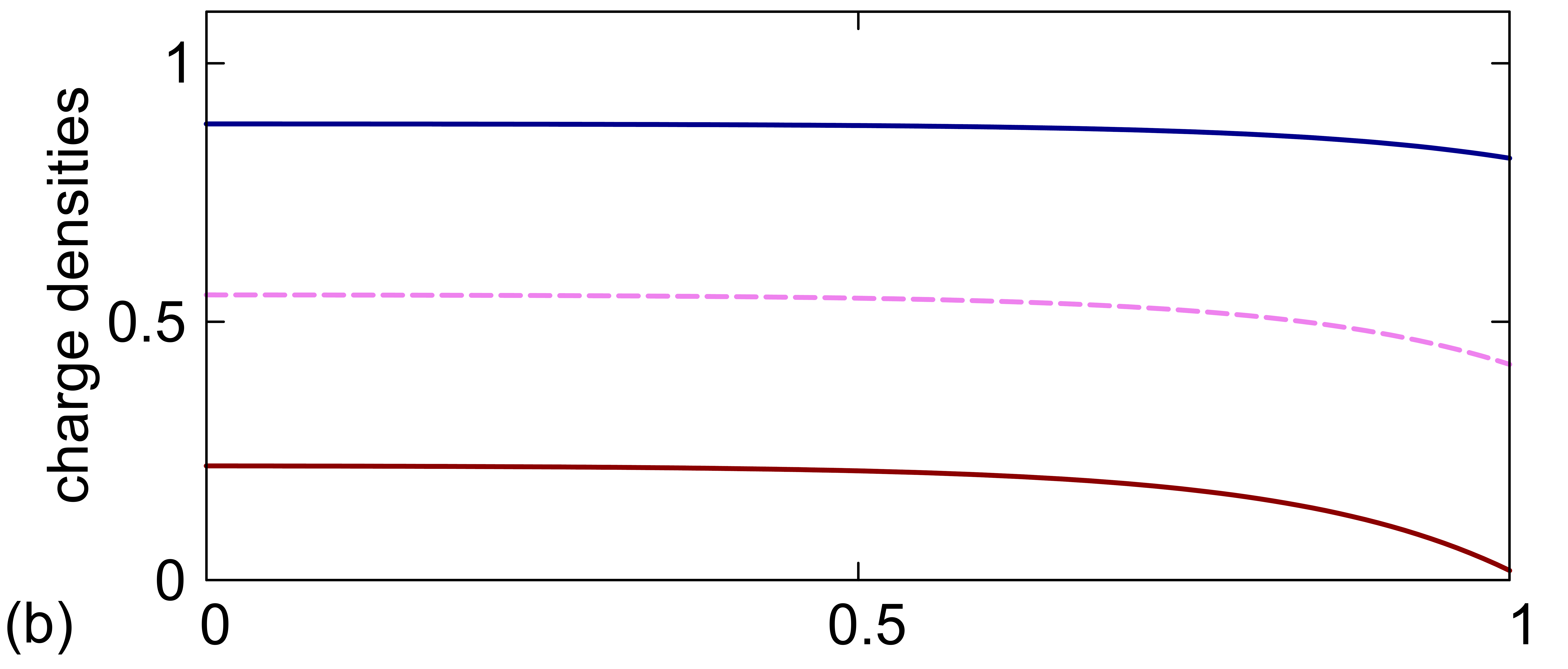}
\includegraphics[width=0.95\columnwidth]{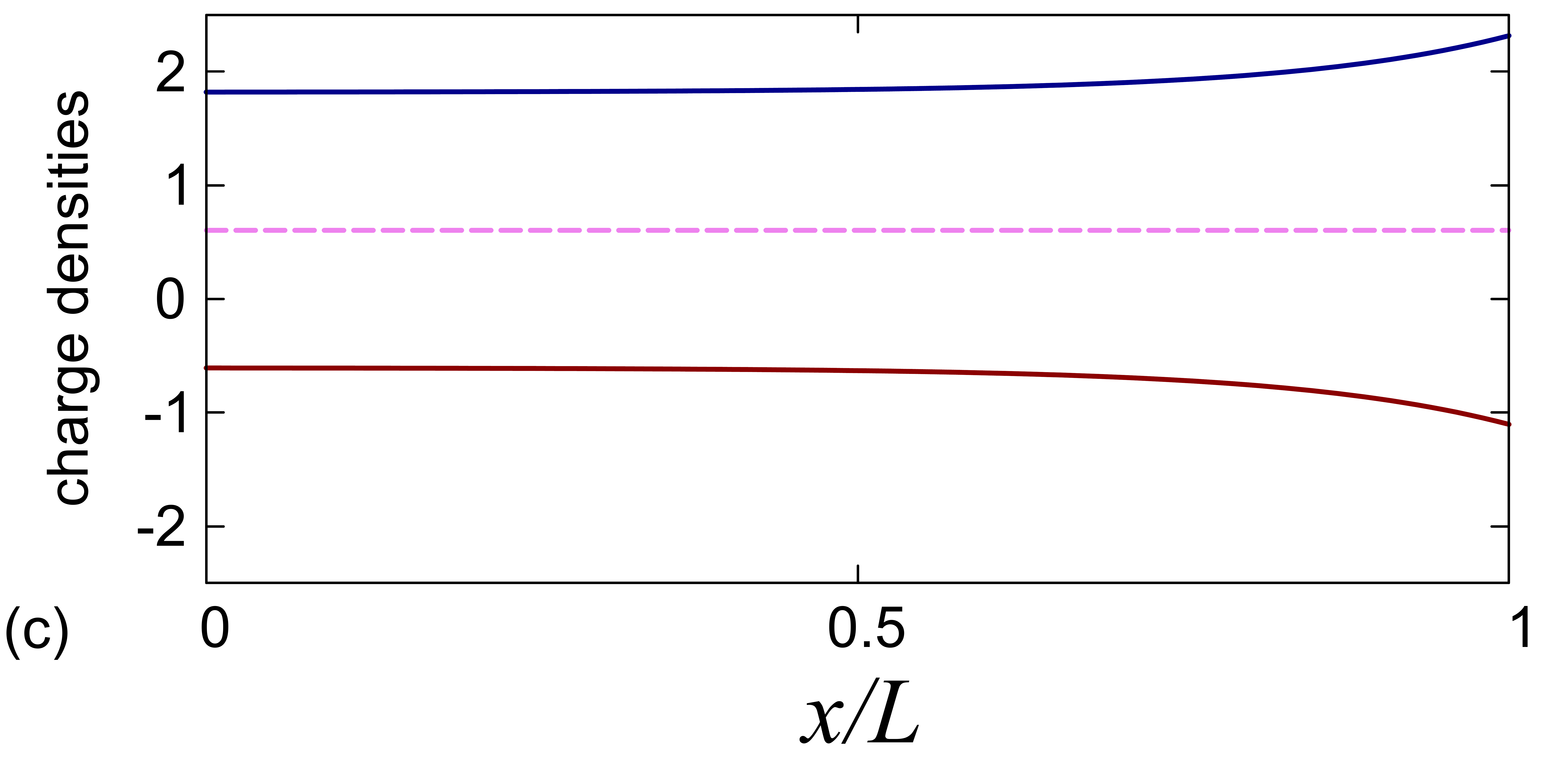}
\caption{Evolution of the distribution of the charge densities $n_1$ (dark blue) and $n_2$ (dark red) for channels 1 and 1/3, respectively, and the distribution of the total density $n=n_1+n_2$ (dashed violet, shown is $n/2$) [Eq.~(\ref{134})], as the strength of interchannel interaction increases. Top to bottom: $\Delta=2$ (a), $1.5$ (b), $1$ (c). The charge densities are shown in units of $\mu_L/2\pi v_1$ for the upper part of the edge in Fig.~\ref{f3a} for $\mu_R=0$, $v_2=0.8 v_1$, $\Delta^\textrm{eff}_c=1.75$, and $\gamma L=6$.}
\label{f9}
\end{figure}

Apart from the spatial distribution of the eigenmode currents for a long edge with $\gamma L\gg 1$ [Eq.~(\ref{129})], it is worthwhile to also comment on that of the densities in channels 1 and 1/3. For arbitrary strength of interchannel interaction, $\bar{\textbf n}$ in the upper part of the edge is represented in terms of $v_{12}$ (more compactly than in terms of $\Delta$) as
\begin{align}
\bar{\textbf n}&=\frac{1}{6\pi v_+v_-(e^{\gamma L}-\chi)}\left[\,
(\mu_Le^{\gamma L}-\mu_R\chi)\left(\begin{array}{c}
3v_2-v_{12} \\
v_1-v_{12}
\end{array}\right)\right.\nonumber\\
&\left.\qquad\,\, -\,(\mu_L-\mu_R)e^{\gamma x}\sqrt{\chi}
\left(\begin{array}{c}
v_2-v_{12} \\
v_1-v_{12}/3
\end{array}\right)\,\right]~.
\label{134}
\end{align}
For $\gamma L\gg 1$, Eq.~(\ref{132}) reduces to
\begin{align}
\bar{\textbf n}&\simeq\hat{\bar\Pi}\left(\begin{array}{r}
1 \\
1
\end{array}\right)\mu_L-e^{-\gamma(L-x)}\sqrt{\chi}\nonumber\\
&\times\frac{1}{6\pi v_+v_-}\left(\begin{array}{c}
v_2-v_{12} \\
v_1-v_{12}/3
\end{array}\right)(\mu_L-\mu_R)~,
\label{135}
\end{align}
where the first term represents the densities $n_1$ and $n_2$ equilibrated with each other at the chemical potential $\mu_L$, and the second term describes their equilibration in the vicinity of the right contact [cf.\ Eq.~(\ref{129})]. Note that for $v_{1,2}$ from the interval (\ref{11}), the relative sign of the contributions to $n_1$ and $n_2$ of the first term in Eq.~(\ref{134}) changes---becomes negative---with increasing $v_{12}$ at $v_{12}=v_1$. Similarly, the relative sign of the contributions of the second term changes at $v_{12}=v_2$. The latter means that, for the case of strong interchannel interaction ($v_{12}>v_2$), one of the densities $n_{1,2}$ goes sharply (for $\gamma L\gg 1$) up when approaching the right contact, whereas the other goes sharply down. The evolution of the spatial distribution of $n_{1,2}$ as the strength of interaction increases is illustrated in Fig.~\ref{f9}.

As is seen from Eqs.~(\ref{134}) and (\ref{135}), not only the partial densities $n_{1,2}$ but also the total density $n_1+n_2$ is inhomogeneous for a generic value of $v_{12}$ [Figs.~\ref{f9}(a) and \ref{f9}(b)]. The point to notice is that, for $\Delta=1$ [with $v_{12}$ from Eq.~(\ref{109})], the total charge is exactly homogeneous [Fig.~\ref{f9}(c)], in accordance with the decoupling of the charge and neutral modes, with the equilibration of the neutral mode showing up in a variation of the partial densities on the scale of $\gamma^{-1}$ near the right contact. Specifically, for $\Delta=1$, the ratio of the contribution to $n_1$ to the contribution to $n_2$ is $-3$ in the first term of Eq.~(\ref{134}) and $-1$ in the second [cf.\ Eq.~(\ref{65})].

\subsubsection{Asymmetric setups}
\label{s4c2}

In Sec.~\ref{s4c1}, the contacts were assumed to be identical. Let us now discuss the case of different contacts, i.e., characterized by different values of $\Delta_c^\textrm{eff}$ (recall that, within the line-junction model, this is the only parameter to characterize a contact). This is, arguably, useful to make contact with experiment: in practice, the contacts may be much different from each other, in particular, because of a different strength of disorder beneath them, which may also be different from the strength of disorder outside them (as was already mentioned in Sec.~\ref{s2d}). We will also discuss a more conceptual issue of the contact and bulk resistances for the QH edge.

Assume, as a starting point, that the edge is clean and the left and right contacts in the two-terminal setup [Fig.~\ref{f3a}] are characterized by $\Delta_{cL}^\textrm{eff}$ and $\Delta_{cR}^\textrm{eff}$, respectively. For $\Delta_{cR}^\textrm{eff}\neq\Delta_{cL}^\textrm{eff}$, Eqs.~(\ref{53})--(\ref{55}) generalize to
\begin{align}
j_1(\delta)&=\frac{1}{2\pi}\,\frac{(1-c_L)\mu_L+c_L(1-3c_R)\mu_R}{1-3c_Lc_R}~,
\label{136}\\
j_2(\delta)&=-\frac{1}{2\pi}\,\frac{c_R(1-c_L)\mu_L+\frac{1}{3}(1-3c_R)\mu_R}{1-3c_Lc_R}~,
\label{137}
\end{align}
and
\be
j_{1,2}(-\delta)=j_{1,2}(\delta)\!\left\vert_{\begin{subarray}{c}\mu_L\leftrightarrow\mu_R\\c_L\leftrightarrow c_R\end{subarray}}\right.~,
\label{138}
\ee
with $\Delta_{cL,R}^\textrm{eff}$ expressed in terms of $c_{L,R}$ by Eq.~(\ref{57}) with a substitution of $c_{L,R}$ for $c$. The conductance $G$ then reads
\be
G=\frac{2}{3}\,\frac{1+\sqrt{\chi_L\chi_R}}{1-\sqrt{\chi_L\chi_R}}~,
\label{139}
\ee
where $\chi_{L,R}$ is given by Eq.~(\ref{128}) with $\Delta_c^\textrm{eff}$ substituted by $\Delta_{cL,R}^\textrm{eff}$, or, explicitly in terms of $\Delta_{cL,R}^\textrm{eff}$:
\be
\!\!G=\frac{2}{3}\,\frac{\Delta_{cL}^\textrm{eff}+\Delta_{cR}^\textrm{eff}}{1+\Delta_{cL}^\textrm{eff}\Delta_{cR}^\textrm{eff}
-\sqrt{\left[\left(\Delta_{cL}^\textrm{eff}\right)^2-1\right]\!\left[\left(\Delta_{cR}^\textrm{eff}\right)^2-1\right]}}~.
\label{140}
\ee
For $\Delta_{cR}^\textrm{eff}=\Delta_{cL}^\textrm{eff}$, Eq.~(\ref{140}) reduces to Eq.~(\ref{56}). Note that $G=2/3$ if one of the parameters $\Delta_{cR}^\textrm{eff}$ and $\Delta_{cL}^\textrm{eff}$ is equal to 1 (strong backscattering beneath the contact), irrespective of the value of the other.

It is worth briefly discussing, in connection with the setup asymmetry, a question of whether the notion of a ``contact resistance" is generically applicable to QH edges. Namely, the question is if the two-terminal resistance in the clean case $R_c$---which is only nonzero because of dissipation in the contacts---is representable as a sum of two terms, one of which, $R_{cL}$, depends only on the properties of the left contact and the other, $R_{cR}$, only on the properties of the right contact. For a single-channel QH edge, or a clean LL quantum wire for that matter, $R_c=R_{cL}+R_{cR}$ is trivially the case, with each of the terms being the contact resistance for one of the contacts. As follows from Eq.~(\ref{139}), this is not so for the $\nu=2/3$ edge.

To exemplify the Landauer-type picture for a QH edge with counterpropagating channels, $G$ from Eq.~(\ref{139}) can be represented for arbitrary $\chi_{L,R}$ as a sum of two ``nonchiral" terms: $G=G_1+G_2$, where $G_1$ is the contribution of the upper and lower parts of channel 1, and $G_2$ of the upper and lower parts of channel 1/3 (recall that the strength of interchannel interaction drops out from $G$). Specifically, $G_{1,2}=[j_{1,2}(\delta)-j_{1,2}(-\delta)]/(V_L-V_R)$, with $j_{1,2}(\pm\delta)$ from Eqs.~(\ref{136})-(\ref{138}). From this representation, neither the total resistance $1/G$ nor the ``per channel" resistance $1/G_{1,2}$ splits up into a sum of two distinct contact resistances---because of scattering between the nonequivalent channels at the contacts.

The idea of a contact resistance in the above sense (``per contact") is thus not applicable to a QH edge with counterpropagating channels. In fact, this is also generically true for an edge with copropagating channels, which contribute in parallel to the total resistance regardless of their mutual chirality. However, an important difference between the edges with counterpropagating and copropagating channels is that a long line-junction contact in the latter case equilibrates channels with each other \cite{Kane1995}. As such, it is characterizable by a universal contact resistance, irrespective of possibly different microscopic details of electron scattering beneath the right and left contacts. Note also that, fundamentally, it is chirality of the edge that makes it impossible to represent $R_c$ as $R_{cL}+R_{cR}$ for $G$ from Eq.~(\ref{139}). If the two counterpropagating channels were equivalent [$g_+=g_-$ in Eqs.~(\ref{3}) and (\ref{4})], thus forming a ``nonchiral edge" (a LL ring), the long line-junction contacts would be characterizable by (generically nonuniversal) contact resistances $R_{cL,R}$ with $R_c=R_{cL}+R_{cR}$ (see Appendix).

Turning to the case of a disordered edge, Eq.~(\ref{139}) generalizes to
\be
\!\!G=\frac{2}{3}\,\frac{1-e^{-\gamma_uL_u-\gamma_lL_l}\chi_L\chi_R}{\left(1-e^{-\gamma_uL_u}\sqrt{\chi_L\chi_R}\right)\!\left(1-e^{-\gamma_lL_l}\sqrt{\chi_L\chi_R}\right)}~,
\label{141}
\ee
where, apart from possibly different $\Delta_{cL}^\textrm{eff}$ and $\Delta_{cR}^\textrm{eff}$, we also assumed, for the sake of generality, different scattering rates $\gamma_u$ and $\gamma_l$ in the upper and lower parts of the edge, respectively. These enter Eq.~(\ref{140}) in the combinations $\gamma_uL_u$ and $\gamma_lL_l$, where $L_{u,l}$ are the distances between the contacts along the upper and lower segments of the edge. Importantly, the cancellation of the strength of interchannel interaction (parameter $\Delta$) outside the contacts occurs separately for the total current in each segment of the edge between them. That is, the strength of interaction may also be different in the upper and lower parts of the edge, with no effect on $G$ from Eq.~(\ref{140}), apart from affecting $\gamma_{u,l}$ [Eq.~(\ref{123})]. Note that vanishing of one of the parameters $\chi_L$ or $\chi_R$ yields $G=2/3$, similarly to the clean case, irrespective of the strength of disorder.

It is instructive to represent Eq.~(\ref{141}) as
\be
G=G^c_++G^c_--\left(G^b_u+G^b_l\right)~,
\label{142}
\ee
where
\be
G^c_+=\frac{2}{3}\,\frac{1}{1-\sqrt{\chi_L\chi_R}}~,\quad G^c_-=\frac{2}{3}\,\frac{\sqrt{\chi_L\chi_R}}{1-\sqrt{\chi_L\chi_R}}~,
\label{143}
\ee
and
\be
G^b_{u,l}=\frac{2}{3}\,\frac{\sqrt{\chi_L\chi_R}}{1-\sqrt{\chi_L\chi_R}}\,\frac{1-e^{-\gamma_{u,l}L_{u,l}}}{1-e^{-\gamma_{u,l}L_{u,l}}\sqrt{\chi_L\chi_R}}~.
\label{144}
\ee
The four terms in Eq.~(\ref{142}) relate $j_{u,l}$ to $\mu_{L,R}$ by
\be
\left(\begin{array}{c}
j_u \\
-j_l
\end{array}\right)=\left(\hat{G}^c-\hat{G}^b\right)\left(\begin{array}{r}
\mu_L \\
-\mu_R
\end{array}\right)
\label{145}
\ee
(with the total current between the contacts $j=j_u-j_l$), where the conductance matrices $G^{c,b}$ read
\be
\hat{G}^c=\left(\begin{array}{ll}
G^c_+ & G^c_- \\
G^c_- & G^c_+
\end{array}\right)~,\quad \hat{G}^b=\left(\begin{array}{ll}
G^b_u & G^b_u \\
G^b_l & G^b_l
\end{array}\right)~.
\label{146}
\ee
The matrix $\hat{G}^c$ characterizes the contacts, with the sum $G^c_++G^c_-$ giving the two-terminal conductance $G$ of a clean edge from Eq.~(\ref{139}). The other matrix, $\hat{G}^b$, describes disorder-induced interchannel scattering in the bulk of the edge. Note that the entries of $\hat{G}^b$ are the same for a given row, i.e., the backscattering-induced current in either segment of the edge (upper or lower) is separately gauge invariant (proportional to $\mu_L-\mu_R$).

Related to the above discussion of the contact resistance for a multichannel edge, another subtle question concerns the notion of a disorder-induced ``bulk" resistance $R_b$, defined---in the conventional sense---by representing the total resistance $R$ as a sum $R=R_c+R_b$, where $R_c$ characterizes dissipation in the contacts. For $R$ equal to $1/G$ from Eq.~(\ref{142}) and $R_c$ to $1/G=1/(G^c_++G^c_-)$ from Eq.~(\ref{139}), $R_b$ is given by
\be
R_b=R_c\frac{G^b_u+G^b_l}{R_c^{-1}-\left(G^b_u+G^b_l\right)}~.
\label{147}
\ee
When using the term ``bulk resistance" for $R_b=R-R_c$, the notion in the back of our minds is that $R_b$ depends only on the scattering processes in the bulk, but not at or inside the contacts. By definition, $R_b$ indeed vanishes to zero for $\gamma_{u,l}\to 0$. However, as straightforwardly follows from Eq.~(\ref{147}), $R_b$ for the $\nu=2/3$ edge is a function of $\chi_{L,R}$, i.e., explicitly depends on the properties of the contacts.

\begin{figure}[h]
\centering
\includegraphics[width=0.95\columnwidth]{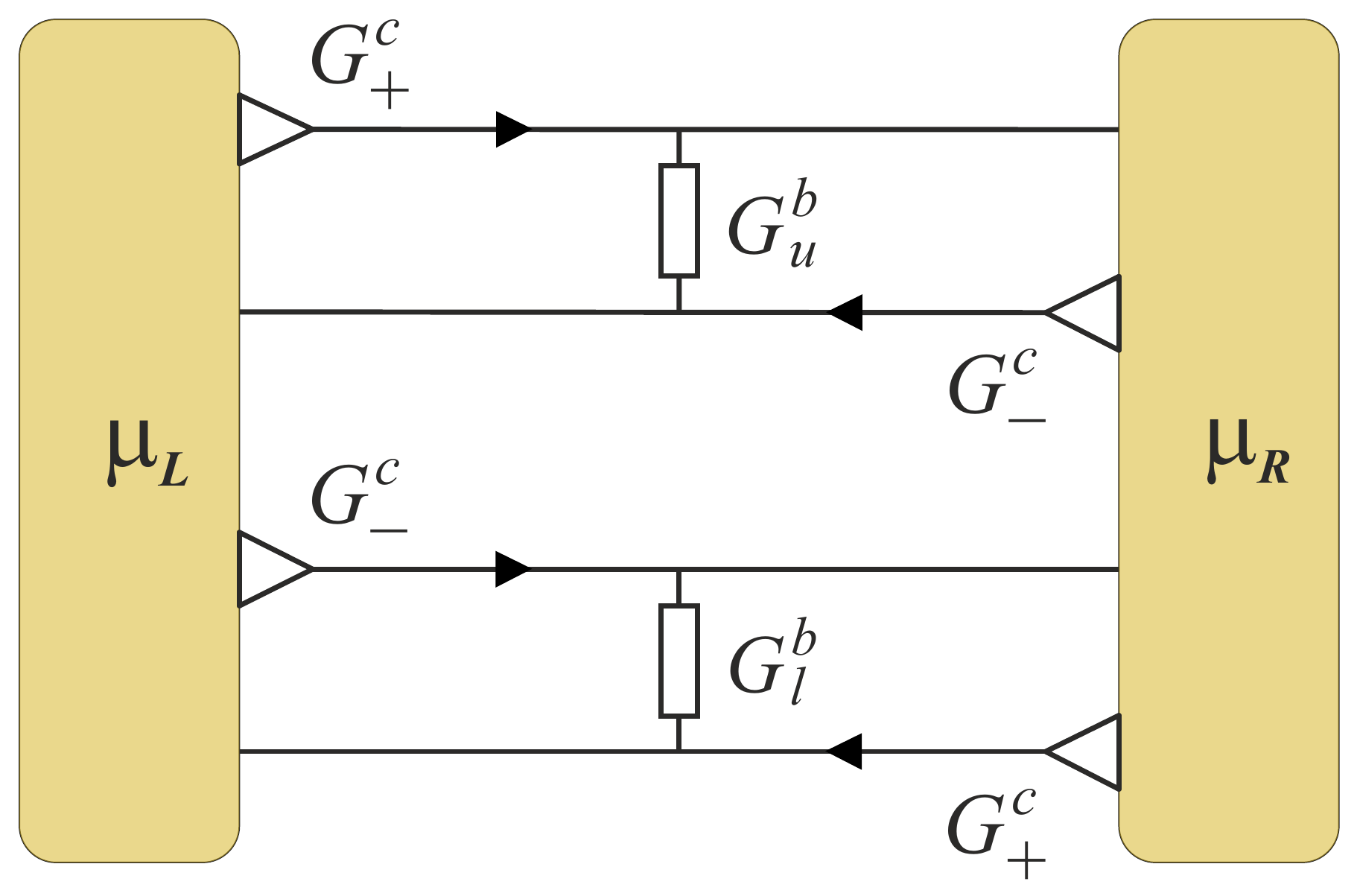}
\caption{Effective two-terminal electrical circuit for a disordered $\nu=2/3$ edge. The triangles denote the conductances $G^c_\pm$ that give the chiral currents emitted by the contacts, the rectangles stand for the resistances $1/G^b_{u,l}$ that shunt the chiral currents in the upper and lower parts of the edge [Eq.~(\ref{145})].}
\label{f10}
\end{figure}

Equation (\ref{145}) corresponds, effectively, to the electrical circuit in Fig.~\ref{f10}. The conductances $G^b_u$ and $G^b_l$ backscatter the current in the upper and lower parts of the edge, respectively, with the conductances $G^c_\pm$ determining the chiral currents in the absence of backscattering. Similarly to the ``nonsplitting" of $R_c$ into a sum of two contact resistances, as discussed above, the dependence of the bulk resistance $R_b=R-R_c$ on the properties of contacts is a direct consequence of chirality of the edge (see Appendix). Note also that $R_b$ is finite in the limit of large $\gamma_uL_u$ and $\gamma_lL_l$ [namely, for the $\nu=2/3$ edge, $R_b\to 3\sqrt{\chi_L\chi_R}/(1+\sqrt{\chi_L\chi_R})$ and $G=1/(R_b+R_c)\to 2/3$], in contrast to the nonchiral case, where $R_b$ diverges in this limit.

\section{Conclusion}
\label{s5}

In closing, let us recapitulate briefly on the main points of our analysis. We have given a detailed discussion of how charge transport in an edge with counterpropagating channels, exemplified by the $\nu=2/3$ edge, is affected by interchannel interaction, both beneath the contacts and outside them. One conceptually important aspect of multichannel transport that we have highlighted is that the conductance does not explicitly depend on the interaction strength either beneath the side-attached contacts or in the bulk of the edge. This is true for both a clean and a disordered edge, apart from the interaction-induced renormalization of the scattering rates beneath and, in the disordered case, outside the contacts. The screening properties of the side-attached contact to the chiral edge, which can vary depending on the mutual position of the contact and the edge, thus do not show up in the conductance directly.

If this were the whole story, it might be viewed as a generalization of the conventional train of thought about transport in the LL model to the chiral case. However, we have also specifically studied long line-junction contacts to the edge and shown that these are characterized by an effective interaction strength, which depends on the specifics of scattering beneath them. This effective interaction does affect the conductance. Another way to formulate it is that such a contact is at thermal equilibrium with a certain set of edge modes, which are generically not the eigenmodes either beneath or outside it and with respect to which it is ideal. Encoding the peculiarities of a given contact into a single parameter---the effective strength of interaction---allows for a universal characterization (also at the phenomenological level) for a set of contacts to the edge. This discussion may have many points of contact with experiments on multichannel edges, particularly because, apparently, often not much is known with sufficient certainty about the microscopic structure of the contacts to the edge.

We have presented a theory of fractionalization-renormalized tunneling between edge channels. The fractionalization framework is, in fact, crucial for the understanding of the effect of disorder on the edge eigenmodes---for an arbitrary strength of interchannel interaction. We have demonstrated how fractionalization determines the outcome of charge scattering off a single tunneling link and used the solution of the scattering problem to describe transport through the disordered edge. Particular attention has been paid to charge equilibration between the channels, which establishes the universal conductance quantization, for the line-junction contacts characterized by the effective---possibly different for different contacts---interaction strength. On a separate but related note, we have discussed the inapplicability---for the chiral edge with counterpropagating channels---of the concept, conventionally used to describe the two-terminal resistance, that represents the resistance as a sum of two contact resistances and the bulk resistance.

\acknowledgments

We thank M. Heiblum, A.D. Mirlin, and J. Park for discussions. This work was supported by DFG Grants Nos.\ MI 658/10-1 (CS and YG) and 658/10-2 (YG), by German-Israeli Foundation Grant No.\ I-1505-303.10/2019 (YG and IG), by CRC 183 of the DFG (project C01) (YG), by the Minerva Foundation (YG), by the Italia-Israel QUANTRA grant (YG), by the Helmholtz International Fellow Award (YG), and by the Russian Foundation for Basic Research (RFBR), Grant No.\ 18-02-01016 (IG).

\appendix*

\section{COMPARISON TO THE NONCHIRAL MODEL}
\label{a}

The purpose of Appendix is to concisely discuss the differences and similarities between the QH edge with two counterpropagating channels, drawing on the example of the considered model of the $\nu=2/3$ edge [Eqs.~(\ref{1}) and (\ref{2})], and the nonchiral LL, primarily in two respects: (i) the line-junction model of the contacts and (ii) fractionalization-renormalized tunneling. The framework for this discussion is provided in the main text, so that we mostly only need to highlight the points at which the changes are made.

The nonchiral LL model is defined in terms of the charge densities by Eqs.~(\ref{1}) and (\ref{2}) with $\delta\nu_1=-\delta\nu_2=1$ and $3v_2\to v_2$ (for $v_{1,2}$ being the channel velocities in the absence of interchannel interaction, generically different), or, equivalently, by Eqs.~(\ref{3}) and (\ref{4}) with
\be
g_+=g_-=K~,
\label{a1}
\ee
where $K=\sqrt{(1-\alpha)/(1+\alpha)}$, with
\be
\alpha=2v_{12}/(v_1+v_2)~,
\label{a2}
\ee
is the Luttinger constant. The equality $g_+=g_-$ signifies ``nonchirality." The eigenmode velocities are given by Eq.~(\ref{7}) with $\alpha$ from Eq.~(\ref{a2}). The stability condition reads $v_{12}^2\leq v_1v_2$. For repulsive interaction, $K$ varies in the interval
\be
|\sqrt{v_1}-\sqrt{v_2}|/(\sqrt{v_1}+\sqrt{v_2})\leq K\leq 1
\label{a3}
\ee
within the stability domain. In contrast to the $\nu=2/3$ model from Sec.~\ref{s2a}, $g_\pm$ from Eq.~(\ref{a1}) are single-valued functions of $\alpha$.

For the nonchiral case, the matrix $\hat\lambda$ [Eq.~(\ref{20})], which fixes the generalized boundary conditions (\ref{22}) and (\ref{23}), has a simple form:
\begin{align}
\hat\lambda&=\frac{1}{2K_c}\left(\begin{array}{ll}
K_c+K & \,K_c-K \\
K_c-K & \,K_c+K
\end{array}\right)~,
\label{a4}
\end{align}
where $K_c$ is the Luttinger constant that characterizes the modes with which the thermal reservoir is at equilibrium. For this boundary condition, the two-terminal conductance $G$ in the clean case is given by
\be
G=K_c
\label{a5}
\ee
(assuming identical contacts). The value of $G$ in Eq.~(\ref{a5}) is fully determined by the strength of interaction between the modes that are at equilibrium with the contacts, similarly to Eq.~(\ref{32}) for the $\nu=2/3$ edge.

\subsection{Line-junction contact}

For the LL case, the line-junction model of a contact is described by the equation of motion for the currents $j_1=v_1n_1+v_{12}n_2$ and $j_2=-(v_2n_2+v_{12}n_1)$ beneath the contact ($|x|<\delta$):
\begin{align}
&\partial_xj_1+\frac{1}{2}\gamma_c j-\gamma_1\left(\frac{\mu}{2\pi}-j_1\right)=0~,
\label{a6}\\
&\partial_xj_2-\frac{1}{2}\gamma_c j-\gamma_2\left(\frac{\mu}{2\pi}+j_2\right)=0~,
\label{a7}
\end{align}
where $j=j_1+j_2$. The meaning of the inverse scattering lengths $2\gamma_c^{-1}$ and $\gamma_{1,2}^{-1}$ is similar to that for Eqs.~(\ref{36}) and (\ref{37}). An essential difference compared to the $\nu=2/3$ case is that the backscattering term, proportional to $\gamma_c$, relaxes in Eqs.~(\ref{a6}) and (\ref{a7}) the total current.

The analog of Eqs.~(\ref{40})-(\ref{42}) simplifies to
\be
\hat\upgamma_c=\frac{1}{2}\left(\begin{array}{cc}
\gamma_c+2\gamma_1 & \gamma_c \\
-\gamma_c & -\gamma_c-2\gamma_2
\end{array}\right)~,
\label{a8}
\ee
with the eigenvalues $k_{1,2}$ of $\hat\upgamma_c$ written as
\be
k_{1,2}=\frac{\gamma_1-\gamma_2}{2}\pm \frac{Q}{2}~,
\label{a9}
\ee
where
\begin{align}
Q=(\gamma_1+\gamma_2)\left(\,1+\frac{2\gamma_c}{\gamma_1+\gamma_2}\,\right)^{1/2}~.
\label{a10}
\end{align}
The leaky current-scattering matrix $\hat{\textrm{W}}$, which obeys Eq.~(\ref{43}) and defines the current through the contact by Eq.~(\ref{45}), now reads
\begin{align}
\hat{\textrm{W}}=\,\,\frac{1}{Q \cosh(Q\delta)+\left(\gamma_1+\gamma_2+\gamma_c\right)\sinh(Q\delta)}&\nonumber\\
\times\left[\begin{array}{cc}
Qe^{-(\gamma_1-\gamma_2)\delta} & \gamma_c \sinh(Q\delta)\\
\gamma_c\sinh(Q\delta) & Qe^{(\gamma_1-\gamma_2)\delta}
\end{array}\right]&~.
\label{a11}
\end{align}
Similarly to Eqs.~(\ref{49}) and (\ref{50}), $\hat{\textrm{W}}$ is diagonal in the limit of $\gamma_c\to 0$:
\be
\hat{\textrm{W}}\rightarrow\left(\begin{array}{cc}
e^{-2\gamma_1 \delta} & 0 \\
0 & e^{-2\gamma_2 \delta}
\end{array}\right)~,
\label{a12}
\ee
and purely nondiagonal in the limit of $\delta\to\infty$:
\be
\hat{\textrm{W}}\to c\left(\begin{array}{cc}
0 & 1 \\
1 & 0
\end{array}\right)~,
\label{a13}
\ee
where the single parameter $c$ characterizing the long contact is given by
\be
c=\frac{1}{1+\zeta+\sqrt{\zeta^2+2\zeta}}
\label{a14}
\ee
with
\be
\zeta=\frac{\gamma_1+\gamma_2}{\gamma_c}~.
\label{a15}
\ee
For the nonchiral case, $c$ varies within the interval $0\leq c\leq 1$, differently from $0\leq c\leq 1/3$ in Eq.~(\ref{51}).

For two long line-junction contacts [Eq.~(\ref{a13})] at the chemical potentials $\mu_L$ and $\mu_R$, the currents $j_{1,2}$ at the left contact are then obtained, for the clean case, as
\begin{align}
j_1(\delta)&=-j_2(-\delta)=\frac{1}{2\pi}\,\frac{\mu_L+c\mu_R}{1+c}~,
\label{a16}\\
j_2(\delta)&=-j_1(-\delta)=-\frac{1}{2\pi}\,\frac{c\mu_L+\mu_R}{1+c}
\label{a17}
\end{align}
(assuming that both contacts are characterized by the same parameter $c$). In contrast to Eqs.~(\ref{53}) and (\ref{54}) for the $\nu=2/3$ edge, the currents in Eqs.~(\ref{a16}) and (\ref{a17}) obey $j_{1,2}(\pm\delta)=-j_{2,1}(\mp\delta)$ [which is a symmetry relation additional to that in Eq.~(\ref{55})]. As a consequence, if we were to consider two-terminal transport through a closed loop made up of the nonchiral edge [as in Fig.~\ref{f3a}], the conductance $G$ would simply be doubled compared to transport through a single segment of the edge between the contacts. Below, $G$ will be understood as the conductance of the single segment of the edge between the long $(\delta\to\infty)$ line-junction contacts. Short contacts side-attached to the LL quantum wire (or the nonchiral edge for that matter) should be described by Eq.~(\ref{a11}).

In similarity to Eq.~(\ref{56}), the conductance $G=2\pi[j_1(\delta)+j_2(\delta)]/(\mu_L-\mu_R)$ is given by [cf.\ Eq.~(\ref{a5})]
\be
G=K_c^\textrm{eff}~,
\label{a18}
\ee
where the effective Luttinger constant characterizing the (identical) contacts
\be
K_c^\textrm{eff}=\frac{1-c}{1+c}~.
\label{a19}
\ee
Again, similarly to the $\nu=2/3$ edge, the line-junction contact with $K_c^\textrm{eff}=K$ is the incarnation of the notion of an ideal contact, now---for the nonchiral system. Importantly, this is so irrespective of the strength of interaction beneath the contact.

If the left and right contacts are characterized by different constants $K^\textrm{eff}_{cL}$ and $K^\textrm{eff}_{cR}$, respectively, $G$ is obtained as
\be
G=\frac{2\,K^\textrm{eff}_{cL}\,K^\textrm{eff}_{cR}}{K^\textrm{eff}_{cL}+K^\textrm{eff}_{cR}}~.
\label{a20}
\ee
Note that the structure of Eq.~(\ref{a20}) signifies the applicability of the notion of a contact resistance, in contrast to Eq.~(\ref{139}) for the $\nu=2/3$ edge. Indeed, the resistance $1/G$, with $G$ from Eq.~(\ref{a20}), is a sum of two contact resistances $1/2K^\textrm{eff}_{cL}$ and $1/2K^\textrm{eff}_{cR}$, each of which depends only on the properties of the corresponding contact.

\subsection{Fractionalization-renormalized tunneling}

We now turn to electrostatics of tunneling in the nonchiral case. From Eq.~(\ref{a4}), the inverse of the matrix $\hat\Lambda$, defined analogously to that in Eq.~(\ref{59}) for the $\nu=2/3$ edge, is now symmetric and reads:
\be
\hat{\Lambda}^{-1}=\frac{1}{2K}
\left(\begin{array}{rr}
1+K & \,-1+K\\
-1+K & \,1+K
\end{array}\right)~.
\label{a21}
\ee
The screening coefficients $\eta_\pm$ [Eqs.~(\ref{60}) and (\ref{61})] are, in contrast to the chiral case, the same in both eigenmodes:
\be
\eta_\pm=-\frac{1-K}{1+K}~.
\label{a22}
\ee
If $v_1=v_2$, the limit of $K\to 0$, where screening is perfect $(\eta_\pm\to -1)$, is reachable with increasing strength of interaction [Eq.~(\ref{a3})]. Either of the eigenmodes is then ``neutral," analogously to the $-$ mode for the $\nu=2/3$ edge at $\Delta\to 1$.

The matrices $\hat q$ and $\hat{\bar q}$, which describe charge fractionalization, resolved with respect to both chirality and channels, upon insertion of a unit charge into the channels $n_1$ and $n_2$, respectively, are written as
\begin{align}
\hat{q}=\frac{1}{4K}\left[\begin{array}{rr}
(1+K)^2 & \,-1+K^2 \\
-(1-K)^2 & 1-K^2
\end{array}\right]
\label{a23}
\end{align}
and
\begin{align}
\,\,\,\hat{\bar q}=\frac{1}{4K}\left[\begin{array}{rr}
1-K^2 & \,-(1-K)^2 \\
-1+K^2 & (1+K)^2
\end{array}\right]~.
\label{a24}
\end{align}
Note that both $q_{\pm,1}=\bar{q}_{\mp,2}$ and $q_{\pm,2}=\bar{q}_{\mp,1}$ in Eqs.~(\ref{a23}) and (\ref{a24}), in contrast to Eqs.~(\ref{67}) and (\ref{68}) for the chiral case, where only the former is true.

Similarly to the partial (for $\Delta\neq 1$) cancellation, in the $\nu=2/3$ edge, of the screening charge $\bar{q}_{+,1}$ for the fractionalized electron tunneled to channel 2 and the charge $-q_{+,1}$ of the fractionalized hole left behind in channel 1, the factor $q_{+,1}-\bar{q}_{+,1}$ [Eq.~(\ref{77})] with $\hat{q}$ and $\hat{\bar q}$ from Eqs.~(\ref{a23}) and (\ref{a24}) decreases with increasing strength of repulsive interaction. Specifically,
\be
q_{+,1}-\bar{q}_{+,1}=\bar{q}_{-,2}-q_{-,2}=\frac{1+K}{2}~,
\label{a25}
\ee
where we also took into account that, in the nonchiral case, the first factors in Eqs.~(\ref{77}) and (\ref{78}) are the same. In contrast to the case of $\Delta=1$ for the $\nu=2/3$ edge, however, the mutual compensation of the two terms in the factor $q_{+,1}-\bar{q}_{+,1}$ is never exact. The second factors in Eqs.~(\ref{77}) and (\ref{78}), with $W_{1,2}^\pm$ from Eqs.~(\ref{82}) and (\ref{83}), are also the same in the system obeying chiral symmetry, as are, altogether, the reflection coefficients from the left and from the right:
\be
R_+=R_-=|\tilde{t}_0|^2\frac{K}{v_+v_-}~.
\label{a26}
\ee
The equality $R_+=R_-$ is a direct consequence of $g_+=g_-$ [cf.\ Eq.~(\ref{85})]. Within the picture of fractionalization-renormalized tunneling, fractionalization is represented by the factor of $K$ in Eq.~(\ref{a26}). As already noted below Eq.~(\ref{85}), the factors $|\tilde{t}_0|^2$ in the chiral and nonchiral cases, i.e., in Eqs.~(\ref{84}) and (\ref{a26}), respectively, are inherently related to each other.

The equation of motion for the densities $n_\pm$ in the nonchiral case is Eq.~(\ref{99}) with $\Gamma_\pm$ from Eq.~(\ref{101}), where $R_\pm$ are given by Eq.~(\ref{a26}). Similarly to Eq.~(\ref{105}), the collision terms $\bar{I}_t$ and $I_t$ in the equations of motion for $n_{1,2}$ and $n_\pm$, respectively, are proportional to each other. The relation between them now reads
\be
\bar{I}_t=\frac{1}{K}I_t~.
\label{a27}
\ee
The collision integral describes relaxation of the total current [cf.\ Eq.~(\ref{108})]:
\be
\bar{I}_t=\gamma_0j~,
\label{a28}
\ee
with $\gamma_0$ from Eq.~(\ref{107}).
The partial scattering rates $\Gamma_{1,2}$, which define $\bar{I}_t$ in terms of the density relaxation [Eq.~(\ref{104})], are written as
\be
\Gamma_{1,2}=\gamma_0(v_{1,2}-v_{12})~,
\label{a29}
\ee
Note that Eq.~(\ref{a29}) satisfies the general constraint on the relation between $\Gamma_1$ and $\Gamma_2$ from Eq.~(\ref{114}).

One of the relaxation rates $\Gamma_{1,2}$ may become negative,
\be
\Gamma_1/\Gamma_2<0~,
\label{a30}
\ee
for sufficiently strong interaction, similarly to the $\nu=2/3$ edge (Sec.~\ref{s4a}), but, in contrast to the chiral case, only if $v_1\neq v_2$. Specifically, the condition for the inequality (\ref{a30}) to hold is
\be
\min\{v_1,v_2\}<v_{12}\leq\sqrt{v_1v_2}~,
\label{a31}
\ee
where the upper limit is, as mentioned above, the instability threshold for the LL model. Similarly to the case of the $\nu=2/3$ edge, this does not violate causality, which only requires that $\Gamma=\Gamma_1+\Gamma_2\geq 0$ [Eq.~(\ref{111})] be satisfied.

For the nonchiral case, both roots of the characteristic equation to Eq.~(\ref{99}) in the dc limit are zero, in contrast to Eq.~(\ref{123}), as a result of which the two-terminal resistance $1/G$ is a linear function of $L$ [cf.\ Eq.~(\ref{141})]:
\be
\frac{1}{G}=\frac{1}{2K_{cL}^\textrm{eff}}+\frac{1}{2K_{cR}^\textrm{eff}}+\gamma_0L~.
\label{a32}
\ee
The disorder-induced contribution to $1/G$ in Eq.~(\ref{a32}) trivially does not depend on the properties of the contacts, in stark contrast to the $\nu=2/3$ edge, as discussed at the end of Sec.~\ref{s4c2}. In similarity to the chiral case, $G$ from Eq.~(\ref{a32}) does not explicitly depend on the strength of interaction outside the contacts, apart from the renormalization of $\gamma_0$. The resistivity $\gamma_0$ in Eq.~(\ref{a32}) corresponds to charge diffusion with the diffusion coefficient $v_+v_-/\Gamma$ [Eq.~(\ref{119}) with $v_\rho=0$].

\bibliography{qhe_contact_subm}

\begin{thebibliography}{106}
\bibitem [{rem()}]{remark1}%
  \BibitemOpen
  \href@noop {} {}\bibinfo {note} {Note that $v_1$, $v_2$, $v_{12}$ are treated
  within the model (\ref{1}) as independent parameters, while they should be
  all determined by the interplay of confinement and electrostatics in more
  specific ``microscopic" approaches (which is, however, beyond the scope of
  this paper).}\BibitemShut {Stop}%
\bibitem [{che()}]{chem-electrochem}%
  \BibitemOpen
  \href@noop {} {}\bibinfo {note} {It is important to distinguish the chemical
  potential difference for electrons inside the edge $\mu_1-\mu_2$ and the
  electrochemical potential difference for them. The latter is given by the
  chemical potential difference for the eigenmodes $\mu_+-\mu_-$, where
  $\mu_\pm=\pm 2\pi j_\pm/g_\pm$. The two are linearly related to each other as
  $\mu_1-\mu_2=\sqrt{(\Delta^2-1)/3}\,(\mu_+-\mu_-)$. This is analogous to the
  relation $\mu_1-\mu_2=K(\mu_+-\mu_-)$ for the nonchiral LL model, where $K$
  is the Luttinger constant.}\BibitemShut {Stop}%
\bibitem [{asy()}]{asymptW}%
  \BibitemOpen
  \href@noop {} {}\bibinfo {note} {More specifically, the limit of large
  $\delta$, in which $\hat{\textrm W}$ is given by Eq.~(\ref{50}), is:
  $(\min\{\gamma_1,\gamma_2\}+\gamma_c)\delta\gg\ln\zeta$ for $\zeta\gg 1$
  (weak backscattering) and $\max\{\gamma_1,\gamma_2\}\delta\gg 1$ for
  $\zeta\alt 1$ (strong backscattering).}\BibitemShut {Stop}%
\bibitem [{sin()}]{sing}%
  \BibitemOpen
  \href@noop {} {}\bibinfo {note} {Because of the vanishing of $g_-$, $j_-$,
  $\lambda_{-+}$, and $\lambda_{--}$ at $\Delta=1$, care should be taken when
  solving Eqs.~(\ref{22}) and (\ref{23}) for $j_+$ at $\Delta=1$---by first
  assuming $\Delta-1$ finite.}\BibitemShut {Stop}%
\bibitem [{fou()}]{fourterm}%
  \BibitemOpen
  \href@noop {} {}\bibinfo {note} {In the four-terminal configuration with both
  voltage probes on the same side of the Hall bar, the source-drain conductance
  is also equal to 2/3 in the limit of full charge equilibration between the
  terminals, whereas the voltage between the voltage probes vanishes in this
  limit.}\BibitemShut {Stop}%
\end{thebibliography}
\end{document}